\documentclass[reprint,onecolumn,aip]{revtex4-1}

\usepackage{amsmath,amssymb,graphicx,suffix,rotating,pstool,subcaption, fullpage}

\usepackage{epstopdf}
\usepackage{tikz}
\usepackage{pgf,pgfplots}
\usepackage{pgffor}
\usepgfmodule{shapes}
\usepgfmodule{plot}
\usetikzlibrary{decorations}
\usetikzlibrary{arrows,through}
\usetikzlibrary{decorations.markings}
\tikzset{->-/.style={decoration={
  markings,
  mark=at position #1 with {\arrow{stealth}}},postaction={decorate}}}

\usepackage[]{hyperref}
\hypersetup{pdfborder={0 0 0}}

\usepackage{breakurl}

\usepackage[notref,notcite,final]{showkeys}

\numberwithin{equation}{section}

\usepackage[noabbrev,capitalise]{cleveref}

\newcommand{\axislabels}{\node [rotate=0,align=center] at (0.05,0.7) {\footnotesize $H(y)$\\[-.4em] \footnotesize $+z$};\node at (0.7,0.02) {\footnotesize $y$};}

\newcommand{\threewide}{0.31}
\newcommand{\twowide}{0.4}
\newcommand{\onewide}{0.5}

\renewcommand{\vec}[1]{\mathbf{#1}}
\renewcommand{\d}{\mathrm d}
\newcommand{\bv}[1]{\hat{\mathrm{\mathbf{#1}}}} 
\newcommand{\dd}[2]{\frac{\mathrm d #1}{\mathrm d #2}}

\newcommand{\pd}[2]{\frac{\mathrm \partial #1}{\mathrm \partial #2}}

\renewcommand{\Re}{\ensuremath{\mathrm{Re}}} 
\newcommand{\Fr}{\ensuremath{\mathrm{Fr}}} 

\newcommand{\grad}{\boldsymbol{\nabla}}

\newcommand{\ys}[1]{#1}
\newcommand{\bl}[1]{#1}

\begin{document}

\title{Particle-laden thin-film flow in helical channels with arbitrary shallow cross-sectional shape}

\author{D. J. Arnold}\email{darnold@math.ucla.edu}
\affiliation{Department of Mathematics, University of California Los Angeles, Los Angeles, CA 90095, USA}
\affiliation{School of Mathematical Sciences, The University of Adelaide, Adelaide, SA 5005, Australia}
\author{Y. M. Stokes}\email{yvonne.stokes@maths.adelaide.edu.au}
\author{J. E. F. Green}
\affiliation{School of Mathematical Sciences, The University of Adelaide, Adelaide, SA 5005, Australia}

\begin{abstract}
Particle-laden flows in helical channels are of interest for their applications in spiral particle separators used in the mining and mineral processing industries. In this paper, we extend the previous work of \citet{LSB2013} by studying thin-film flows of \ys{mono-disperse} particle-laden fluid in helically-wound channels of arbitrary centreline curvature and torsion, and arbitrary cross-sectional shape. In the case where the particles are uniformly distributed through the \ys{depth of the} film, significant analytic progress can be made: the governing equations reduce to a single nonlinear ordinary differential equation, which is readily integrated numerically to obtain the solution subject to appropriate boundary conditions. Motivated by possible application to the design of spiral separators, we consider the effects of changing the channel centreline geometry, the cross-sectional shape and the particle density on the resulting flows \ys{and the radial distribution of particles}. Our results support the findings in \citet{ASG2016} regarding the effect of channel centreline geometry and cross-sectional shape on flows in particle-free regions. In particle-rich regions, similar effects are seen, although the primary velocity is lower due to increased effective mixture viscosity. \ys{Of key interest, is the effect of channel geometry on the focusing of the particles, for given fluxes of fluid and particles. We find that introducing a trench into the channel cross-section, a feature often used in commercial spiral particle separators, leads to smaller radial width of the particle-rich region, i.e. sharper focusing of the particles, which is consistent with experiments showing that channel geometry influences particle separation in a spiral separator.} 
\end{abstract}

\maketitle

\section{Introduction}
\begin{figure}[t]
\centering
\begin{tikzpicture}
    \node[anchor=south west,inner sep=0] (image) at (0,0) {{\includegraphics[width=\onewide\textwidth,trim=65 0 0 40]{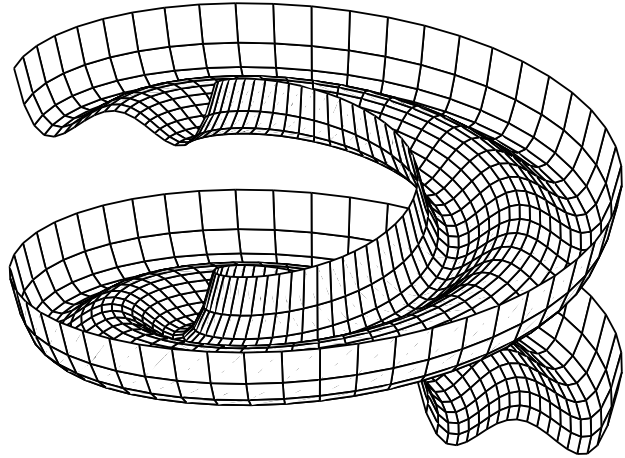}}};
	\begin{scope}[x={(image.south east)},y={(image.north west)}]
	\end{scope}
\end{tikzpicture}
\caption{A helical channel with quartic cross-sectional profile.}
\label{fig:channel}
\end{figure}

The work described in this paper is motivated by the aim of improving our understanding of the operation of spiral particle separators, devices used in the
mining and mineral-processing industries to separate ores and clean coal.\citep{MT2015,Michaud2016}
They consist of a helically-wound channel (as shown in \cref{fig:channel}), down which a slurry
of crushed rock/ore and water flows, driven by gravity. The complex secondary motion of the fluid,
along with the particles' inertia and tendency to settle under gravity, acts
to sort the particles across the channel width according to their density, assuming milling has produced particles of uniform size.
Thus, particles of different densities can be separated from each other by using vertical splitter plates at the end of the channel to 
split the flow at appropriate positions across its width. \ys{It is known that} the effectiveness of spiral particle separators is quite sensitive to the composition of
the input slurry, and small changes to the size, density, or concentration of
the particles can lead to dramatically impaired separation efficiency. \ys{Conversely, small changes to the separator geometry can greatly affect separation efficiency for a given feed slurry.} Often,
spiral particle separators operate in a regime very close to beaching, where
particles are deposited on the channel bottom at the edge of the region occupied by fluid, and thus fail to reach the end of the channel \citep{HB1995a}.

Spiral particle separators are an established technology; an 1899 US patent \citep{Pardee1899}
describes such a device. Yet they did not see
widespread adoption until the late 1940's and 1950's, with the Humphrey's
Spiral Concentrator \citep{Humphreys1943,RHHB1985}. Despite their widespread use, and experimental work undertaken by \citet{HB1975,HB1989} in the 1970s, 1980s and 1990s, mathematical modelling of the flow in spiral separators has been limited, and the design process remains largely experimental; \citet{HB1995} gives a summary of the process. However, experimentation with spiral separators has, in general, proven difficult. Estimates of the errors in the measurement of flow velocities can be as high as $30\%$. Nevertheless, a series of experiments \citep{Holtham1990,Holtham1992,HBH1991} showed that a complex secondary flow exists, that a fully-developed steady state flow profile appears relatively quickly (within 2--3 turns down the spiral), and that the fluid depth is typically very shallow (1--12mm) relative to the channel width (250--350mm). These last two observations in particular, suggest that, in mathematical models, the steady flow and thin film assumptions can be exploited, resulting in significant simplification of the governing equations. Recently, \citet{Boucher2014,Boucher2015} performed experiments using positron emission particle tracking to obtain more precise information about the motion of particles in spiral separators. Their results allow particles to be tracked in the radial and axial directions, but they do not track the vertical velocity of particles \ys{and/or fluid, nor give the fluid depth profile and the vertical particle distribution}. Improving mathematical modelling of flow in spiral separators has the potential to aid the design process by providing quantitative predictions of the separation performance of different channel configurations.

\ys{Single-phase} flows of Newtonian fluid \ys{(i.e. excluding particles)} in helically wound pipes 
have been extensively studied by \citet{Wang1981,Germano1982,Berger1983} and many more. In the case of \ys{an open helical channel} 
however, \ys{the flow has} a free-surface, which makes the problem more complex. Early progress came in the form of empirical formulae, notably from \citet{HBH1991}. They performed \ys{single-phase flow}
experiments on spiral particle separators and tried to explain the
results by using empirical equations based on a Manning equation near
the inner wall and a free-vortex in the main section of the flow. \Citet{Stokes01,Stokes01a} mathematically investigated flow in a helical channel with small centreline torsion
and curvature, and with a semi-circular cross-section \ys{filled} with fluid.
Following this, efforts were made to find solutions valid under the more general
conditions of non-small curvature and torsion,  shallow fluid depth, and with different channel
cross-sectional shapes. In particular, exploiting the thin-film approximation, which pertains to flows in spiral separators, has allowed
significant progress for rectangular channels \citep{SDW2004,SDW13}, although all of these
studies required some limits on the channel centreline, be it small torsion,
small curvature, or small centreline slope. Subsequent work \citep{ASG2015} showed that significant analytical progress was possible even when these assumptions were relaxed in the case of rectangular channels. Results for the more general case of an arbitrary channel cross-sectional shape have recently been obtained \citep{ASG2016}.

The mechanics of particle-laden fluids has been a topic of research for many years.
\Citet{Einstein1906} developed a leading order approximation for the increase
in effective viscosity of a dilute mixture of particles in a viscous fluid. The second order term was found in the 1970s by \citet{BG1972} and \citet{Batchelor1977}. For dilute suspensions, the particles
have a relatively minor effect on the bulk flow, but as the volume fraction of
particles increases, the particles have a larger effect. To describe
more concentrated suspensions, models that allow for two-way coupling
between the fluid and particle motion are required. Such models can be
split into two main categories: discrete and continuum. Discrete
models track individual particles, and are very computationally intensive
\citep{MFP1998}. Continuum models treat the fluid and particles
as two separate \ys{continuous} phases, with 
coupled equations for each. 

Two extensively-studied continuum models for suspension
flows are the diffusive-flux model introduced by \citet{LA1987} and  the suspension-balance model of \citet{NB1994}, \ys{which are discussed and compared in some detail by \citet{Murisicetal2013}} The diffusive-flux model gives a simple framework to study various phenomena acting on the particle
phase, such as Brownian diffusion, settling under the effect
of gravity, and shear-induced migration. The suspension-balance model is based
on rigorously averaging the mass, momentum, and energy balance equations, is
more general, and actually contains the diffusive-flux model \ys{as a special case} \citep{NB1994}. \ys{As shown by \citet{NB1994}, for 2D channel flow and pipe-flow the suspension-balance model predicts a plug-like region in the centre of the channel/pipe, in keeping with experimental observation, and overcomes the non-physical infinite viscosity and cusp in the particle volume fraction at the centreline predicted by the diffusive-flux model (which demands that the particle fraction reach the maximum packing fraction there due to zero shear). For other flows, in particular wide-gap Couette flow, the suspension-balance model reduces to the diffusive-flux model.} 
\citet{Bertozzi2011,Murisicetal2013} reported excellent \ys{agreement between experiments and a formulation of the diffusive-flux model similar to the well-known \citet{Phillips1992} model, 
for free-surface \ys{monodisperse particulate} flows on inclined planes (neglecting side boundaries), for which zero shear at the free surface does not necessarily correspond to the maximum particle volume fraction and there can be no cusp in the particle volume fraction.}  
More recently \citet{WB2016} extended the diffusive-flux model to bidensity suspensions \ys{on inclined planes}, finding good agreement with experimental results. \ys{The flows of interest in this present paper} 
differ from \ys{these studies of flow on inclined planes in just two respects:} 
\ys{the bottom of the channel is not necessarily flat and the channel} 
is curved rather than straight. Therefore, we expect that the diffusive-flux model should provide reasonable predictions for our work \ys{and, as a simpler particle-flux model for coupling with a more complex flow,  is worth investigation.}

The diffusive-flux model has been used in several papers \citep{CBH2008,Bertozzi2011,Murisicetal2013} to study particle laden thin films flowing down
straight inclined planes. \Citet{Bertozzi2011,Murisicetal2013} performed experiments, and
observed three distinct behaviours depending on the slope of the channel
and the particle volume fraction. The particles can settle to the bottom of the fluid, rise to the surface, or disperse evenly throughout the channel
depth. For shallow angles of inclination and low particle concentrations, the particles settle to the bed and the suspending liquid flows over the top of the particles with higher speed. For steeper angles of inclination and higher particle volume fractions, the particles collect near the surface of the film, flowing at higher velocity over the top of the suspending liquid. Between
the settled and ridged regimes is the well-mixed regime, where the particles
disperse uniformly throughout the film thickness, and the fluid and particle
velocities are similar. \ys{The diffusive-flux model performed extremely well in correctly predicting the flow behaviour; even where parameter values led to the maximum volume fraction being reached at the free surface (ridged regime) or at the channel bed (some, but not all, instances of the settled regime) the model predictions agreed closely with experiments.}

\Citet{LSB2013} used the diffusive-flux model in a study of \ys{monodisperse} particle-laden flows in helical channels of small slope and found solutions in the well-mixed regime. \ys{In this paper we} 
extend their work to channels of arbitrary centreline curvature and torsion, and arbitrary (but shallow) cross-sectional shape, and demonstrate a method for coupling particle transport with fluid flow in helically-wound channels. \ys{Although, unlike in the work of \citet{LSB2013}, our model does not dictate the well-mixed regime, we will require this as a simplifying assumption (to be discussed later), which precludes the particle volume fraction from reaching its maximum value and reinforces our expectation that the diffusive-flux particle-transport model is a reasonable choice for the present work. Then, motivated by spiral particle separators in which, assuming all particles in the flow are of the same size, the best separation of particles of different densities is achieved by the focusing of different particles into distinct non-overlapping horizontal bands, we use our model to investigate the effect of channel geometry on the horizontal distribution of particles and, in particular, the focusing of the particles for a given monodisperse feed, where sharper focusing is interpreted to mean a particulate region of smaller horizontal extent for the same total particle and fluid fluxes along the channel.
Thus, we aim to use mathematical modelling to gain insight into the critical geometrical parameters that control particle focusing and so increase the understanding needed for improved separation.}

\bl{Studies on the effects of varying channel cross-sectional shape on resulting flows have been performed for unsteady viscous gravity currents in straight channels, for example by \citet{HT2007} and \citet{HG2010}. Likewise, we aim to undertake a systematic study of the impact of the channel geometry on particle focusing. 
Our model involves a number of governing parameters additional to cross-sectional shape, including slurry composition and channel centreline shape, leading to a large configuration space. Hence, we restrict ourselves to studying three basic types of channel cross-section which, nevertheless, provide a good understanding of how cross-sectional geometry may be used to improve particle focusing, namely rectangular, parabolic, and rectangular with a trench.} 

This paper is structured as follows. We begin by describing the geometry of the spiral channel and define a coordinate system in \cref{sec:geom}. We then introduce the fluid/particle model, and derive two conservation equations in \cref{sec:flowmodel}. We nondimensionalise and scale the system of governing equations in \cref{sec:thinfilm}, and present the solution in a special case in \cref{sec:solution}. We present and discuss results in \cref{sec:results} and make conclusions in \cref{sec:conc}. A table summarising the notation is provided in \cref{notation}.

\section{Mathematical formulation}
\subsection{Geometry and coordinate system}\label{sec:geom}
We consider a helically wound channel with arbitrary shallow cross-section, such as shown in \cref{fig:channel}. The choice of coordinate system is very important in studying flows in helical geometries. We are interested in the final configuration of fluid and particles \ys{near} the \ys{channel exit} 
and hence study the fully-developed steady-state, helically-symmetric flow. Helical symmetry necessitates the use of a non-orthogonal coordinate system \citep{ZM1998}, and we choose to use a body-fitted helicoidal coordinate system as used in \citet{ASG2016}, which is similar to that used by \citet{MC2000,LSB2013} and \citet{ASG2015} A point $\vec x$ in the fluid domain is represented by its radius, $r$, from the vertical axis, angular position, $\beta$, measured from a reference angle, and a vertical displacement, $z$, from the channel bottom, as
\begin{equation}
\vec x(r,\beta,z)=r\cos\left(\beta\right)\bv i+r\sin\left(\beta\right)\bv j+\left(P\beta+H(r)+z\right)\bv k, \label{eq:coorddef}
\end{equation}
where $2\pi P$ is the pitch of the channel centreline, and $H(r)$ is the channel cross-sectional profile. The final parameter required to specify the channel geometry is $A$, the radius of the channel centreline. \Cref{eq:coorddef} gives a non-orthogonal coordinate system, and tensor calculus must be used in order to derive component forms of differential operators. \Citet{Simmonds1994} is a good introduction into the techniques used.

The velocity components in this coordinate system are denoted $v^r$, $v^\beta$, and $v^z$, in the $r$, $\beta$, and $z$ directions, respectively.

\subsection{Particle-laden flow model}\label{sec:flowmodel}
We consider a monodisperse suspension of uniformly-sized spherical particles with diameter $d$. The channel half-width is denoted $a$, and $a\delta$ is a characteristic fluid depth where $\delta\ll1$, as the fluid depth is very small relative to the channel width. We require $d\ll a\delta$ so that the particles are small relative to the fluid depth, and the slurry can be treated as a continuum. We will later give a further restriction on $d$ to ensure that the particles remain large enough to avoid the colloidal limit where Brownian diffusion dominates gravitational settling.

Following \citet{Phillips1992}, \citet{Bertozzi2011}, and \citet{LSB2013}, we assume that the slurry can be treated as a viscous fluid, and denote its velocity, $\vec v (\vec{x}, t) $. We introduce the local particle volume fraction, $\phi (\vec{x}, t) $, which can vary from zero (no particles) to a maximum value $\phi_m$ ($0<\phi_m<1$) which corresponds to the close-packing limit for randomly-arranged spheres, where the mixture can no longer be modelled as a Newtonian fluid. We assume the densities of the particles and fluid are constant, and so the local effective mixture density is given by,
\begin{equation} 
\rho(\phi)=\phi\rho_p+(1-\phi)\rho_f, 
\end{equation}
where $\rho_p$ and $\rho_f$ are the densities of the particles and fluid, respectively. Note that the effective mixture viscosity, $\mu$, will vary with $\phi$ (i.e., $\mu = \mu(\phi)$) in a manner which will be discussed later. 

We assume the motion of the particles involves two contributions: advection with the mixture velocity $\vec v$, and motion relative to the mixture due to other physical effects, denoted by the particle-flux vector $\vec J$.
Conservation of mass for the particles and fluid, respectively, then gives
\begin{equation}
\pd{}{t}\left(\phi\rho_p\right)+\grad\cdot\left(\vec v\rho_p\phi+\rho_p\vec J\right)=0,\label{eq:ca}
\end{equation}
and
\begin{equation}
\pd{}{t}\left((1-\phi)\rho_f\right)+\grad\cdot\left(\vec v\rho_f(1-\phi)\right)=0.\label{eq:cb}
\end{equation}
Summing \eqref{eq:ca} and \eqref{eq:cb} gives the conservation equation of the total mixture,
\begin{equation}
\pd{\rho}{t}+\grad\cdot\left(\rho\vec v+\rho_p\vec J\right)=0,
\end{equation}
using $\rho=\phi\rho_p+(1-\phi)\rho_f$. By taking \eqref{eq:ca}$/\rho_p+$\eqref{eq:cb}$/\rho_f$, we obtain
\begin{equation}
\grad\cdot\left(\vec v+\vec J\right)=0,\label{eq:divvj}
\end{equation}
and taking \eqref{eq:cb}$/\rho_f$,
\begin{equation}
\frac{\mathrm D\phi}{\mathrm Dt}=(1-\phi)\grad\cdot\vec v,\label{eq:partrans}
\end{equation}
\ys{where $D/Dt=\partial/\partial t+\vec v\cdot\grad$ is the usual material derivative.}
We use equations \eqref{eq:divvj} and \eqref{eq:partrans} to impose conservation of mass for the two phases. Thus, in place of the usual continuity equation for incompressible flow, $\grad\cdot\vec v=0$, we have that $\vec v+\vec J$ is solenoidal. 

We obtain an equation for $\vec v$ by considering the momentum balance for the mixture. Assuming the flow is steady, and treating the slurry as a viscous fluid (but allowing its viscosity and density to vary in space) yields the steady Navier-Stokes equations, which can be written as,
\begin{equation} 
\rho\left(\vec v\cdot\grad\vec v\right)=-\grad p+\mu\grad\cdot\mathcal S+\grad\mu\cdot\mathcal S-\rho g\vec e_z,\label{eq:pNS} 
\end{equation}
where $p$ is the pressure, $\rho=\rho(\phi)$ is the effective density, $\mu=\mu(\phi)$ is the effective viscosity, $\mathcal S = \grad \vec v + \grad \vec v ^T $ is the rate-of-strain tensor, $g$ is the acceleration due to gravity and $\vec e_z$ is a unit vector pointing in the vertical direction.  

The local effective viscosity, $\mu$, of the fluid is dependent on the particle volume fraction and we use an empirical model to describe this dependence. Note that many such models exist (see \citet{Murisicetal2013} for a discussion), however, where a specific functional form is required,  we follow \citet{Bertozzi2011} and adopt the Maron-Pierce equation\citep{MP1956}
\begin{equation} 
\mu(\phi)=\mu_f\left(1-\frac{\phi}{\phi_m}\right)^{-2},\label{eq:mu} 
\end{equation}
\ys{which is simply the Krieger-Dougherty equation with the power rounded from $-1.8$ to $-2$;}
 $\mu_f$ is the viscosity of the fluid without particles. 
Various values of $\phi_m$, \ys{the maximum packing volume fraction,} are reported in the literature, and we will use $\phi_m=0.61$, again as in \citet{Bertozzi2011,Murisicetal2013} 
\ys{As $\phi\rightarrow\phi_m$ we have $\mu\rightarrow\infty$ so that this equation is strictly not valid for $\phi$ very near $\phi_m$. The Krieger-Dougherty equation was obtained by fitting data for  $0.01<\phi<0.5$ \cite{Phillips1992}, while the models of \citet{Bertozzi2011,Murisicetal2013} for suspension flows on inclined planes, which utilise the Maron-Pierce equation, have been validated against experiments for feeds with $0.2<\phi<0.5$ and local solids volume fraction as high as 0.6. Should $\phi$ anywhere increase above 0.6, the viscosity there will increase rapidly and the flow will go to zero, which can be intepreted as beaching and, consequently, a channel geometry that is not fit for purpose.
We note that in the feed to spiral separators, which motivate our study, the total solids volume fraction of particles of all densities is typically 0.25--0.35\cite{Michaud2016}.}

It remains to specify the particle-flux vector, $\vec J$, which models the movement of particles relative to the suspending fluid. This motion is assumed to occur due to the effects of gravitational sedimentation and shear-induced migration, and hence we use the same form as \citet{Bertozzi2011} and \citet{LSB2013}: 
\begin{equation} 
\vec J=\frac{2d^2\phi(\rho_p-\rho_f)}{9\mu(\phi)}\omega(\phi)\vec g-K_cd^2\phi\grad(\dot\gamma\phi)-K_vd^2\frac{\phi^2\dot\gamma}{\mu}\dd{\mu}{\phi}\grad\phi, 
\end{equation}
where $K_c$ and $K_v$ are diffusion coefficients and $\dot\gamma = \sqrt{\frac{1}{2}\mathcal{S}^{i j} \mathcal{S}^{kl} g_{ik} g_{jl}} $ is the local shear rate ($\mathcal S$ is the rate-of-strain tensor and $g$ is the metric tensor). This particle-flux vector is valid when the diffusion due to shear is much larger than the Brownian diffusion of particles in the solvent. It is valid for unidirectional shear \citep{Phillips1992}, which, although not true in general for the problem of interest here, is the case at leading order. The first term in the expression for $\vec J$ corresponds to sedimentation, the settling of particles on the channel bottom due to the influence of gravity, and the second and third terms correspond to shear-induced migration, the tendency of particles to move away from areas of high shear and high particle volume-fraction (such as near the channel bottom) towards areas of low shear and particle volume-fraction (such as the free surface near the centre of the channel). The function $\omega(\phi)$ is a hindrance function that models the resistance to motion of a single particle based on the local concentration of particles. There are many different models for the hindrance function, however here we use the simple function $\omega(\phi)=1-\phi$, valid in the presence of shear \citep{SAZ1990} and used in \citet{Murisicetal2013} and \citet{LSB2013}.

We note that in our model, the velocity $\vec v$ is not solenoidal. This changes some aspects of the derivation of the Navier-Stokes equations in our body-fitted coordinate system from that undertaken in \citet{ASG2015,ASG2016}, for example, where the usual continuity equation was used to help simplify the equations. The derivation is lengthy and, because of the similarity to \citet{LSB2013} and \citet{ASG2016}, is not presented here, however, the full helically-symmetric, steady-state Navier-Stokes and conservation equations are given in \cref{sec:fullparticleequations}. We further remark that in other studies using the shear-induced migration model, the velocity has been assumed to be solenoidal. \Citet{Phillips1992} and \citet{Murisicetal2013} required that the velocity $\vec v$ be divergence-free, and an equation equivalent to \cref{eq:partrans} was also given. \Citet{LSB2013} gave two conservation equations equivalent to those we have derived above, and also required $\vec v$ to be divergence-free. Thus two equations were added to the clear-fluid model, but only one variable (the particle volume fraction $\phi$), implying that the system of governing equations obtained was overdetermined. The model was only self-consistent if $\nabla\cdot\vec J=0$, which is not necessarily true. This is an important difference between these previous models and that presented here.
\section{Thin-film equations}\label{sec:thinfilm}
We here nondimenionalise the steady-state, helically-symmetric governing equations given in \cref{sec:fullparticleequations}. As defined earlier, $a$ is a representative channel half-width, $a\delta$ is a representative fluid depth, and $A$ is the radius of the channel centreline. Since we are concerned with thin-film flow, we assume $0<\delta\ll1$. Using carets to denote dimensionless quantities, we write
\begin{align}
\left(\hat r,\hat\beta,\hat z\right)=\left(\frac{r}{a},\delta\beta,\frac{z}{a\delta}\right),\quad (\hat v,\hat u,\hat w)=&\left(\frac{v^r}{U\delta},\frac{v^\beta}{U},\frac{v^z}{U\delta^2}\right),\nonumber\\
\left(\hat J^y,\hat J^\beta,\hat J^z\right)=\frac{a^2\delta^2}{d^2U}\left(\frac{J^r}{\delta},0,J^z\right),&\quad \hat p=\frac{a\delta}{U\mu_f}p, \nonumber\\
\hat \mu(\phi)=\frac{\mu(\phi)}{\mu_f},\quad\hat\rho(\phi)=\frac{\rho(\phi)}{\rho_f}=\rho_s\phi+&1,\quad\hat y=\frac{r-A}{a},
\end{align}
where $U$ is a characteristic velocity scale and $\rho_s=(\rho_p-\rho_f)/\rho_f$ is the density difference between particles and fluid, relative to the fluid density. Note the new radial direction variable $y$, measured from the channel centreline. The dimensionless slope of the channel centreline is $\lambda=P/A$ (the pitch of the centreline divided by its radius), and we define a curvature-like parameter $\epsilon=a/A$. Then
\begin{equation}
\Lambda=\frac{\lambda}{1+\epsilon\hat y}
\end{equation}
gives the local slope of the channel bottom at any point across its width, and $\Upsilon=1+\Lambda^2$ is used for notational convenience.

The Reynolds and Froude numbers are defined as,
\begin{equation} \Re=\frac{\rho_f Ua\delta}{\mu_f},\quad\Fr=\frac{U}{\sqrt{ga\delta}},\end{equation}
and, as in \citet{ASG2016}, we set,
\begin{equation}
\frac{\Re}{\Fr^2}\frac{\lambda}{\left(1+\lambda^2\right)^{3/2}}=1\quad\text{ and }\quad\frac{6}{35}\Re\frac{\epsilon\lambda}{\left(1+\lambda^2\right)^{3/2}}=1. \label{eq:scales}
\end{equation}
The first of these relationships implies a significant gravity-driven axial flow at the channel centreline $y=0$ and, along with the definition of the Reynolds and Froude numbers, gives an equation for the axial velocity scale $U$,
\begin{equation}
U=\frac{\lambda}{\left(1+\lambda^2\right)^{3/2}}\frac{(a\delta)^2g\rho_f}{\mu_f}.
\end{equation}
The second relation in \cref{eq:scales} ensures sufficient cross-sectional flow for an interesting (non-flat) free-surface profile.

The scaling of the particle-flux vector $\vec J$ includes the particle diameter $d$, which must be small, but we need to compare it to the fluid depth scaling, $\delta$, in order to compare terms in the perturbation expansions of the governing equations. Previously we stated the requirement $d\ll a\delta$ so that the particles are small relative to the fluid depth, justifying the continuum modelling approach. If the particles are too small, however, Brownian diffusion becomes a dominant effect and settling due to gravity is negligible. To ensure this does not occur, and as in \citet{LSB2013} and \citet{Murisicetal2013}, we require the settling distance along the channel, $L_{settle}$, of the particles to be small relative to the axial length scale of the channel. We assume that the channel is much longer than it is wide, and set the axial length scale as $a/\delta$. Considering the film thickness, settling velocity, and axial velocity scales in the spirit of equation~(2.3) in \citet{Murisicetal2013}, we obtain,
\begin{align}
L_{settle}&\approx\frac{\text{fluid depth}}{\text{settling velocity}}\times\text{axial velocity}\nonumber\\
&\approx\frac{a\delta}{\frac{2d^2g(\rho_p-\rho_f)}{9\mu_f}}\frac{\rho_fg(a\delta)^2\lambda}{\mu_f\left(1+\lambda^2\right)^{3/2}}=\frac{(a\delta)^3}{d^2}\frac{9\rho_f\lambda}{2\left(\rho_p-\rho_f\right)\left(1+\lambda^2\right)^{3/2}}.
\end{align}
For this settling length to be small relative to the axial length scale, we require
\begin{equation}
\frac{(a\delta)^3}{d^2}\ll\frac{a}{\delta},
\end{equation}
or $(d/a)^2\gg\delta^4$. Combining this with the condition $d\ll a\delta$, we obtain,
\begin{equation}
\delta^4\ll\left(\frac{d}{a}\right)^2\ll\delta^2
\end{equation}
asymptotically as $\delta\rightarrow0$, and so we set $(d/a)^2=O(\delta^3)$ when comparing terms in the asymptotic expansions of the governing equations.

Using Maple we obtain the following system of equations at leading order in $\delta$, where we have, from here on, omitted the carets on all dimensionless variables. The Navier-Stokes equations \eqref{eq:NSFr}--\eqref{eq:NSFz} yield
\begin{equation}
\pd{}{z}\left(\mu\pd{v}{z}\right)=-\frac{\Re\,\epsilon\rho}{(1+\epsilon y)\Upsilon^2}u^2+\frac{1}{\Upsilon}\pd{p}{y}-\frac{1}{\Upsilon}\dd{H}{y}\pd{p}{z}-\frac{2\mu\epsilon\Lambda}{(1+\epsilon y)\Upsilon^{3/2}}\pd{u}{z},\label{eq:nsR}
\end{equation}
\begin{equation}
\pd{}{z}\left(\mu\pd{u}{z}\right)=\frac{\Re}{\Fr^2}\frac{\rho\Lambda}{\Upsilon^{3/2}},\label{eq:nsB}
\end{equation}
\begin{equation}
\pd{p}{z}=-\frac{\Re}{\Fr^2}\frac{\rho}{\Upsilon},\label{eq:nsZ}
\end{equation}
the conservation equations \eqref{eq:co1} and \eqref{eq:co2} give
\begin{equation}
\pd{J^z}{z}=0,\label{eq:Jeq}
\end{equation}
\begin{equation}
v\pd{\phi}{r}+w\pd{\phi}{z}=\left(1-\phi\right)\left(\pd{v}{r}+\pd{u}{z}+\frac{v}{r}\right),\label{eq:cont}
\end{equation}
and the boundary conditions \eqref{eq:ans}--\eqref{eq:PARTkinematic} give
\begin{equation} u=v=w=J^z=0\quad\text{at $z=0$,} \end{equation}
\begin{equation} p=\pd{v}{z}=\pd{u}{z}=0,\quad w=\dd{h}{y}v,\quad\text{at $z=h(y)$.} \end{equation}

Although these equations are significantly simplified compared to the full Navier-Stokes equations, they cannot be solved analytically as, in general, the particle volume fraction will vary in both the $y$ and $z$ directions, and hence $\mu$ and $\rho$ are functions of both $y$ and $z$.

\section{The thin-film solution for $\phi=\phi(y)$}\label{sec:solution}
Experimental work by \citet{ZDBH2005}, \citet{Bertozzi2011} and others on particle-laden thin film flows on inclined planes found three qualitatively different behaviours depending on the angle of inclination and initial particle volume fraction. The three regimes are termed settled, well-mixed, and ridged. In the settled regime, particles tend to settle quickly and clear fluid flows more quickly over the top of the particles. In the ridged regime, particles move towards the surface of the film, move more quickly than the surrounding fluid, and amass at the fluid front. Finally, in the well-mixed regime, the fluid and particle velocities are roughly the same, and the particles do not tend to congregate at the bottom or top of the fluid, but disperse uniformly throughout the film thickness. \ys{As is readily seen from the figures in \citet{Bertozzi2011}, for a plane at a given angle of inclination, the settled regime applies for smaller initial particle volume fraction and the ridged regime for larger initial particle volume fraction, with the well-mixed regime applying at intermediate values of the initial particle volume fraction, between the settled and ridged regimes. The range of particle volume fraction associated with the settled regime decreases as the inclination angle increases.}

In \citet{LSB2013}, it was found that the only solutions that their thin-film governing equations could admit were of the well-mixed type. However, as discussed in \cref{sec:flowmodel}, we believe their system of equations was overdetermined, and unnecessarily constrained the particle volume fraction $\phi$ to be independent of depth. For our model, this restriction need not necessarily apply. Nevertheless, motivated by the existence of well-mixed flows down inclined planes, we here seek solutions in helically-wound channels in the well-mixed regime, by assuming $\phi=\phi(y)$ so that the particle volume-fraction is independent of depth. 
This has the benefit of making the governing equations significantly more tractable, enabling further analytic progress. \ys{This modelling approach is also reasonable given the present lack of experimental data on particle distribution in the vertical direction with which to compare, and the need for a model to aid separator design that gives understanding of the radial distribution of particles. 
Spiral separators work because of a radial separation of particles of different densities, while vertical distribution appears not to be important.} We \ys{thus} consider whether there is a region in the parameter space in which such solutions exist and, if so, whether they differ significantly from those found by \citet{LSB2013} for channels of small slope. The more general case where the particle volume-fraction varies with both depth and radial position is left for future work.

\begin{figure}
    \centering
	\begin{tikzpicture}
    	\node[anchor=south west,inner sep=0] (image) at (0,0) {\input{ppPhi.tex}\includegraphics[width=\onewide\textwidth]{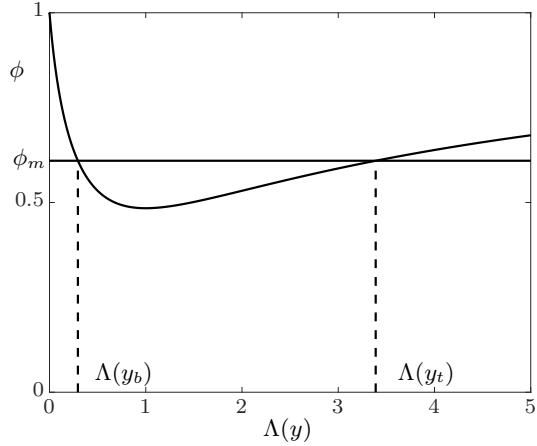}};
		\begin{scope}[x={(image.south east)},y={(image.north west)}]
			\node [rotate=0,align=center] at (0.09,0.8) {$\phi$};
			\node at (0.52,0.03) {$\Lambda(y)$};
			\node at (0.26,0.15) {$\Lambda(y_b)$};
			\node at (0.74,0.15) {$\Lambda(y_t)$};
			\node at (0.11,0.61) {$\phi_m$};
		\end{scope}
	\end{tikzpicture}
	\caption{Particle volume fraction $\phi$ plotted as a function of local slope $\Lambda(y)$ with $\rho_s=1.5$, $K_c=0.41$, and $\phi_m=0.61$.}\label{fig:phip}
\end{figure}

The steps outlined below are broadly similar to \citet{LSB2013}, although the equations are different. 
With $\phi=\phi(y)$, both the density and viscosity become functions of $y$ only, and this allows the thin-film governing equations to be solved by sequential integration. \Cref{eq:nsZ} can be solved directly to give
\begin{equation}
p=-\frac{\Re}{\Fr^2}\frac{\rho}{\Upsilon}(z-h(y)),
\end{equation}
and \cref{eq:nsB} gives
\begin{equation}
u=\frac{\Re}{\Fr^2}\frac{\rho\Lambda}{\mu\Upsilon^{3/2}}\frac{z}{2}\left(z-2h(y)\right).\label{eq:u}
\end{equation}
Note this equation is used to set the first of the scalings in \cref{eq:scales} to ensure sufficiently high axial velocity at the channel centreline.
Now \cref{eq:Jeq} can be integrated with respect to $z$ to give $J^z=C(y)$ for some $C(y)$, and the boundary condition $\left.J^z\right|_{z=0}=0$ gives $J^z=0$, so gravitational settling is balanced by shear-induced migration. We first note that at leading order in the thin-film limit, the shear rate is
\begin{equation}
\dot\gamma=\sqrt{\Upsilon}\pd{u}{z}.
\end{equation}
Then, writing the $z$-component of the particle-flux vector explicitly gives an equation involving $\phi$, $\mu$, and $u$,
\begin{equation}
0=-\frac{\Re}{\Fr^2}\frac{2d^2\phi\rho_s}{9\mu}\omega(\phi)+K_cd^2\phi\pd{}{z}\left(\sqrt{\Upsilon}\phi\pd{u}{z}\right)+K_vd^2\frac{\phi^2\sqrt{\Upsilon}}{\mu}\pd{u}{z}\dd{\mu}{\phi}\pd{\phi}{z},
\end{equation}
which can be rearranged as
\begin{equation}
\frac{2\rho_s\Re}{9\Fr^2K_c\sqrt{\Upsilon}}\omega(\phi)-\phi\pd{}{z}\left(\mu \pd{u}{z}\right)-\left[1+2\frac{K_v-K_c}{K_c}\frac{\phi}{\phi_m-\phi}\right]\pd{\phi}{z}\left(\mu \pd{u}{z}\right)=0.
\end{equation}
Now $\partial\phi/\partial z=0$ by assumption, we use \eqref{eq:nsB} to replace $\mu \partial u/\partial z$,
and substitute the hindrance function $\omega(\phi)=1-\phi$, to give
\begin{equation}
\frac{{2}\Upsilon\rho_s}{9\Lambda K_c}(1-\phi)-\phi(1+\rho_s\phi)=0.\label{eq:phieqn}
\end{equation}
This is a quadratic equation for $\phi(y)$
with solution
\begin{equation}
\phi=-\left(\frac{\Upsilon}{9\Lambda K_c}+\frac{1}{2\rho_s}\right)+\sqrt{\left(\frac{\Upsilon}{9\Lambda K_c}+\frac{1}{2\rho_s}\right)^2+\frac{2\Upsilon}{9\Lambda K_c}}.\label{eq:phiprofile}
\end{equation}
The particle volume fraction given in \cref{eq:phiprofile} is a function of $\Lambda(y)$ (the local slope of the channel bottom at position $y$) and must satisfy $0\leq\phi<\phi_m=0.61$ in the particle-rich region in order for our model to be valid. \Cref{fig:phip} shows a typical profile for $\phi$ in terms of $\Lambda(y)$. Note that $\phi(y)\rightarrow1$ as $\Lambda\rightarrow0$ (corresponding to $y\rightarrow\infty$) and as $\Lambda\rightarrow\infty$ (corresponding to $y\rightarrow-1/\epsilon^+$), and takes a minimum value at $\Lambda=1$ (where $y=(\lambda-1)/\epsilon$). In the limit $\Lambda^2\rightarrow0$, when $\Upsilon=1$, \eqref{eq:phiprofile} is equivalent to Equation~(11) in \citet{LSB2013}. This is to be expected; their model was derived in the small-slope limit $\lambda^2\rightarrow0$, when $\Lambda^2$ is also small (so long as $\epsilon$ is not too close to 1). Their $\phi$ profiles were monotonic increasing, but now allowing a region of higher shear at the inside of the channel (with non-small $\lambda$), a higher concentration of particles can be maintained there. Increasing the local fluid velocity therefore allows for a higher concentration of particles to be sustained in the well-mixed regime.

To ensure $\phi(y)<\phi_m$ wherever particles occur, the particle-rich region must lie fully within \ys{the open interval} $(y_b,y_t)$, the end-points of which satisfy $\phi(y_b)=\phi_m$ and $\phi(y_t)=\phi_m$. We find
\begin{equation}
y_b=\frac{1}{\epsilon}\left\{-1+{\lambda}\left({\chi-\sqrt{\chi^2-1}}\right)\right\}\quad\text{and}\quad y_t=\frac{1}{\epsilon}\left\{-1+{\lambda}\left({\chi+\sqrt{\chi^2-1}}\right)\right\},
\label{particle_limits}
\end{equation}
where 
\begin{equation}
\chi=\frac{9K_c\phi_m(1+\rho_s\phi_m)}{4\rho_s(1-\phi_m)},
\end{equation}
and, for example, $\chi=1.8421$ with $K_c=0.41$, $\rho_s=1.5$ and $\phi_m=0.61$.
This restricts our results somewhat, \ys{which} 
will be discussed further in \cref{sec:results}. Finally we note that there is a maximum $\rho_s$ above which $\phi(y)\geq\phi_m$ for all $y$. To have any particles in the flow, we require points $y_b$ and $y_t$ to exist and hence require $\chi>1$.
This is satisfied as long as $\rho_s<12.04$ (with the given $K_c$ and $\phi_m$). By way of comparison, glass particles with $\rho_s\approx 2.5$ and ceramic particles with $\rho_s\approx 3.8$ have been used in experiments\citep{Murisicetal2013}.

Having determined $\phi$, we next find the radial velocity, $v$, by integrating \cref{eq:nsR} twice, yielding,
\newcommand{\ai}{\frac{\Re^3\,\epsilon\rho^3\Lambda^2}{\Fr^4(1+\epsilon y)\mu^3\Upsilon^5}}
\newcommand{\aii}{\left(-\frac{\Re}{\Fr^2}\frac{1}{\mu\Upsilon^2}\dd{\rho}{y}-\frac{\Re}{\Fr^2}\frac{4\epsilon\rho\Lambda^2}{(1+\epsilon y)\mu\Upsilon^{3}}\right)}
\newcommand{\aiii}{\frac{\Re}{\Fr^2}\frac{\rho}{\mu\Upsilon^2}\left(\dd{H}{y}+\dd{h}{y}\right)}
\begin{align}
v=&-\ai\frac{z}{120}\left(z^5-6hz^4+10h^2z^3-16h^5\right)\nonumber\\
&+\aii\frac{(z-h)^3+h^3}{6} \nonumber\\
&+\aiii\frac{z(z-2h)}{2}.
\end{align}

\Cref{eq:cont} may be written as
\begin{equation}
\pd{}{y}\left(\frac{(1-\phi)(1+\epsilon y)v}{\epsilon}\right)+\pd{}{z}\left(\frac{(1-\phi)(1+\epsilon y)w}{\epsilon}\right)=0,
\end{equation}
and integrating with respect to $z$, rearranging using Liebniz's integral rule, and applying boundary conditions, gives
\begin{equation}
\frac{(1-\phi)(1+\epsilon y)}{\epsilon}\int^{h(y)}_{0}v\d z=C,
\end{equation}
where $C$ is a constant which effectively controls the net radial flow at any position $y$. We set $C=0$ so that there is no net flow into or out of the fluid domain, to obtain,
\begin{equation}
\int^{h(y)}_{0}v\,\mathrm dz=0.
\end{equation}
Substituting $v$ into this integral equation and rearranging, gives an equation for the fluid depth $h(y)$,
\begin{equation}
\dd{h}{y}=\frac{6}{35}\frac{\Re^2}{\Fr^2}\frac{\epsilon\rho^2\Lambda^2}{(1+\epsilon y)\mu^2\Upsilon^3}h^4-\frac{3}{2}\frac{\Lambda^2\epsilon}{\Upsilon(1+\epsilon y)}h-\frac{3}{8}\frac{1}{\rho}\pd{\rho}{y}h-\dd{H}{y}.\label{eq:fse}
\end{equation}
This equation differs from the clear-fluid free-surface equation from \citet{ASG2016} in having an extra term, linear in $h$, involving the gradient of the density. As a Chini differential equation, this equation has no analytic solution in general. In the rectangular channel case $H=0$, the equivalent clear-fluid free-surface equation did have an analytic solution \citep{ASG2015}, but \cref{eq:fse} does not, due to the presence of the additional term involving the density gradient.

Substituting the free surface equation into $v$ then yields,
\begin{align}
v=&-\frac{\Re^3}{\Fr^4}\frac{\epsilon\rho^3\Lambda^2}{(1+\epsilon y)\mu^3\Upsilon^5}\frac{z(z-2h)\left[7z(z-2h)(z-2hz-2h^2)-16h^4\right]}{840}\nonumber\\
&-\frac{\Re}{\Fr^2}\frac{1}{\mu\Upsilon^2}\left(\dd{\rho}{y}+\frac{4\epsilon\rho\Lambda^2}{(1+\epsilon y)\Upsilon}\right)\frac{z(8z^2-15hz+6h^2)}{48}.\label{eq:partV}
\end{align}

We now introduce a streamfunction $\psi$, defined by
\begin{equation} \pd{\psi}{z}=(1+\epsilon y)(1-\phi)v,\quad\pd{\psi}{y}=-(1+\epsilon y)(1-\phi)w.\label{eq:psi} \end{equation}
Using \eqref{eq:partV} and requiring $\psi=0$ on the channel bottom $z=0$, we find,
\begin{align}
\psi=&-\frac{\Re}{\Fr^2}\frac{1-\phi}{\mu\Upsilon^2}\left[\frac{\Re^2}{\Fr^2}\frac{\epsilon\rho^3\Lambda^2}{\mu^2\Upsilon^3}\frac{z^2(z-h)(z-2h)^2\left(z^2-2hz-4h^2\right)}{840}\right.\nonumber\\
&\left.+\left((1+\epsilon y)\dd{\rho}{y}+\frac{4\epsilon\rho\Lambda^2}{\Upsilon}\right)\frac{z^2(2z-3h)(z-h)}{48}\right].
\end{align}
The second equation in \eqref{eq:psi} can be used to find the vertical velocity $w$ (not given due to its complexity).

As we are dealing with particle-laden flow, two fluxes are relevant: the total mixture flux and the flux of particles. The total (dimensionless) flux down the channel $Q$ (scaled by $\delta a^2U$) is given by
\begin{align}
Q=&-\int^{y_r}_{y_l}{\int^{h(y)}_{0}{u(z,y)}\,\mathrm dz}\,\mathrm dy \nonumber\\
=&\frac{1}{3}\frac{\Re}{\Fr^2}\int^{y_r}_{y_l}\frac{\rho\Lambda h^3}{\mu\Upsilon^{3/2}}\,\mathrm dy,
\end{align}
where $y_l$ and $y_r$ represent the inner and outer boundaries of the fluid domain respectively. The particle-flux $Q_p$ (scaled by $\delta a^2U$) is
\begin{align}
Q_p=&-\int^{y^*_r}_{y^*_l}{\int^{h(y)}_{0}{\phi(y)u(z,y)}\,\mathrm dz}\,\mathrm dy \nonumber\\
=&\frac{1}{3}\frac{\Re}{\Fr^2}\int^{y^*_r}_{y^*_l}\frac{\rho\phi\Lambda h^3}{\mu\Upsilon^{3/2}}\,\mathrm dy,
\end{align}
where $y^*_l$ and $y^*_r$ represent the boundaries of the particle-rich region of the flow, which do not necessarily correspond to the whole fluid domain. The particle-rich region must be contained in the fluid region, so we must have $y_l\leq y_l^*<y_r^*\leq y_r$. The negative signs in the two fluxes given above are due to the axial coordinate direction pointing up the channel, meaning $u$ is always negative.

\section{Results}\label{sec:results}
Having determined that there is a region of the parameter space in which there is a solution to the thin-film system of equations for particles uniformly dispersed in the vertical direction, we now present and discuss some particular cases. For a given channel geometry, $\epsilon$, $\lambda$, and $H(r)$, we have a set of 8 (not all independent) parameters: $\{y_l,h_l,y_r,h_r,Q,y_l^*,y_r^*,Q_p\}$. To specify a solution, we choose 5 of these parameters: $y_l$ or $h_l$, $y_r$ or $h_r$, $Q$, $y_l^*$, and $Q_p$. From these, we find the remaining parameters: $y_l$ or $h_l$, $y_r$ or $h_r$, and $y_r^*$. For a channel with a vertical wall, we specify the value(s) $y_l$ and/or $y_r$, and for a channel with a curved bottom, we set $h_l=0$ and/or $h_r=0$. Finding the unknown parameters involves a numerical root-finding problem to find the \ys{extents of the partical-laden and clear fluid regions for the chosen total and particle fluxes.} 

Different choices of parameters could be used to specify a solution. For example, we could specify $y_l^*$, $y_r^*$ and three out of the following group of four parameters: $\{y_l,h_l,y_r,h_r\}$. The parameter not picked, as well as the fluxes $Q$ and $Q_p$ could be calculated by solving the governing equations. This choice would be simpler than specifying fluxes because it can be directly calculated by solving the free-surface differential equation in the particle-rich and particle-free regions. Motivated by spiral particle separators, which are often described in terms of the quantity of material that they can process per hour, we choose to specify our solutions in terms of the fluid and particle fluxes. \ys{In addition, this choice enables an examination of how channel geometry affects the radial focusing of particles for given fluid and particle fluxes.}

By considering spiral particle separators, we expect that, in steady state conditions, the particle-rich region of the flow should be near the inside wall of the channel, and extend part or all of the way to the outside channel wall. The volume fraction of particles $\phi$ is dependent on radial position $y$, and the flux of particles, together with $\phi(y)$, determines how wide the particle-rich zone will be. \Citet{LSB2013} provided an intuitive explanation as to why this configuration of particles is expected to be stable. For completeness, we briefly recapitulate their argument here.

\begin{figure}
	\centering
	\begin{subfigure}[t]{.3\textwidth}
		\centering
		\includegraphics[width=.76\textwidth,trim=230 20 130 0,clip]{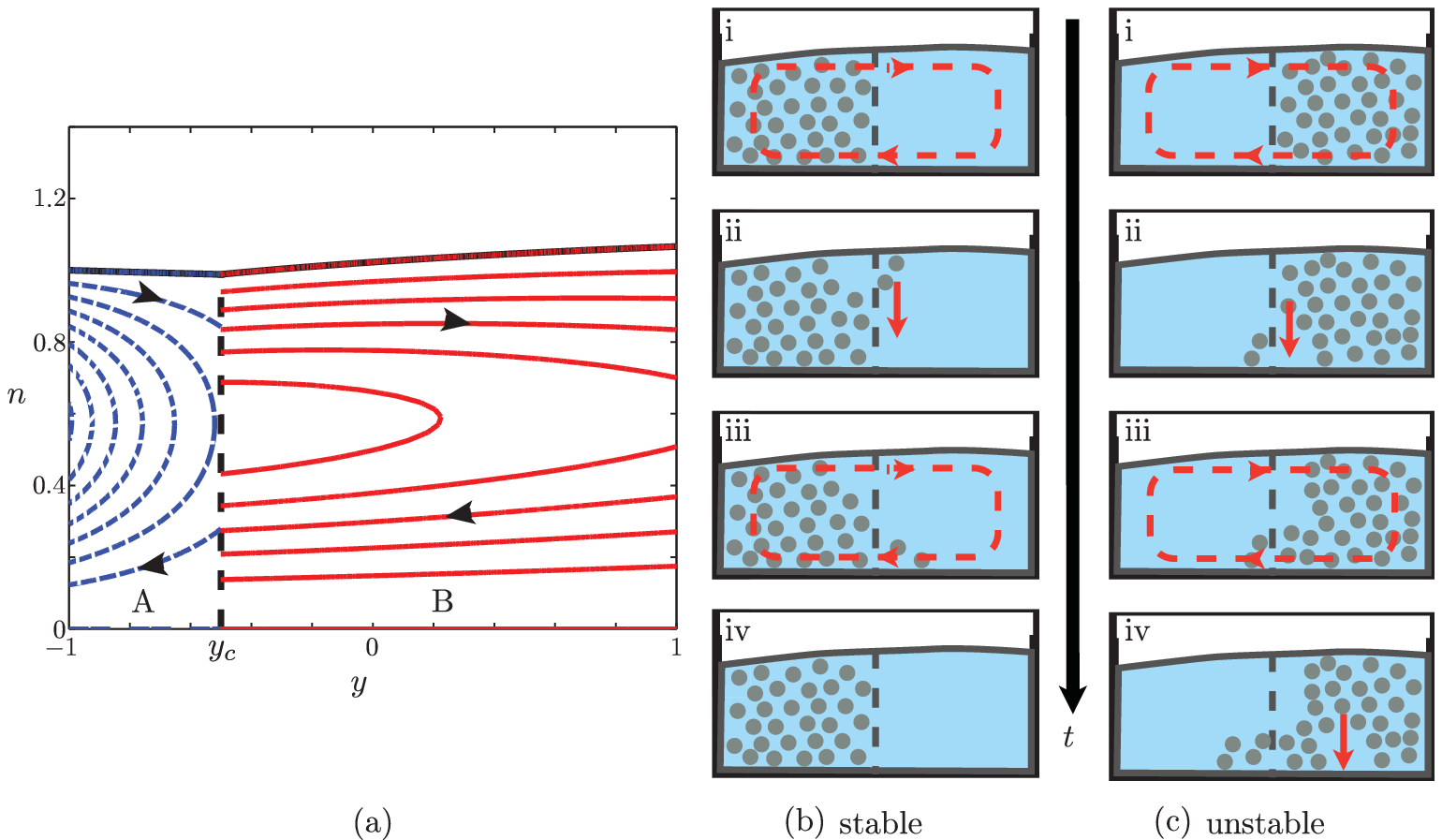}
		\caption{A stable configuration.}
		\label{fig:st}
	\end{subfigure}~
	\begin{subfigure}[t]{0.3\textwidth}
		\centering
		\includegraphics[width=.76\textwidth,trim=350 20 10 0,clip]{fig4}
		\caption{An unstable configuration.}
		\label{fig:unst}
	\end{subfigure}
	\caption{Configurations of particles expected to be (a) stable and (b) unstable. Reprinted from \citet*{LSB2013} with the permission of AIP Publishing.}
	\label{fig:config}
\end{figure}

Let us first assume that the particles collect near the inside channel wall, and there is a clear fluid region in the outer region of the channel cross-section, as in \cref{fig:st} (top image). The outward flow near the free surface will carry particles radially outwards into the particle-free region (second image), where, because \ys{the local slope of the channel bottom will be smaller and} the particle volume fraction will be low, \ys{the settled regime will apply and the particles} 
will rapidly equilibrate to the channel bottom (third image). The equilibration will be fast due to the assumption of short settling length. The secondary flow, which is radially inwards near the channel bottom, then carries the particles back into the particle-rich region (bottom image). So in practice we would expect that if such a configuration of particles was reached, it would remain stable.

Now assume that there is a configuration with a particle-free region near the inside wall and a particle-rich region further outwards, as in \cref{fig:unst} (top image). On the boundary between the particle-rich and particle-free regions, the flow near the channel bottom will carry particles into the particle-free region (second image). Again, \ys{the settled regime will apply due to the low volume fraction of particles so these particles will remain near the channel bottom and continue to be carried to the inside of the spiral by the secondary flow. At the same time, in the outer} 
particle-rich region, \ys{particles} will settle into the vacated area near the channel bottom (second figure). 
The process will continue \ys{over time (third and bottom images) and the particle-rich outer region will first enter the settled regime, when the particle volume fraction reduces sufficiently, and finally disappear.} 
This configuration of particles is therefore not  stable.

\ys{From this we deduce that,} 
over time, the particles will \ys{migrate to} 
a region next to the inside wall of the channel. \ys{Hence,} 
we search for \ys{steady} solutions with the particle-rich region near the inside channel wall and the particle-free region near the outside wall. This means we set $y_l^*=y_l$. \ys{We note that this configuration is consistent with the known fact that heavy particles migrate to the inside wall of a spiral separator.}

We solve the governing equations separately in the particle-rich and particle-free regions, and couple them by requiring that the free-surface be continuous at the boundary between the two regions. At the interface between the particle-rich and particle-free regions of the flow, the fluid properties, \ys{the streamlines (with the exception of the free surface), and the axial flow contours} are discontinuous. In practice \ys{there will be} a thin boundary layer between the two regions, which is not captured by our leading order \ys{model, across which quantities change continuously and which would remove the non-physical cusp in the free surface that is seen in our results at the interface.}

We will be particularly interested in considering how well a channel can focus the particle-rich region of the flow into a narrow band of the channel cross-section. Certainly, for polydisperse slurries, separation of particles of different densities requires the focussing of each particle species into non-overlapping bands in the cross-section. This suggests that minimising the band width is advantageous. \ys{Below we will consider two basic channel cross-sections, rectanglar and parabolic, the first as the simplest case that has vertical side walls at which the fluid depth must be determined, and the second as the simplest case that requires determination of the horizontal extent of the fluid with a zero fluid depth at each side. We then examine the effect of a trench in the channel bottom. As will be seen, this choice of channel cross-sectional geometries provides a good understanding of how cross-sectional geometry may be used to improve particle focusing.}

Throughout this section, we use dashed streamlines for the particle-laden section of the flow, and solid streamlines for the particle-free fluid. A solid vertical line indicates the boundary between particle-free and particle-rich regions at $y=y_r^*$. In all plots we set $K_c=0.41$, which has been empirically determined from experiments \citep{Phillips1992}, and $\phi_m=0.61$ as previously discussed. Unless stated otherwise, we will use $\rho_s=(\rho_p-\rho_f)/\rho_f=1.5$.

\subsection{The effect of centreline geometry}\label{sec:particleresults}
Changing $\epsilon$ and $\lambda$ in particle-free flows has been studied in detail in \citet{ASG2015,ASG2016}, and we now seek to understand the effect of $\epsilon$ and $\lambda$ on particle-rich flows.

\begin{figure}
\centering
    \begin{subfigure}[t]{\threewide\textwidth}
        \begin{tikzpicture}
        		\node[anchor=south west,inner sep=0] (image) at (0,0) {\input{ppRectangleFixLambdaEpsilonp2.tex}\includegraphics[width=\textwidth]{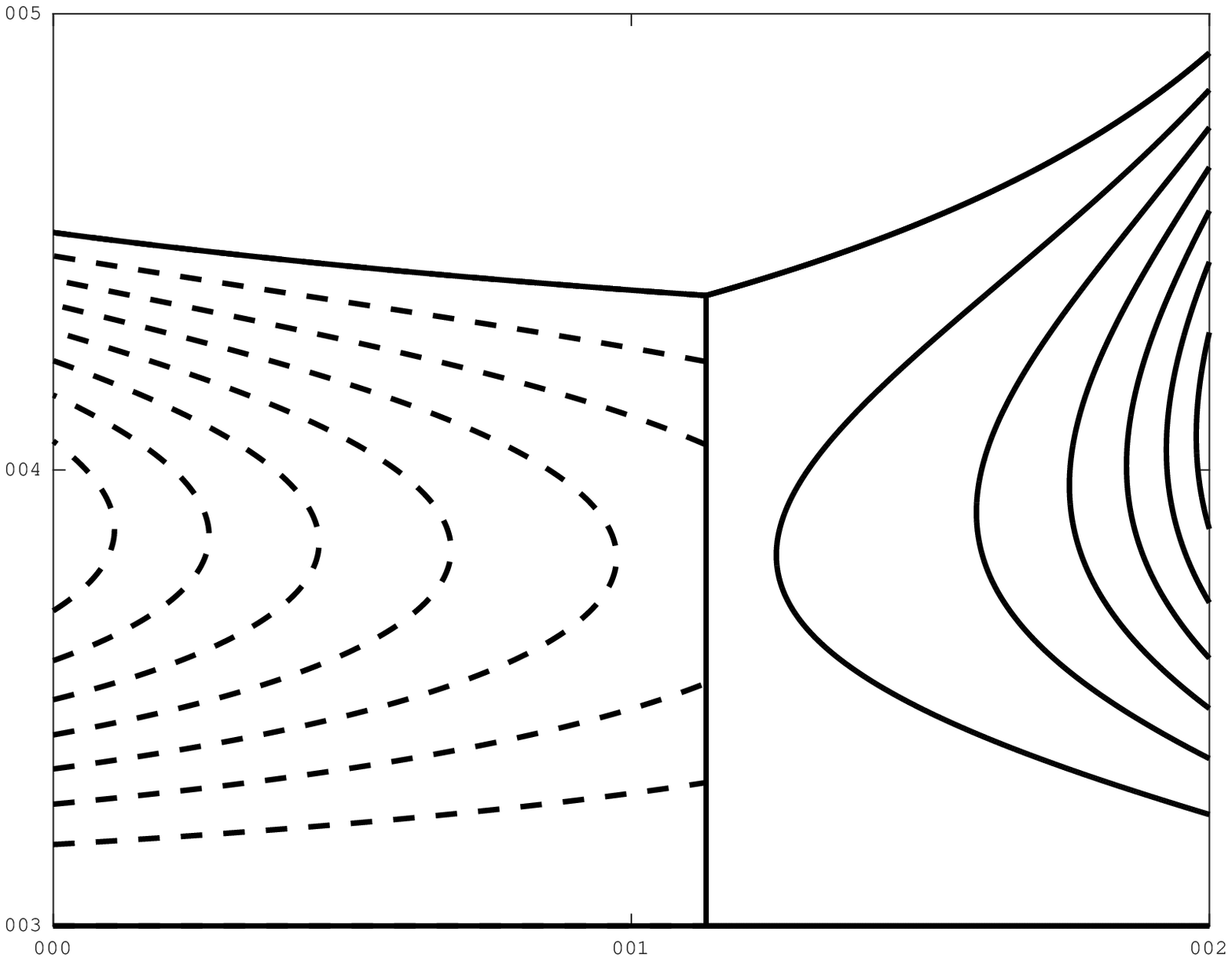}};
		\begin{scope}[x={(image.south east)},y={(image.north west)}]
			\axislabels
    		\end{scope}
	\end{tikzpicture}
	\caption{$\epsilon=0.2$}
    \end{subfigure}%
    \hfill
    \begin{subfigure}[t]{\threewide\textwidth}
        	\begin{tikzpicture}
    		\node[anchor=south west,inner sep=0] (image) at (0,0) {\input{ppRectangleFixLambdaEpsilonp4.tex}\includegraphics[width=\textwidth]{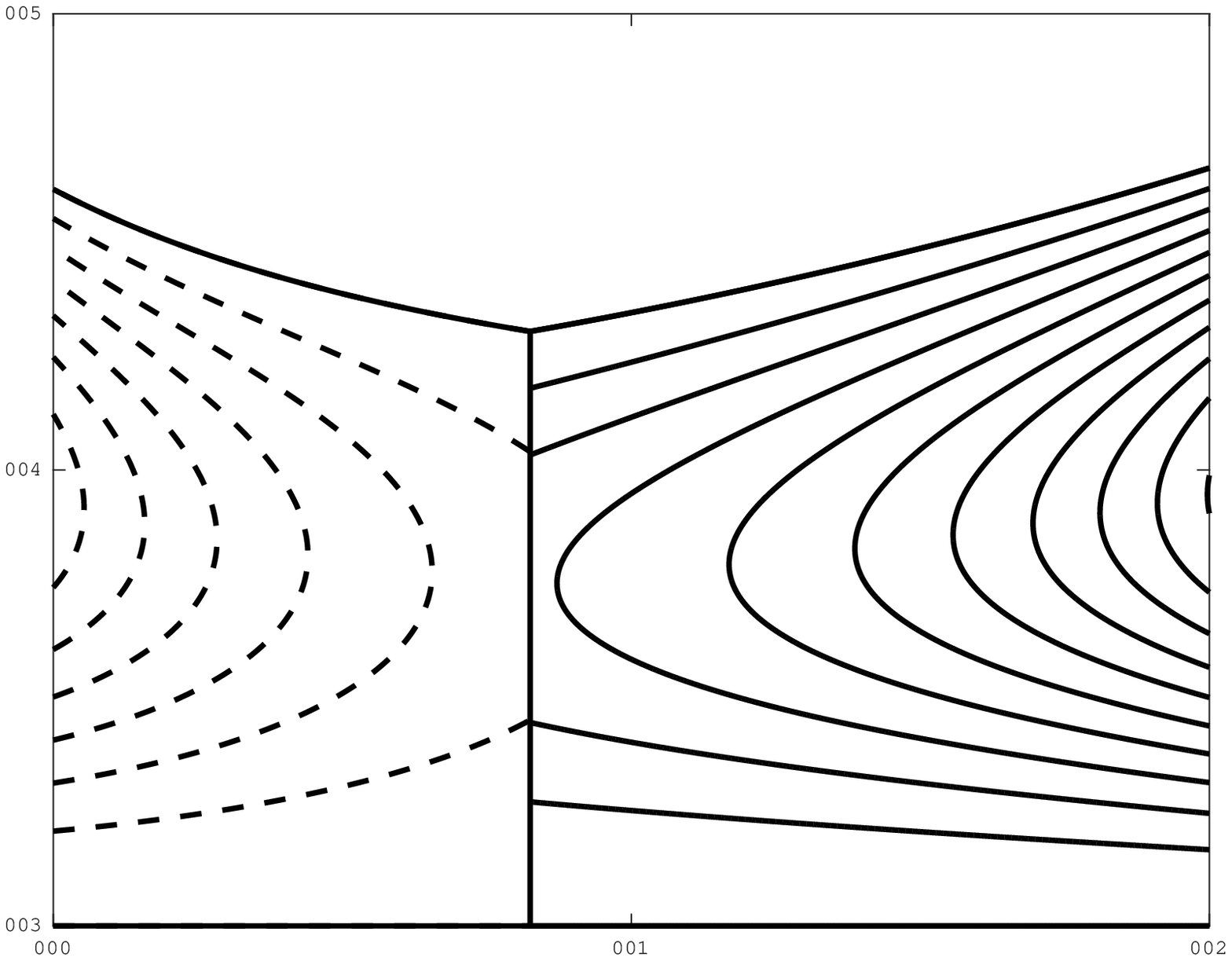}};
		\begin{scope}[x={(image.south east)},y={(image.north west)}]
			\axislabels
	    	\end{scope}
	\end{tikzpicture}
	\caption{$\epsilon=0.4$}
	\end{subfigure}%
    \hfill
    \begin{subfigure}[t]{\threewide\textwidth}
        \begin{tikzpicture}
    	\node[anchor=south west,inner sep=0] (image) at (0,0) {\input{ppRectangleFixLambdaEpsilonp6.tex}\includegraphics[width=\textwidth]{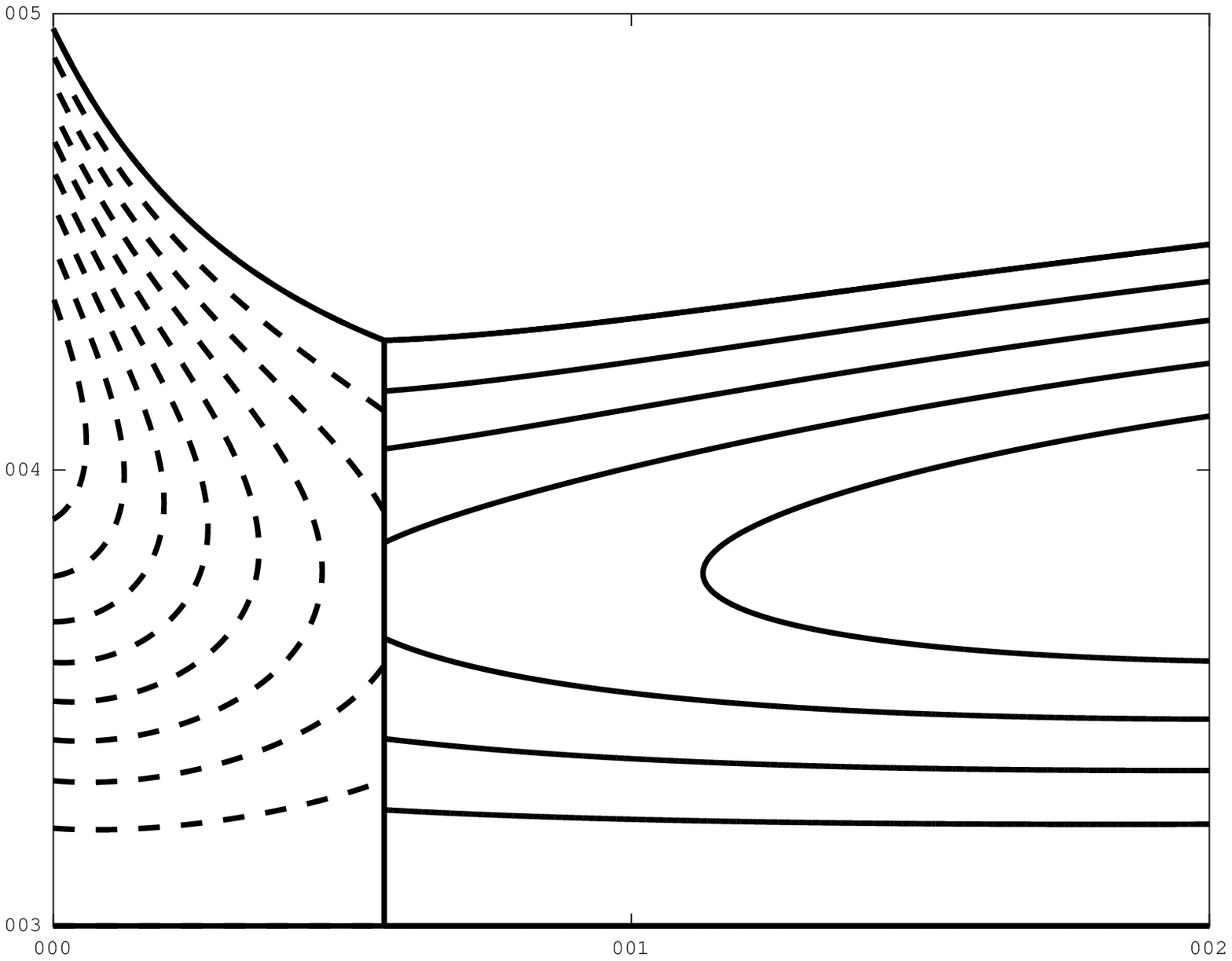}};
		\begin{scope}[x={(image.south east)},y={(image.north west)}]
			\axislabels
    	\end{scope}
	\end{tikzpicture}
	\caption{$\epsilon=0.6$}
	\end{subfigure}
\\
    \begin{subfigure}[t]{\threewide\textwidth}
        \begin{tikzpicture}
    	\node[anchor=south west,inner sep=0] (image) at (0,0) {\input{ppRectanglePhiFixLambdaEpsilonp2.tex}\includegraphics[width=\textwidth]{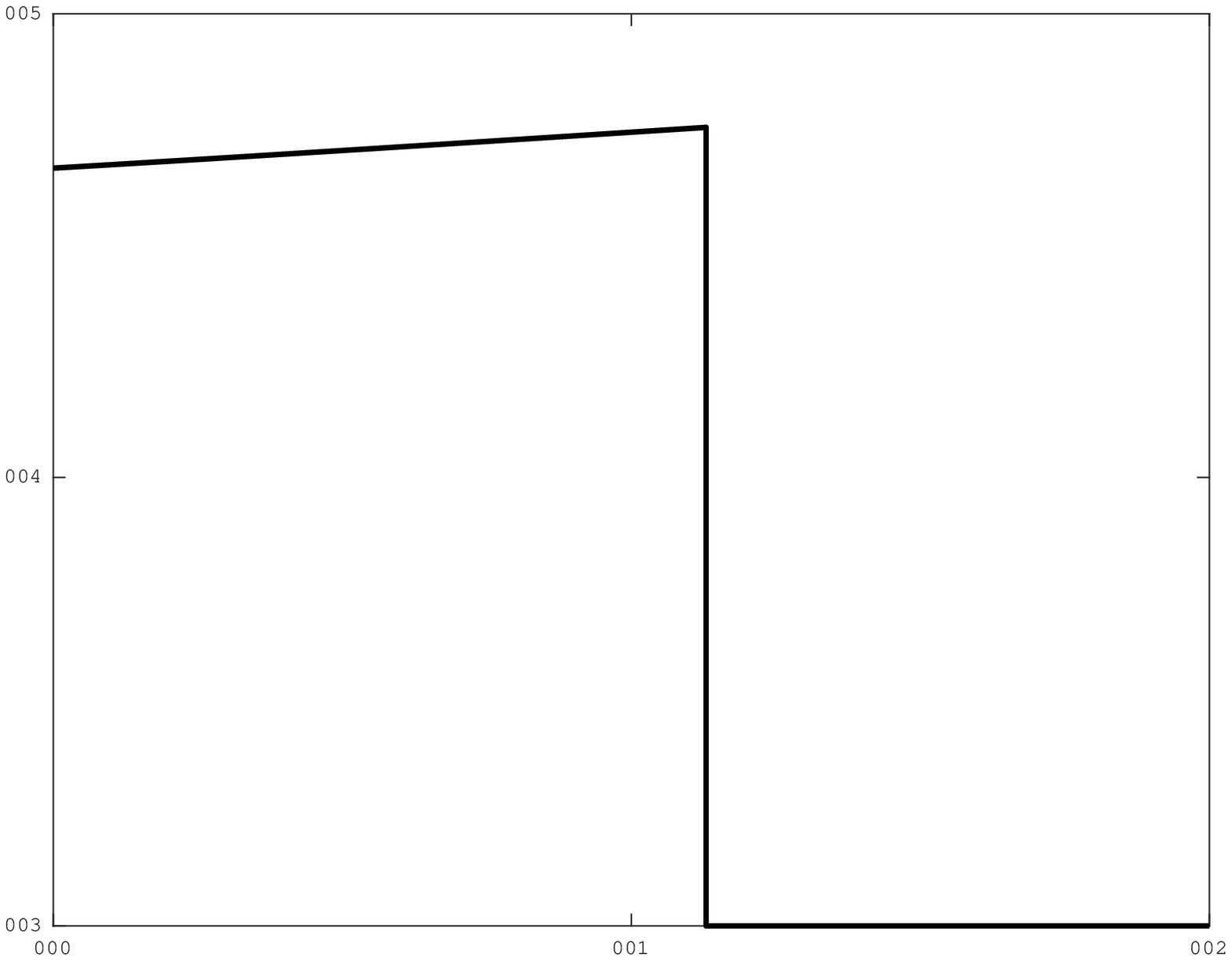}};
		\begin{scope}[x={(image.south east)},y={(image.north west)}]
		\phantom{\axislabels}
			\node [rotate=0,align=center] at (0.05,0.7) {\footnotesize $\phi$};
			\node at (0.7,0.02) {\footnotesize $y$};
    	\end{scope}
	\end{tikzpicture}
	\caption{$\epsilon=0.2$}
	\end{subfigure}
    \hfill
    \begin{subfigure}[t]{\threewide\textwidth}
        \begin{tikzpicture}
    	\node[anchor=south west,inner sep=0] (image) at (0,0) {\input{ppRectanglePhiFixLambdaEpsilonp4.tex}\includegraphics[width=\textwidth]{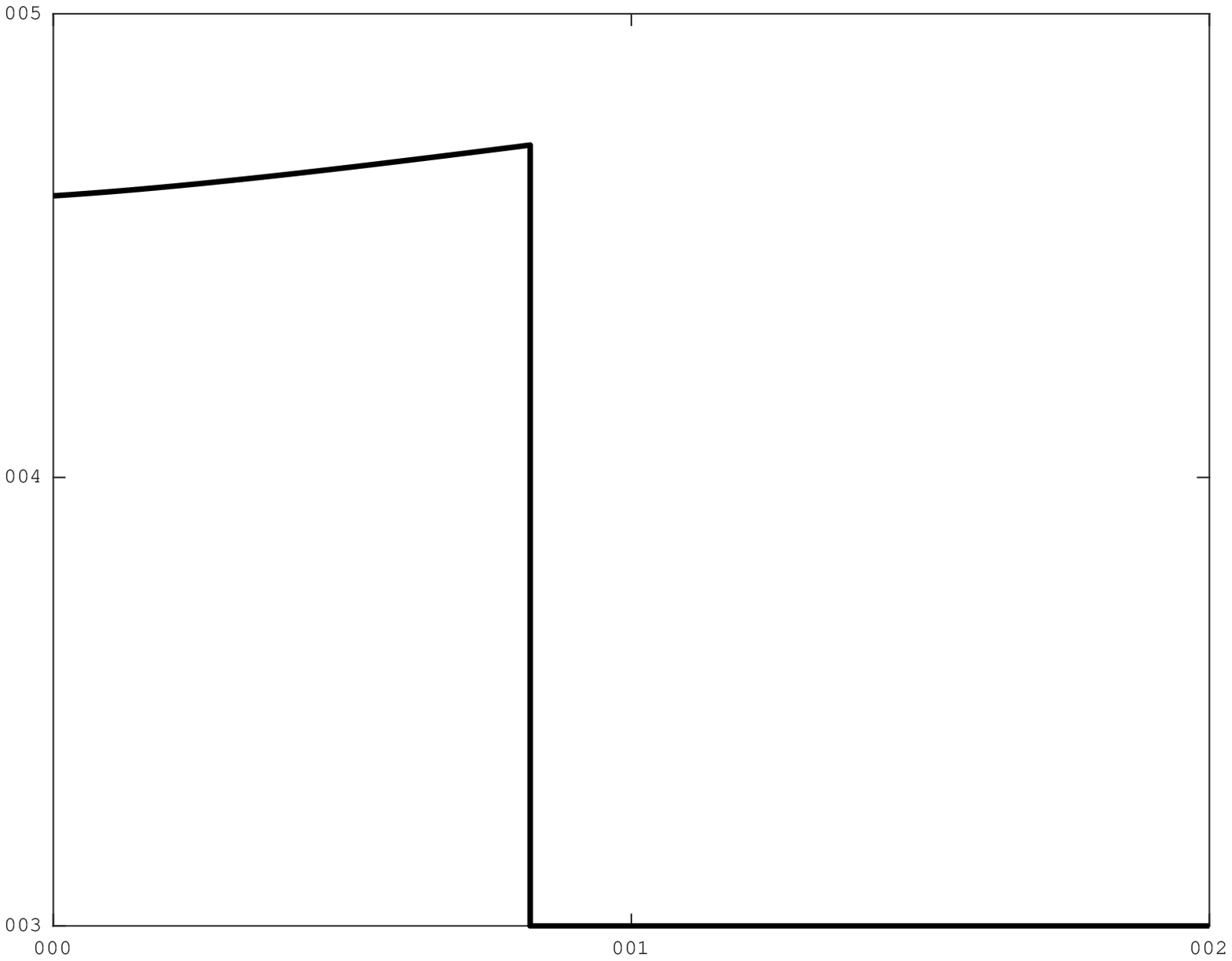}};
		\begin{scope}[x={(image.south east)},y={(image.north west)}]
		\phantom{\axislabels}
			\node [rotate=0,align=center] at (0.05,0.7) {\footnotesize $\phi$};
			\node at (0.7,0.02) {\footnotesize $y$};
    	\end{scope}
	\end{tikzpicture}
	\caption{$\epsilon=0.4$}
	\end{subfigure}
    \hfill
	\begin{subfigure}[t]{\threewide\textwidth}
        \begin{tikzpicture}
    	\node[anchor=south west,inner sep=0] (image) at (0,0) {\input{ppRectanglePhiFixLambdaEpsilonp6.tex}\includegraphics[width=\textwidth]{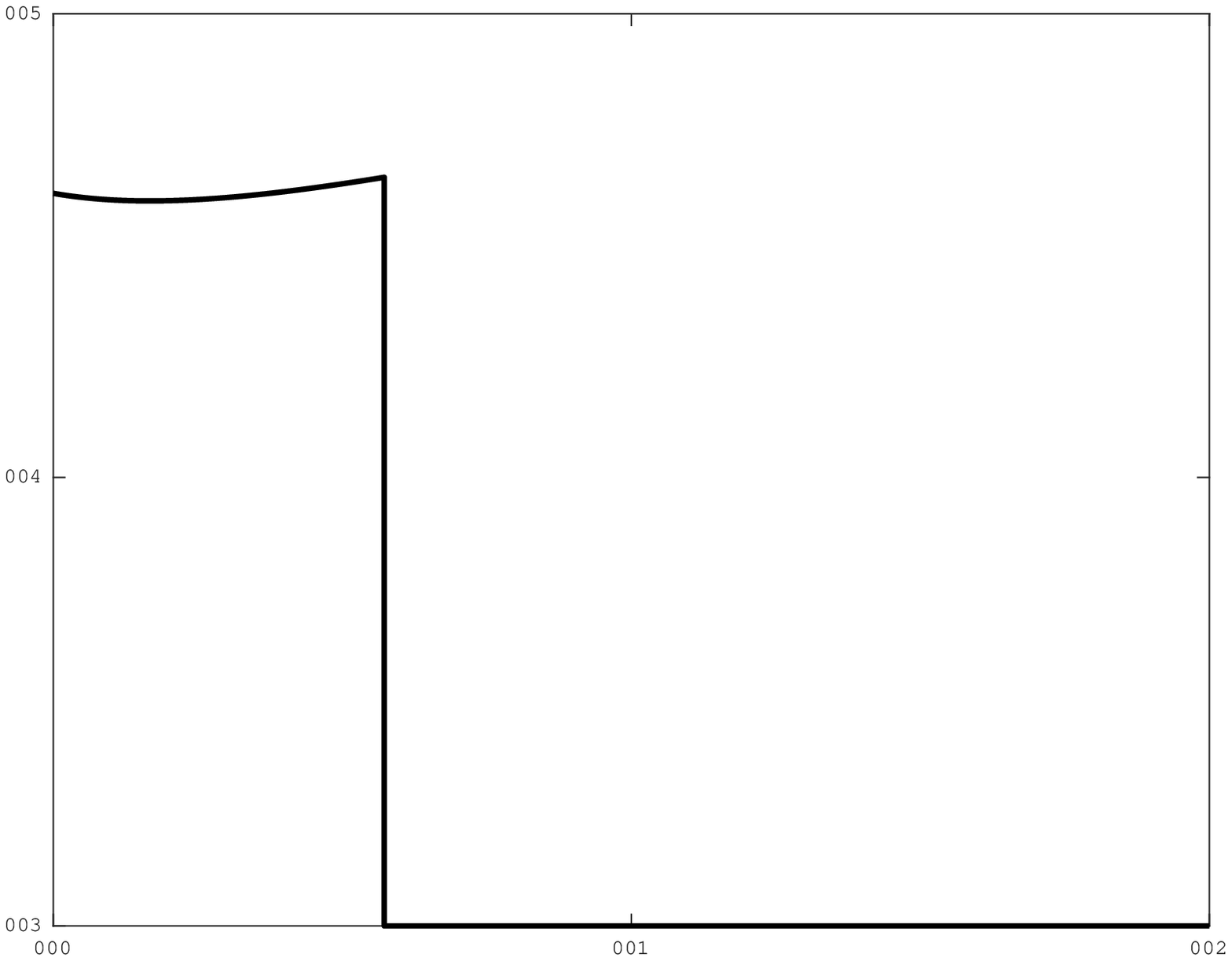}};
		\begin{scope}[x={(image.south east)},y={(image.north west)}]
		\phantom{\axislabels}
			\node [rotate=0,align=center] at (0.05,0.7) {\footnotesize $\phi$};
			\node at (0.7,0.02) {\footnotesize $y$};
    	\end{scope}
	\end{tikzpicture}
	\caption{$\epsilon=0.6$}
	\end{subfigure}
	\\
	\begin{subfigure}[t]{\threewide\textwidth}
        \begin{tikzpicture}
        		\node[anchor=south west,inner sep=0] (image) at (0,0) {\input{ppRectangleUFixLambdaEpsilonp2.tex}\includegraphics[width=\textwidth]{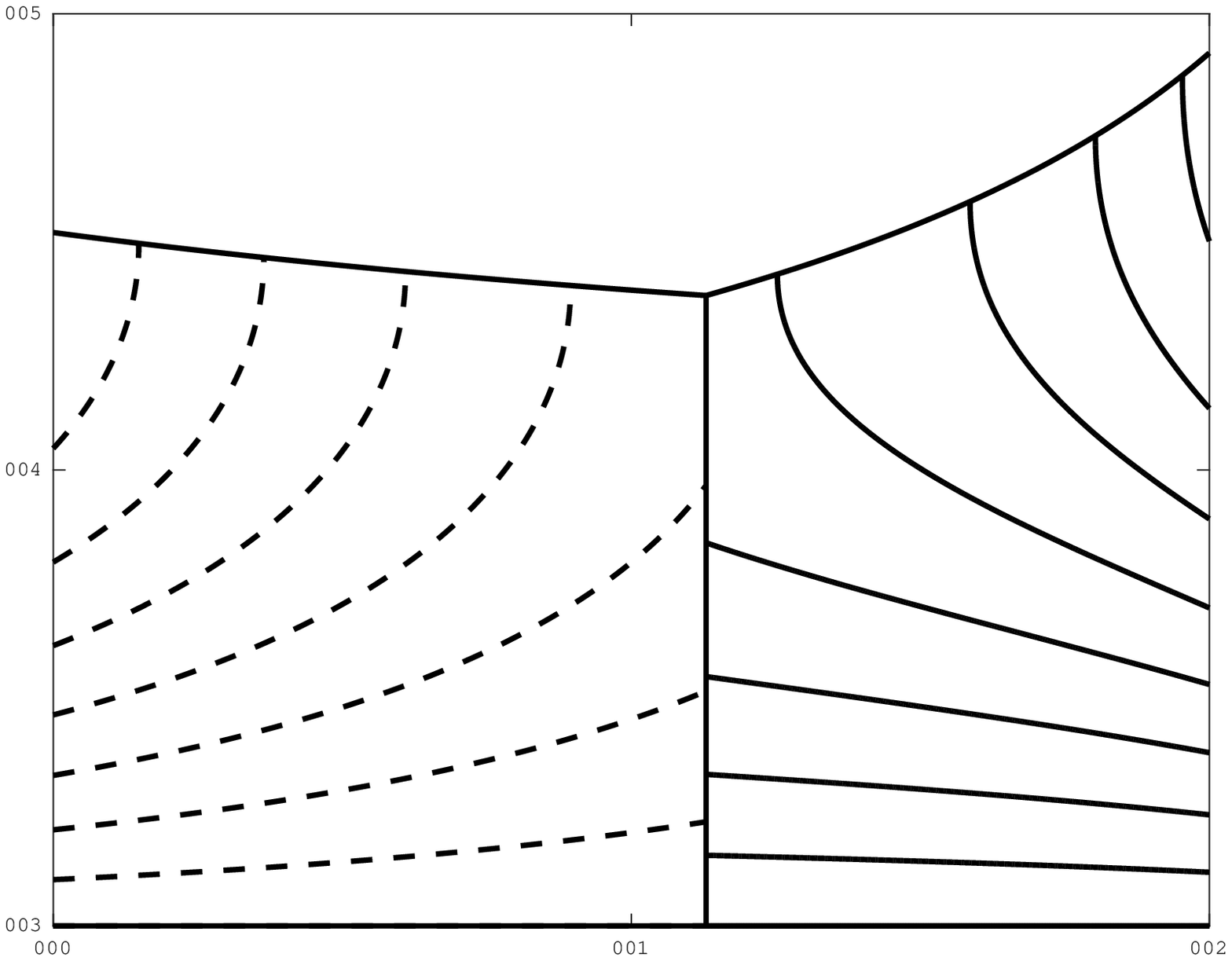}};
		\begin{scope}[x={(image.south east)},y={(image.north west)}]
			\axislabels
    		\end{scope}
	\end{tikzpicture}
	\caption{$\epsilon=0.2$}
    \end{subfigure}%
    \hfill
    \begin{subfigure}[t]{\threewide\textwidth}
        	\begin{tikzpicture}
    		\node[anchor=south west,inner sep=0] (image) at (0,0) {\input{ppRectangleUFixLambdaEpsilonp4.tex}\includegraphics[width=\textwidth]{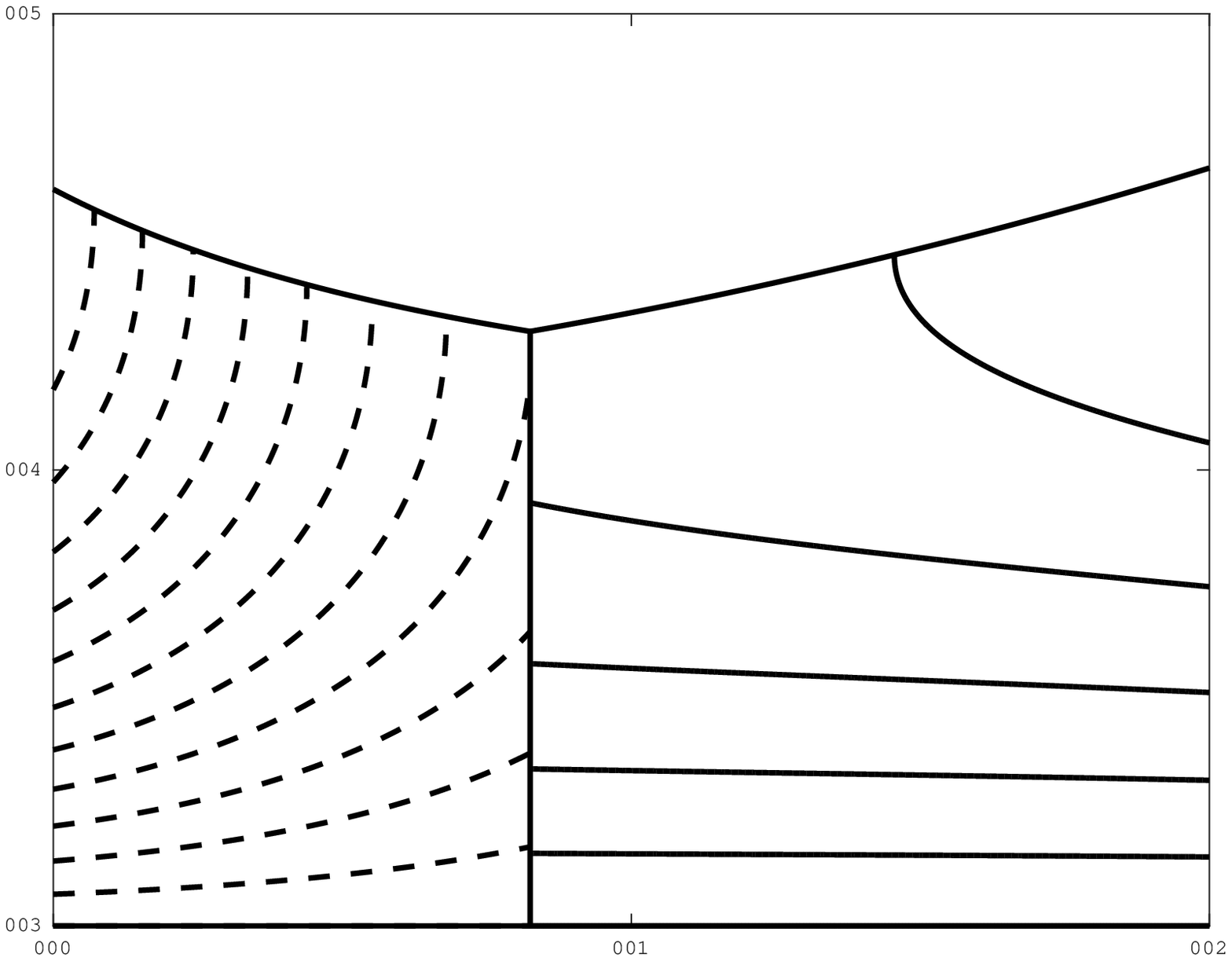}};
		\begin{scope}[x={(image.south east)},y={(image.north west)}]
			\axislabels
	    	\end{scope}
	\end{tikzpicture}
	\caption{$\epsilon=0.4$}
	\end{subfigure}%
    \hfill
    \begin{subfigure}[t]{\threewide\textwidth}
        \begin{tikzpicture}
    	\node[anchor=south west,inner sep=0] (image) at (0,0) {\input{ppRectangleUFixLambdaEpsilonp6.tex}\includegraphics[width=\textwidth]{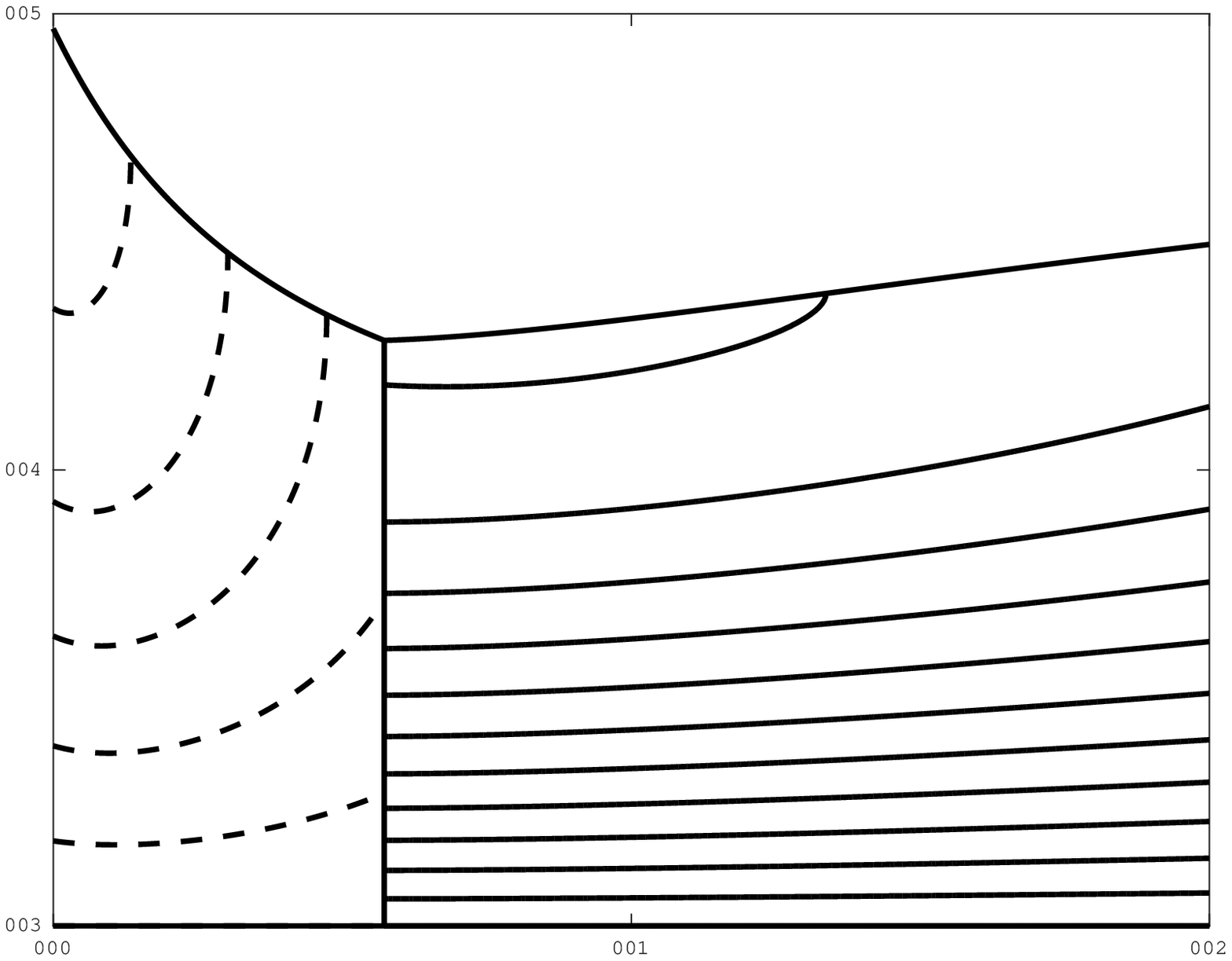}};
		\begin{scope}[x={(image.south east)},y={(image.north west)}]
			\axislabels
    	\end{scope}
	\end{tikzpicture}
	\caption{$\epsilon=0.6$}
	\end{subfigure}
  \caption{Rectangular channels with $\lambda=0.5$ \ys{and values of $\epsilon$ as shown,} 
  for $Q=0.15$ and $Q_p=0.003$. The top row shows the free-surface profile and stream-function of the secondary flow in the particle-rich (dashed lines) and particle-free (solid lines) regions. The middle row shows the particle volume fraction $\phi(y)$. The bottom row shows contours of the axial velocity $u$ in both the particle-rich (dashed lines) and particle-free regions (solid lines). \ys{The scaling of the streamlines and axial-flow contours differs between the regions, with the magnitude of the flow velocity being significantly smaller} 
  in the particle-rich region.}
  \label{fig:RectangleParticlesEpsilon}
\end{figure}

\Cref{fig:RectangleParticlesEpsilon} shows the effect of changing $\epsilon$ on the resulting flow in a channel of rectangular cross-section. The total flux and particle flux are the same in each plot, $Q=0.15$ and $Q_p=0.003$. The addition of the particle-rich zone causes significant differences from clear-fluid solutions, with a sharp change in the free-surface slope at the interface between the particle-rich and particle-free regions, separate secondary flow profiles in the two regions, and much lower axial velocity in the particle rich region. The streamlines in the particle-rich and particle-free regions are not plotted at the same contour levels, as the magnitude of the streamfunction in the particle-free region is larger than in the particle-rich region. There are some similarities with the clear-fluid solutions from \citet{ASG2015,ASG2016}; increasing $\epsilon$ tends to push the mixture towards the inside of the channel, increasing the depth there and decreasing it at the outside of the channel.

The particle volume fraction $\phi(y)$ is also shown in \cref{fig:RectangleParticlesEpsilon}. In all three cases, the particle-rich region of the flow has a high volume fraction of particles, near to $\phi_m=0.61$, but \ys{everywhere} 
below 0.6, \ys{which, as discussed earlier, we interpret to mean no beaching}. The effect of $\epsilon$ on the location of the interface between the particle-rich and particle-free regions of the flow ($y_r^*$) is complex. \ys{The value of $\epsilon$ modifies the local slope $\Lambda(y)=\lambda/(1+\epsilon y)$ at each position $y$ and, therefore, the volume fraction of particles at each $y$ changes in a nonlinear manner according to the relation \eqref{eq:phieqn} plotted in \cref{fig:phip}.} As $\epsilon$ increases,  $y_r^*$ initially decreases, but as $\epsilon$ increases further, $y_r^*$ starts increasing again. Depending on the slope of the channel centreline $\lambda$, we can observe three behaviours: $y_r^*$ decreasing as $\epsilon$ increases (which occurs for large $\lambda$'s), $y_r^*$ increasing as $\epsilon$ increases (which occurs for small $\lambda$'s), and $y_r^*$ decreasing then increasing as $\epsilon$ increases (which occurs for intermediate $\lambda$'s). In \cref{fig:RectangleParticlesEpsilon}, $y_r^*$ decreases as $\epsilon$ increases, but if $\epsilon$ were increased enough, $y_r^*$ would eventually start to increase. 

\begin{figure}
    \centering
    \begin{tikzpicture}
    	\node[anchor=south west,inner sep=0] (image) at (0,0) {\input{pp_ContoursOfYc.tex}\includegraphics[width=\onewide\textwidth]{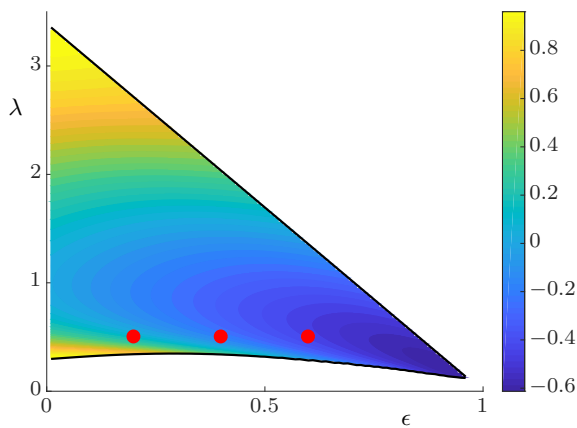}};
		\begin{scope}[x={(image.south east)},y={(image.north west)}]
				\node [align=center] at (0.08,0.72) {$\lambda$};
				\node at (0.7,0.05) {$\epsilon$};
    	\end{scope}
	\end{tikzpicture}
	\caption{Values of $y_r^*$ for a rectangular channel with $Q=0.15$, $Q_p=0.003$. Black curves show the boundary of the \ys{valid region of the $(\epsilon,\lambda)$ parameter space} where $\phi(y)<\phi_m$ everywhere in the particle-rich region, \ys{as given by \eqref{valid_region} with $y_\ell=-1$}. The filled circles correspond to the parameters for \cref{fig:RectangleParticlesEpsilon}.}\label{fig:ycContour}
\end{figure}

\Cref{fig:ycContour} shows the dependence of $y_r^*$ on the channel centreline geometry ($\epsilon$ and $\lambda$), \ys{within the valid portion of this parameter space,} for a channel \ys{of rectangular cross-section} with \ys{the same total and particle fluxes as used for \cref{fig:RectangleParticlesEpsilon} ($Q=0.15$, $Q_p=0.003$).} 
\ys{From \eqref{particle_limits} and noting that $y_\ell^*=y_\ell$ (where $y_\ell=-1$ for the rectangular channel), the valid part of the ($\epsilon,\lambda$) parameter space is given by
\begin{align}
(1+\epsilon y_r^*)(\chi-\sqrt{\chi^2-1})<\lambda<(1+\epsilon y_\ell)(\chi+\sqrt{\chi^2-1}),\label{valid_region}
\end{align}
which is the coloured region; outside of this coloured region there is no solution to the problem that maintains $\phi(y)<\phi_m$ in the particle-rich region.} This plot helps to make sense of \ys{the three behaviours noted above.}  We see that, 
for fixed $\epsilon$, there can be a decrease in $y_r^*$ as $\lambda$ increases, followed by an increase in $y_r^*$ as $\lambda$ increases further. \ys{Further useful data that can be obtained} 
from this plot \ys{are} 
the values of $\epsilon$ and $\lambda$ for which the particle-rich region is smallest, \ys{i.e. $y_r^*$ is minimised}. As mentioned earlier, reducing the radial extent of the particle-rich region could potentially be beneficial in spiral particle separators. For $\rho_s=1.5$, the case considered in \cref{fig:ycContour}, the radial extent of the particle-rich region is minimised when $\epsilon$ is maximised. This agrees with our physical intuition, the larger $\epsilon$, the more gravity \ys{dominates inertia} 
and, hence, particles (and fluid) tend to aggregate near the inside channel wall.

\bl{Also shown in \cref{fig:RectangleParticlesEpsilon} are contours of the axial velocity. Axial velocity is always zero at the channel bottom, but is non-zero on the side walls. This is a result of our thin-film scaling, with does not allow us to impose boundary conditions on the side walls. The only condition we can impose is that there is no net flux through the side walls. In practice we would expect thin boundary layers to form at the side walls, as well as at the interface between the particle-rich and particle-free regions where the fluid properties and velocity are not continuous. The axial velocity contours show that the velocity increases with \ys{height above the channel bottom,} 
but there are other geometric effects that move the point of maximum axial velocity. For example, in the particle-free region, the point of maximum axial velocity is at the outer wall for $\epsilon=0.2$ and $\epsilon=0.4$, but moves to near the middle of the channel for $\epsilon=0.6$.}

\begin{figure}
    \centering
    \begin{subfigure}[t]{\threewide\textwidth}
    \centering
        \begin{tikzpicture}
    	\node[anchor=south west,inner sep=0] (image) at (0,0) {\input{ppParabolaFixEpsilonLambdap4.tex}\includegraphics[width=\textwidth]{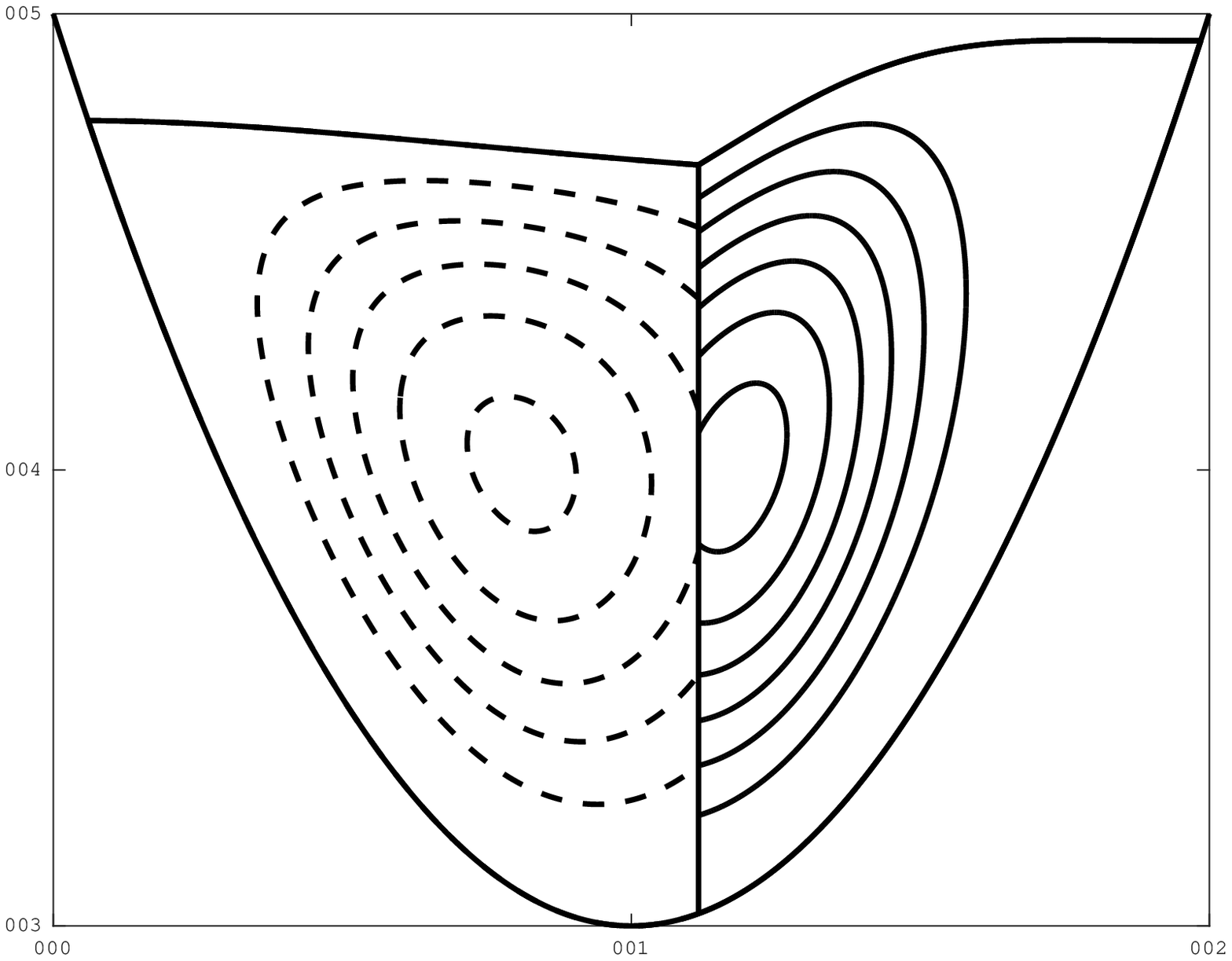}};
		\begin{scope}[x={(image.south east)},y={(image.north west)}]
			\axislabels
    	\end{scope}
	\end{tikzpicture}
	\caption{$\lambda=0.4$}
	\end{subfigure}
	\begin{subfigure}[t]{\threewide\textwidth}
	\centering
        \begin{tikzpicture}
    	\node[anchor=south west,inner sep=0] (image) at (0,0) {\input{ppParabolaFixEpsilonLambdap7.tex}\includegraphics[width=\textwidth]{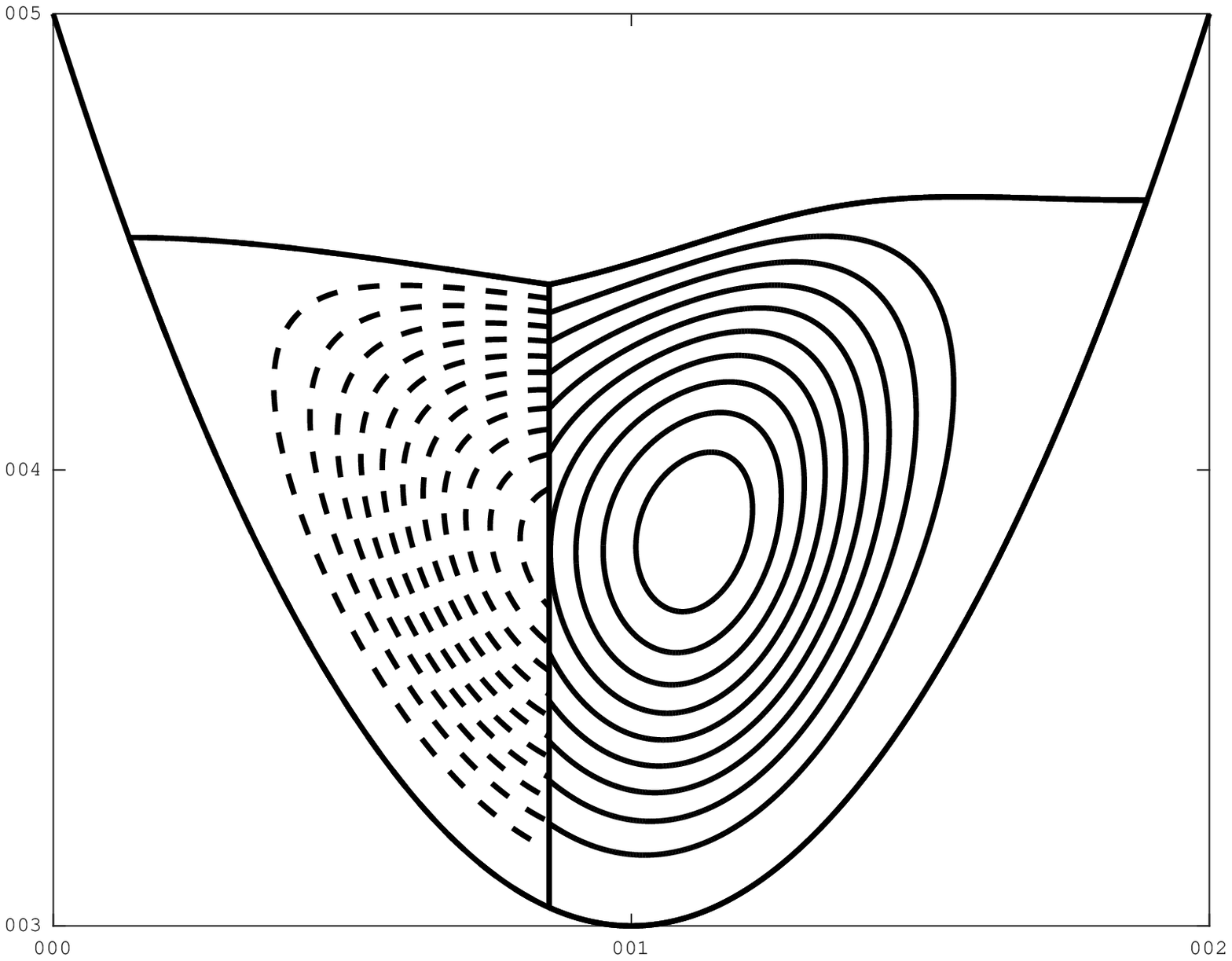}};
		\begin{scope}[x={(image.south east)},y={(image.north west)}]
			\axislabels
    	\end{scope}
	\end{tikzpicture}
	\caption{$\lambda=0.7$}
	\end{subfigure}
\begin{subfigure}[t]{\threewide\textwidth}
    \centering
        \begin{tikzpicture}
    	\node[anchor=south west,inner sep=0] (image) at (0,0) {\input{ppParabolaFixEpsilonLambda1.tex}\includegraphics[width=\textwidth]{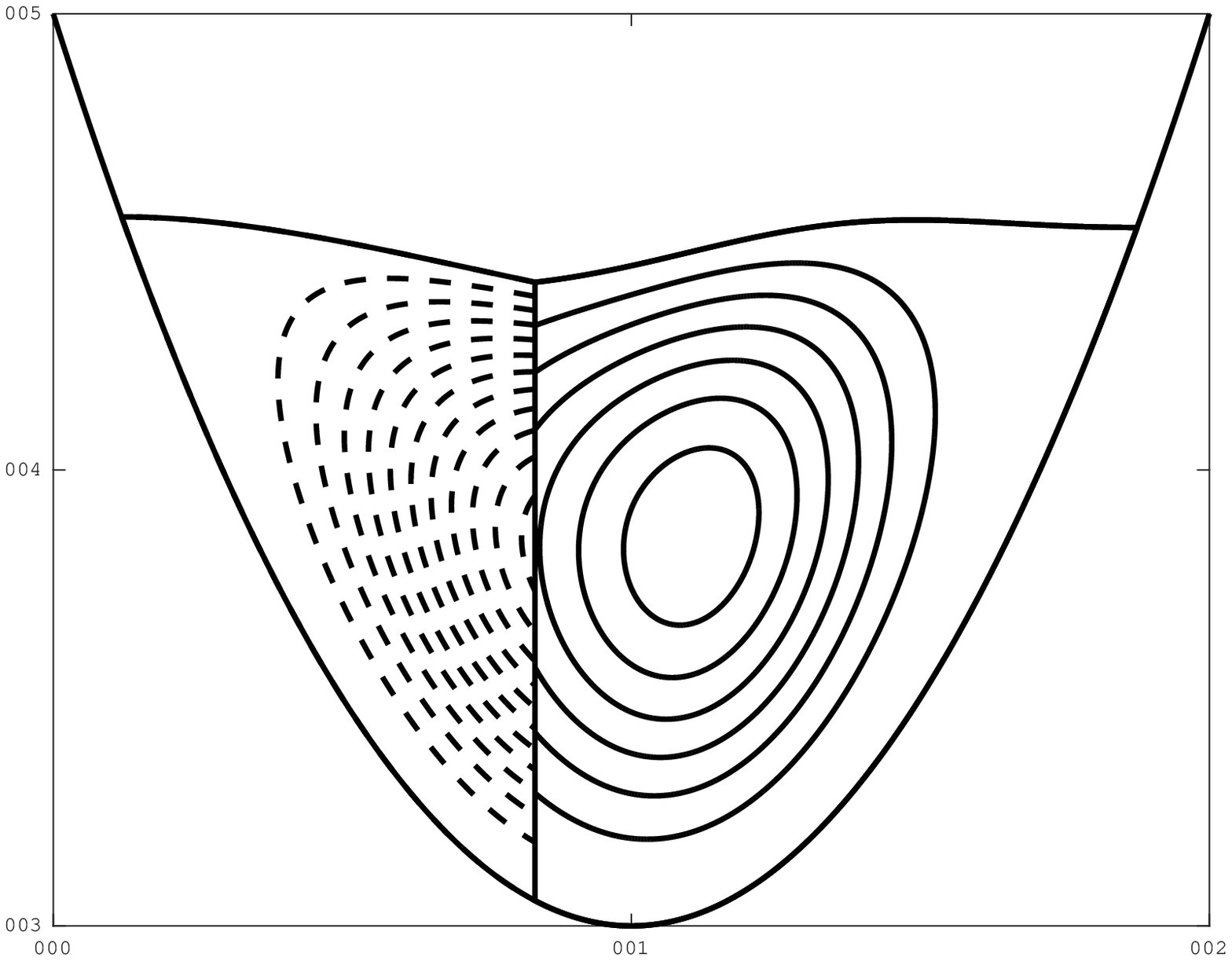}};
		\begin{scope}[x={(image.south east)},y={(image.north west)}]
			\axislabels
    	\end{scope}
	\end{tikzpicture}
	\caption{$\lambda=1$}
	\end{subfigure}
	\\
	\begin{subfigure}[t]{\threewide\textwidth}
    \centering
        \begin{tikzpicture}
    	\node[anchor=south west,inner sep=0] (image) at (0,0) {\input{ppParabolaPhiFixEpsilonLambdap4.tex}\includegraphics[width=\textwidth]{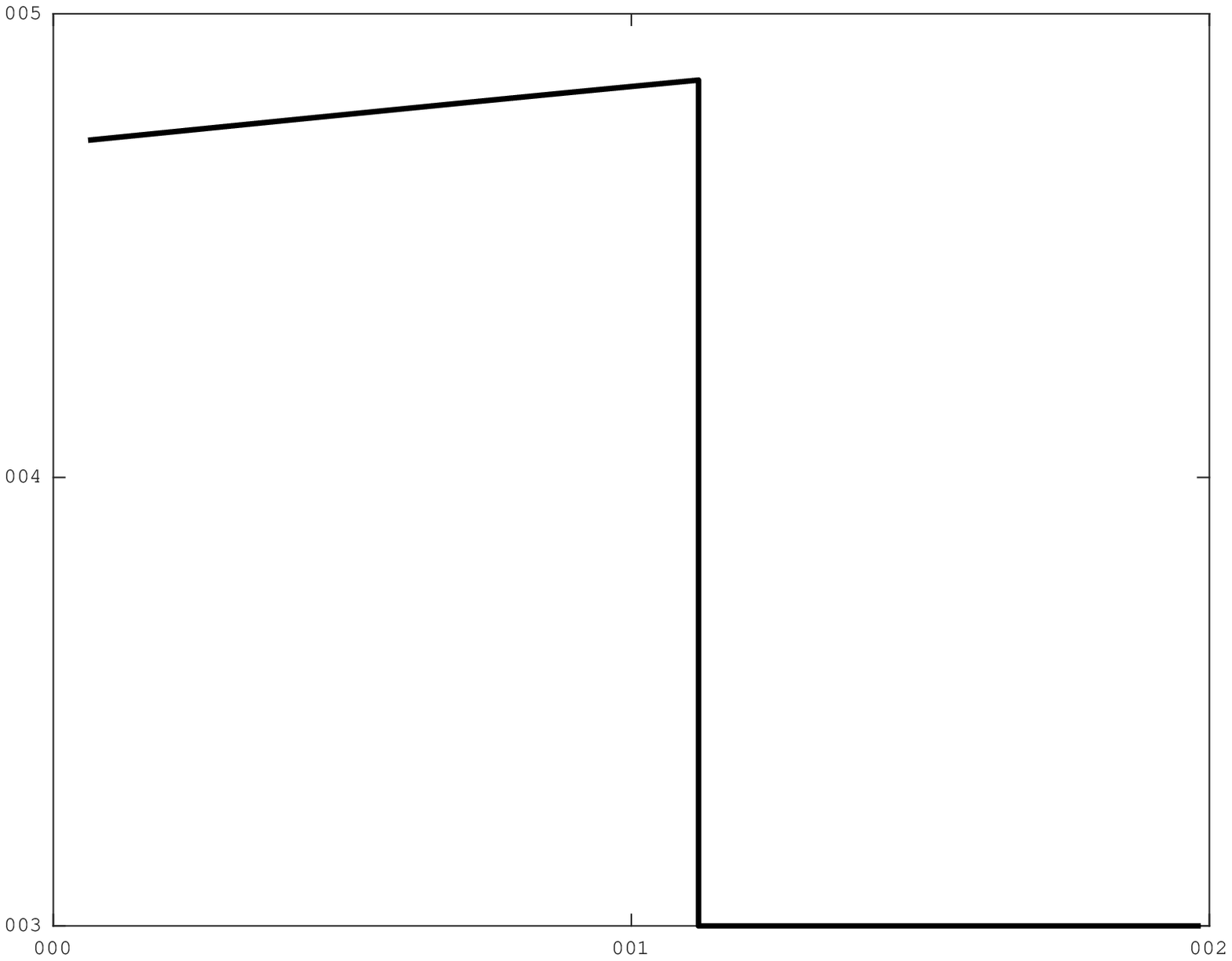}};
		\begin{scope}[x={(image.south east)},y={(image.north west)}]
			\phantom{\axislabels}
			\node [rotate=0,align=center] at (0.05,0.7) {\footnotesize $\phi$};
			\node at (0.7,0.02) {\footnotesize $y$};
    	\end{scope}
	\end{tikzpicture}
	\caption{$\lambda=0.4$}
	\end{subfigure}
	\begin{subfigure}[t]{\threewide\textwidth}
	\centering
        \begin{tikzpicture}
    	\node[anchor=south west,inner sep=0] (image) at (0,0) {\input{ppParabolaPhiFixEpsilonLambdap7.tex}\includegraphics[width=\textwidth]{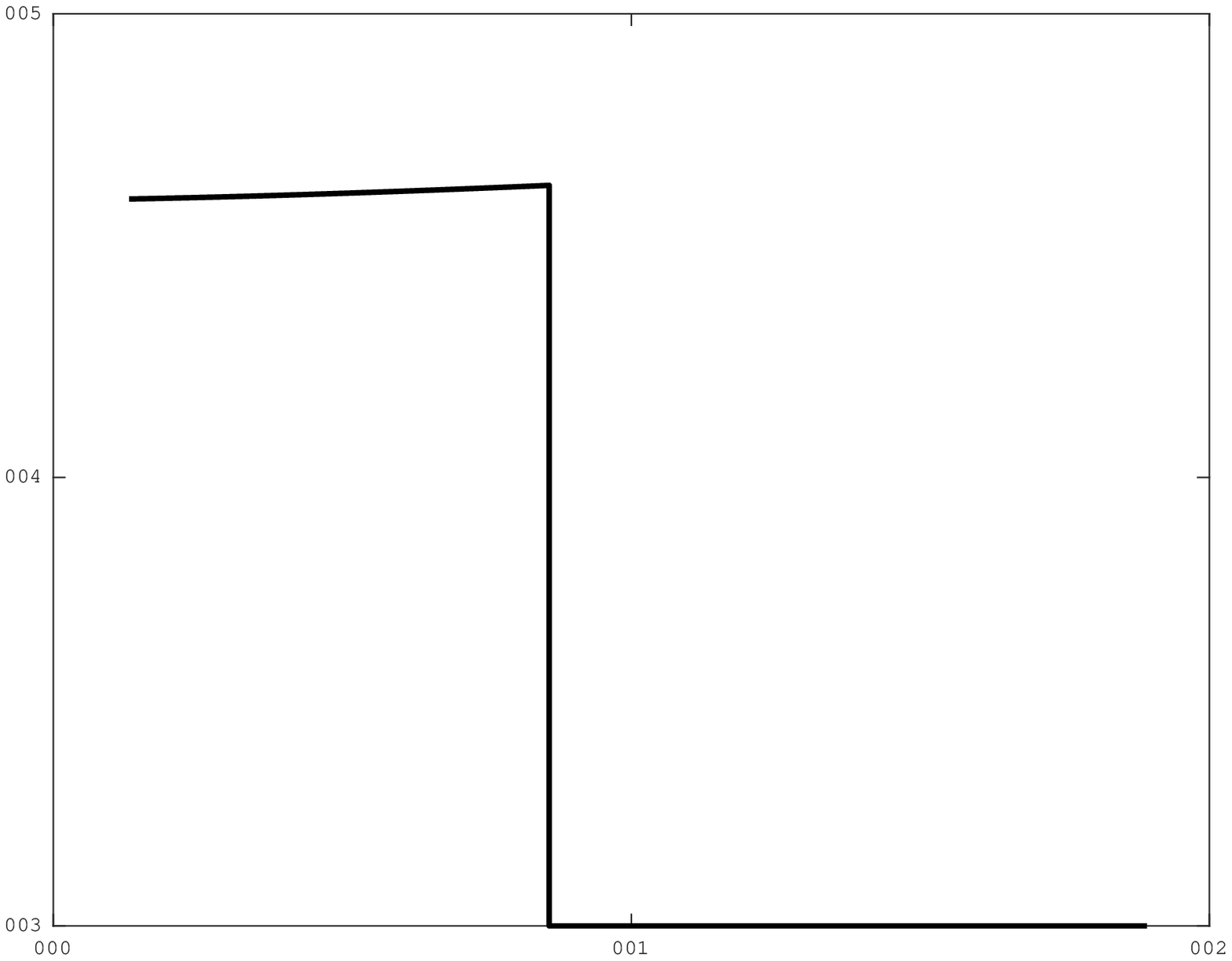}};
		\begin{scope}[x={(image.south east)},y={(image.north west)}]
			\phantom{\axislabels}
			\node [rotate=0,align=center] at (0.05,0.7) {\footnotesize $\phi$};
			\node at (0.7,0.02) {\footnotesize $y$};
    	\end{scope}
	\end{tikzpicture}
	\caption{$\lambda=0.7$}
	\end{subfigure}
\begin{subfigure}[t]{\threewide\textwidth}
    \centering
        \begin{tikzpicture}
    	\node[anchor=south west,inner sep=0] (image) at (0,0) {\input{ppParabolaPhiFixEpsilonLambda1.tex}\includegraphics[width=\textwidth]{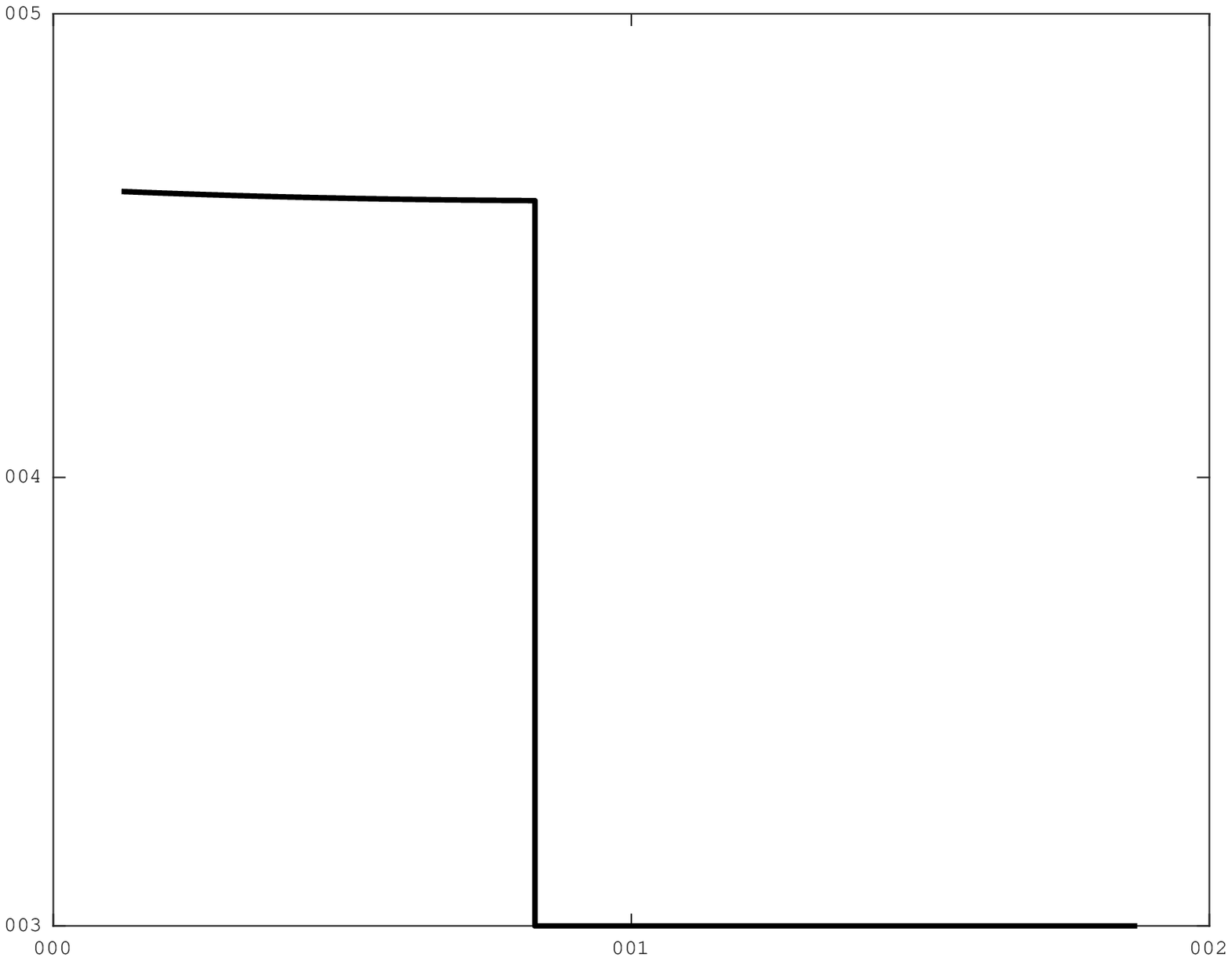}};
		\begin{scope}[x={(image.south east)},y={(image.north west)}]
			\phantom{\axislabels}
			\node [rotate=0,align=center] at (0.05,0.7) {\footnotesize $\phi$};
			\node at (0.7,0.02) {\footnotesize $y$};
    	\end{scope}
	\end{tikzpicture}
	\caption{$\lambda=1$}
	\end{subfigure}
	\\
	\begin{subfigure}[t]{\threewide\textwidth}
    \centering
        \begin{tikzpicture}
    	\node[anchor=south west,inner sep=0] (image) at (0,0) {\input{ppParabolaUFixEpsilonLambdap4.tex}\includegraphics[width=\textwidth]{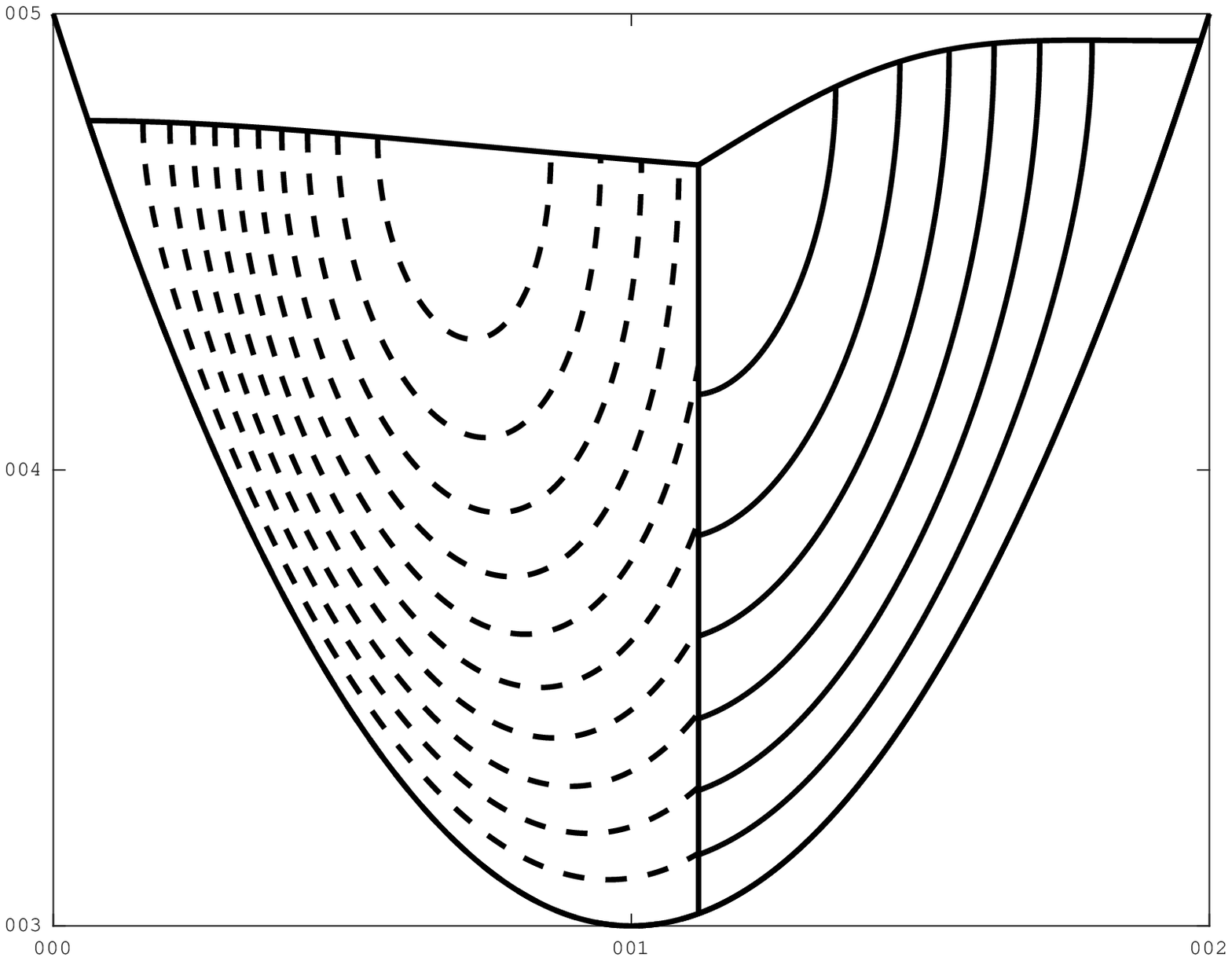}};
		\begin{scope}[x={(image.south east)},y={(image.north west)}]
			\axislabels
    	\end{scope}
	\end{tikzpicture}
	\caption{$\lambda=0.4$}
	\end{subfigure}
	\begin{subfigure}[t]{\threewide\textwidth}
	\centering
        \begin{tikzpicture}
    	\node[anchor=south west,inner sep=0] (image) at (0,0) {\input{ppParabolaUFixEpsilonLambdap7.tex}\includegraphics[width=\textwidth]{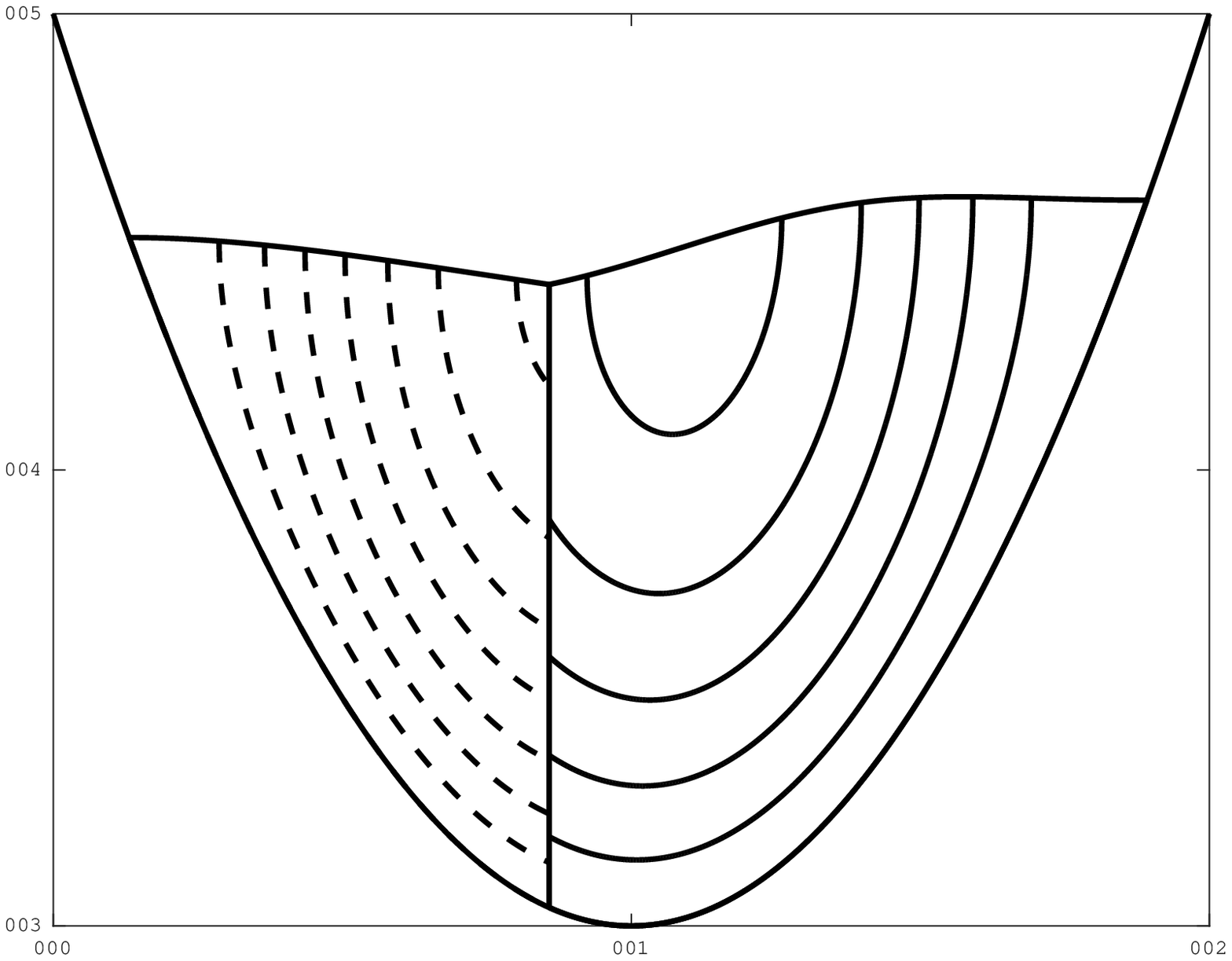}};
		\begin{scope}[x={(image.south east)},y={(image.north west)}]
			\axislabels
    	\end{scope}
	\end{tikzpicture}
	\caption{$\lambda=0.7$}
	\end{subfigure}
\begin{subfigure}[t]{\threewide\textwidth}
    \centering
        \begin{tikzpicture}
    	\node[anchor=south west,inner sep=0] (image) at (0,0) {\input{ppParabolaUFixEpsilonLambda1.tex}\includegraphics[width=\textwidth]{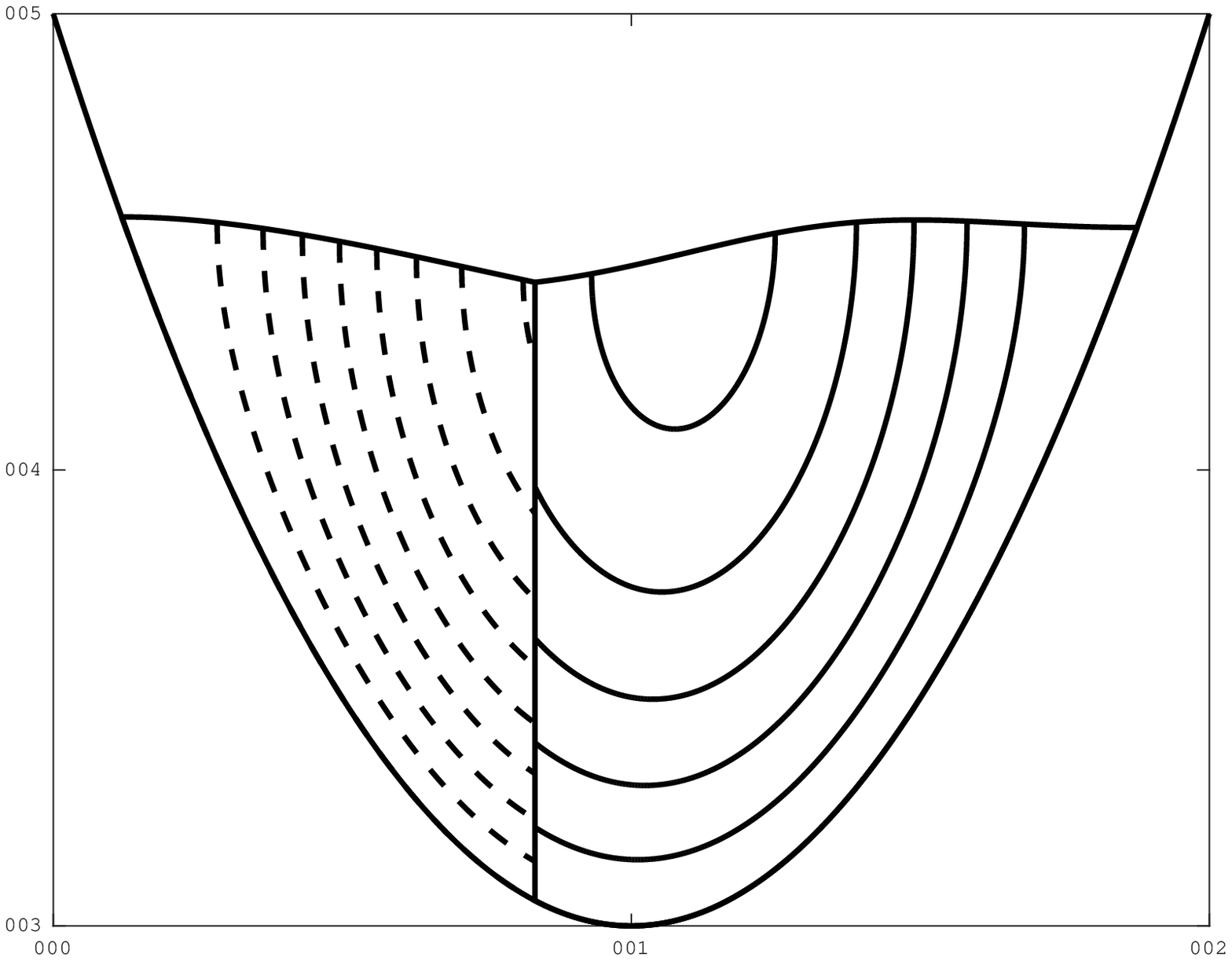}};
		\begin{scope}[x={(image.south east)},y={(image.north west)}]
			\axislabels
    	\end{scope}
	\end{tikzpicture}
	\caption{$\lambda=1$}
	\end{subfigure}
    \caption{Parabolic channels $H(y)=y^2$ with $\epsilon=0.25$ \ys{and values of $\lambda$ as shown,} for $Q=0.08$ and $Q_p=0.001$. The top row shows the free-surface profile and stream-function of the secondary flow in the particle-rich (dashed lines) and particle-free (solid lines) regions. The middle row shows the particle volume fraction $\phi(y)$. \bl{The bottom row shows contours of the \ys{axial velocity $u$ in both the particle-rich (dashed lines) and particle-free regions (solid lines). The scaling of the streamlines and axial-flow contours differs between the regions, with the magnitude of the flow velocity being significantly smaller in the particle-rich region.}}
}\label{fig:ParabolaParticlesLambda}
\end{figure}

\begin{figure}
    \centering
    \begin{tikzpicture}
    	\node[anchor=south west,inner sep=0] (image) at (0,0) {\input{pp_ContoursOfYcParabola.tex}\includegraphics[width=\onewide\textwidth]{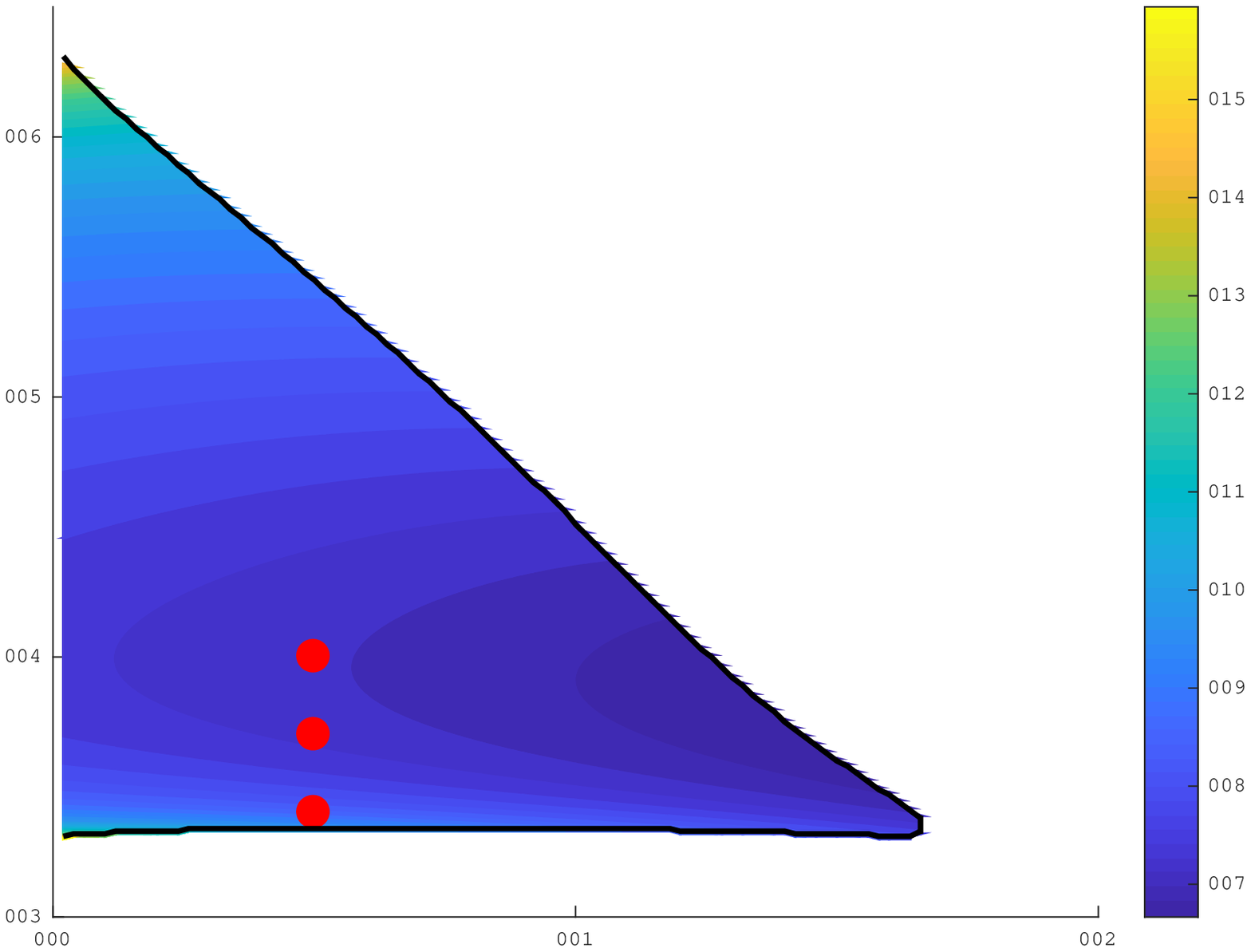}};
		\begin{scope}[x={(image.south east)},y={(image.north west)}]
				\node [align=center] at (0.08,0.72) {$\lambda$};
				\node at (0.7,0.05) {$\epsilon$};
    	\end{scope}
	\end{tikzpicture}
	\caption{Values of $y_r^*$ for a parabolic channel $H(y)=y^2$ with $Q=0.08$, $Q_p=0.001$. \ys{Black curves show the boundary of the valid region of the $(\epsilon,\lambda)$ parameter space where $\phi(y)<\phi_m$ everywhere in the particle-rich region, as given by \eqref{valid_region}}. The filled circles correspond to the parameters for \cref{fig:ParabolaParticlesLambda}.}\label{fig:ContoursOfYcParabola}
\end{figure}

\Cref{fig:ParabolaParticlesLambda} shows the effect of changing $\lambda$ on the resulting flow in a parabolic channel with $H(y)=y^2$ and fixed $\epsilon=0.25$. The particle flux and total mixture fluxes are the same for all three plots, $Q=0.08$ and $Q_p=0.001$. There is little qualitative difference in the solutions except for the change in the location of the interface between the particle-rich and particle-free regions. \bl{The axial velocity contours do change slightly, with the location of maximum axial velocity moving outwards as $\lambda$ increases up to $\lambda=1$.} \Cref{fig:ContoursOfYcParabola} shows \bl{values} of $y_r^*$ \ys{in the valid region of the $(\epsilon,\lambda)$ parameter space as given by \eqref{valid_region}. Note that, the solution process for this channel geometry differs from that for the rectangular channel, in that we set $h_\ell=h_r=0$ and determine the free-surface contact points $y_\ell$ and $y_r$, and the location of the fluid-particle interface $y_r^*$. Hence $y_r^*=y_r$ varies with $\epsilon$ and $\lambda$.} 
Not only is the effect of $\lambda$ relatively minor, but the effect of $\epsilon$ is even smaller, with the contour levels nearly horizontal in much of the parameter space. 
For channels with moderate steepness ($\lambda$ less than roughly 2 or 3), the extent of the particle-rich region is relatively insensitive to both $\epsilon$ and $\lambda$. Compared to the rectangular channel case \cref{fig:ycContour}, $y_r^*$ is less sensitive to $\lambda$ for parabolic channels.
The centreline geometry that minimises the radial extent of the particle rich region is $\epsilon\approx0.7$ and $\lambda\approx0.7$.

A comparison of \cref{fig:RectangleParticlesEpsilon,fig:ParabolaParticlesLambda} suggests that the shape of the channel cross-section has a more significant impact on the secondary flow and particle volume fraction distribution than do the parameters $\epsilon$ and $\lambda$. The parabolic channel allows less freedom for the secondary flow and free-surface profile to change than in the rectangular channel case. \ys{Note too that the no-slip boundary conditions are satisfied on the entire surface of the parabolic channel which, unlike the rectangular channel, has no vertical walls on which the boundary conditions must be relaxed.} In the next section, we will further explore how the channel shape can be changed in order to reduce the radial extent of the particle-rich region.

\subsection{The effect of a trench in the channel bottom}
\begin{figure}
    \centering
    \begin{subfigure}[t]{\threewide\textwidth}
    \centering
        \begin{tikzpicture}
    	\node[anchor=south west,inner sep=0] (image) at (0,0) {\input{ppBumpHeight0.tex}\includegraphics[width=\textwidth]{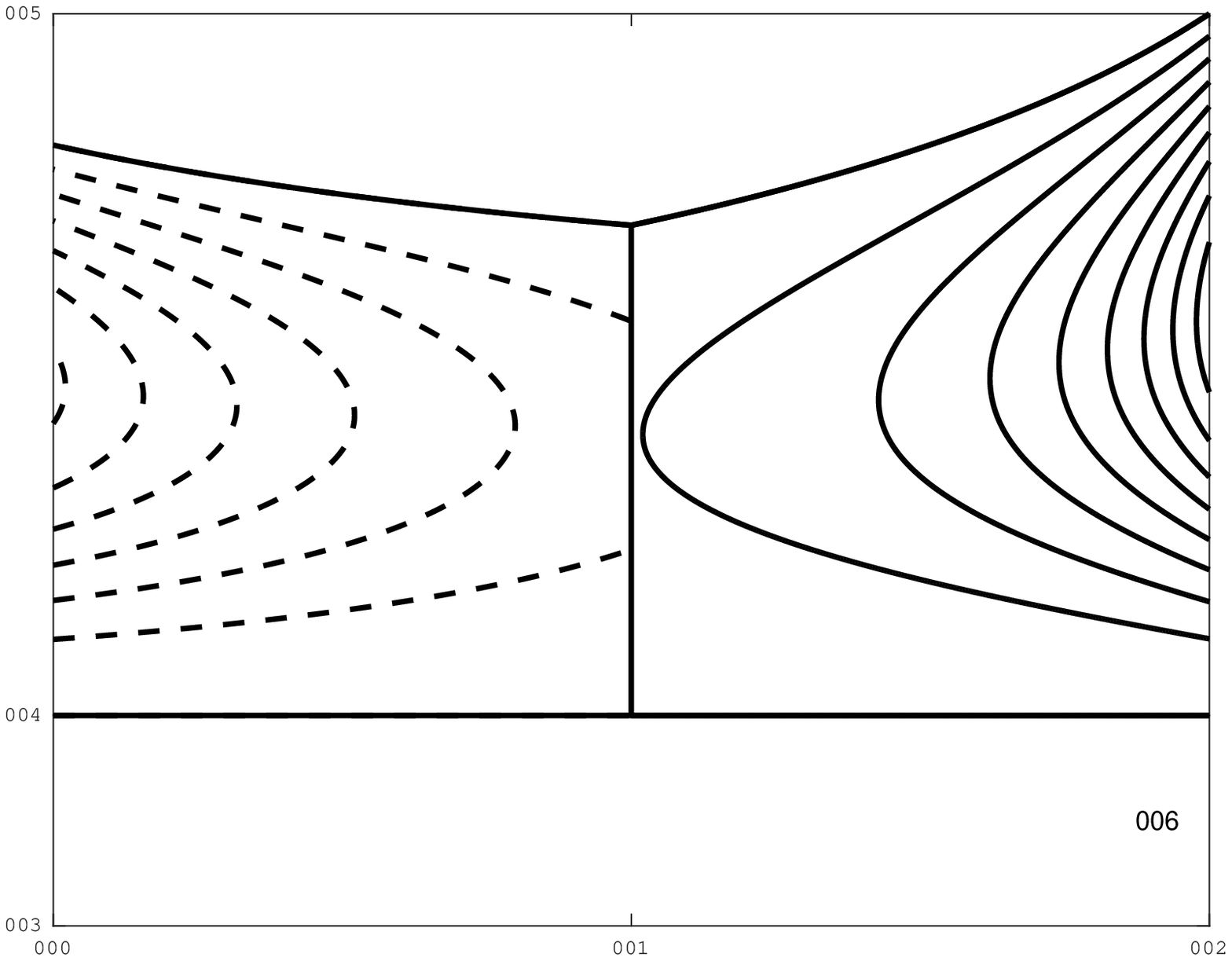}};
		\begin{scope}[x={(image.south east)},y={(image.north west)}]
			\axislabels
    	\end{scope}
	\end{tikzpicture}
	\caption{$k=0$}
	\end{subfigure}
    \begin{subfigure}[t]{\threewide\textwidth}
    \centering
        \begin{tikzpicture}
    	\node[anchor=south west,inner sep=0] (image) at (0,0) {\input{ppBumpHeightp15.tex}\includegraphics[width=\textwidth]{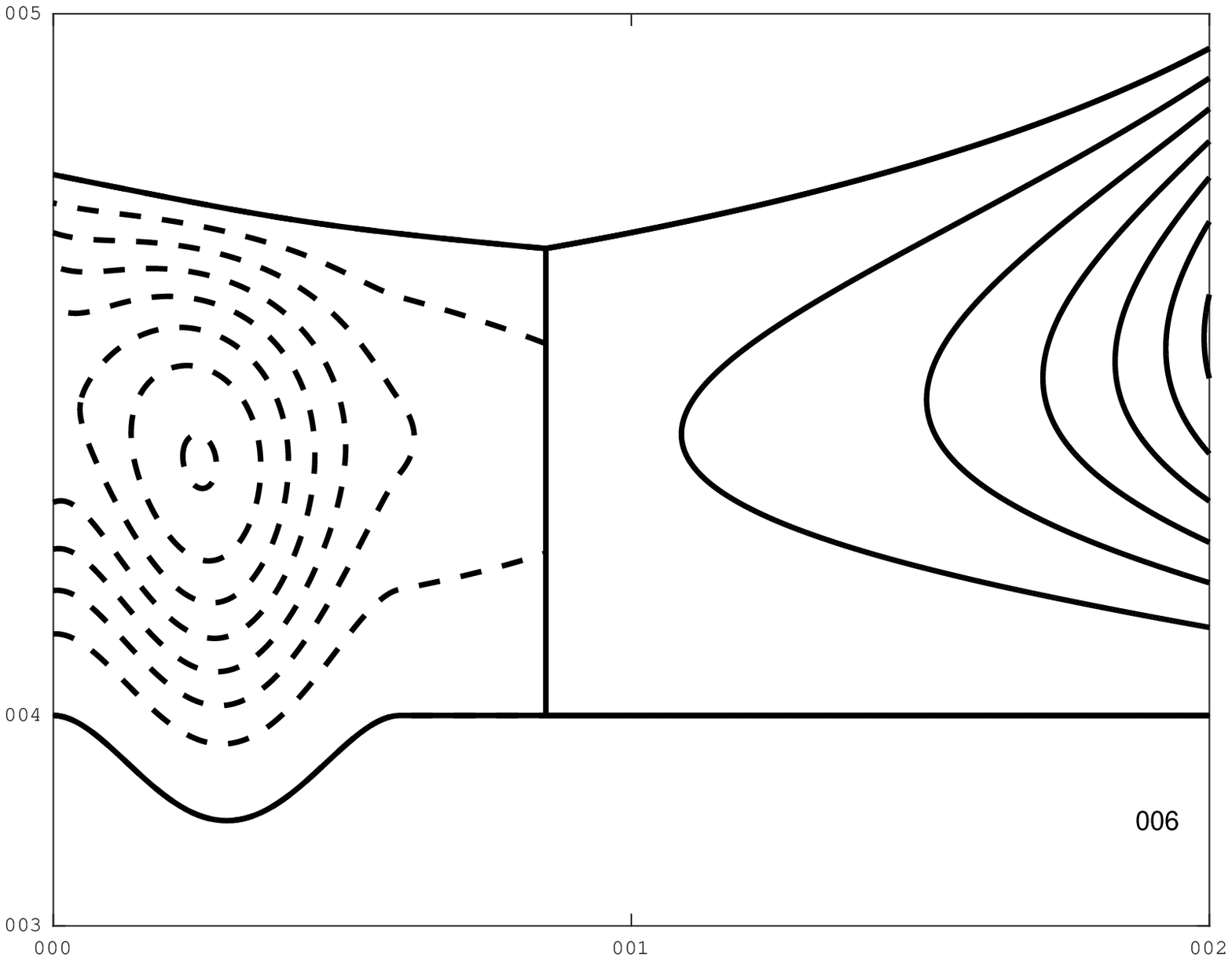}};
		\begin{scope}[x={(image.south east)},y={(image.north west)}]
			\axislabels
    	\end{scope}
	\end{tikzpicture}
	\caption{$k=0.15$}\label{fig:trench1}
	\end{subfigure}
    \begin{subfigure}[t]{\threewide\textwidth}
    \centering
        \begin{tikzpicture}
    	\node[anchor=south west,inner sep=0] (image) at (0,0) {\input{ppBumpHeightp3.tex}\includegraphics[width=\textwidth]{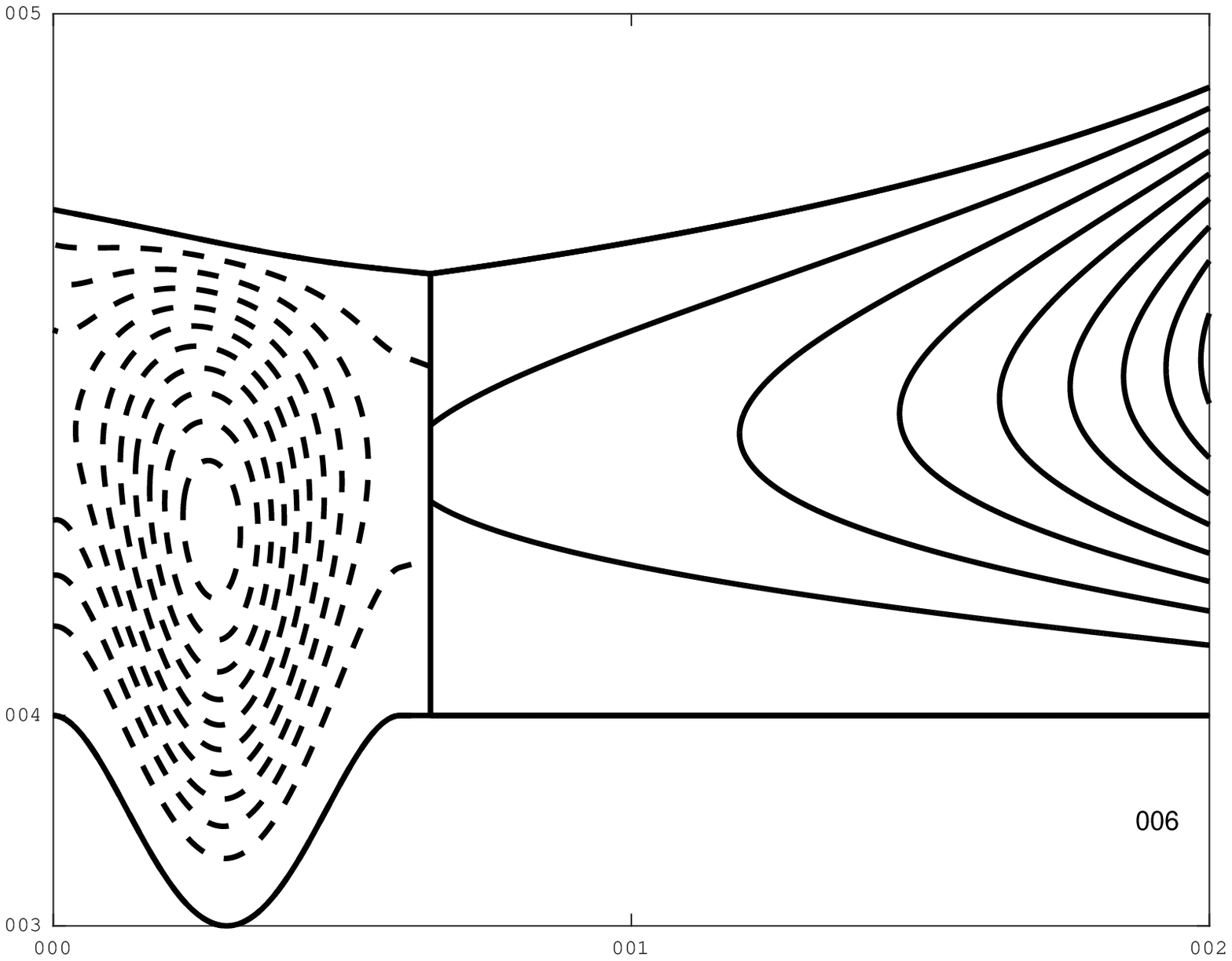}};
		\begin{scope}[x={(image.south east)},y={(image.north west)}]
			\axislabels
    	\end{scope}
	\end{tikzpicture}
	\caption{$k=0.3$}\label{fig:trench2}
	\end{subfigure}
	\\
	\begin{subfigure}[t]{\threewide\textwidth}
    \centering
        \begin{tikzpicture}
    	\node[anchor=south west,inner sep=0] (image) at (0,0) {\input{ppUBumpHeight0.tex}\includegraphics[width=\textwidth]{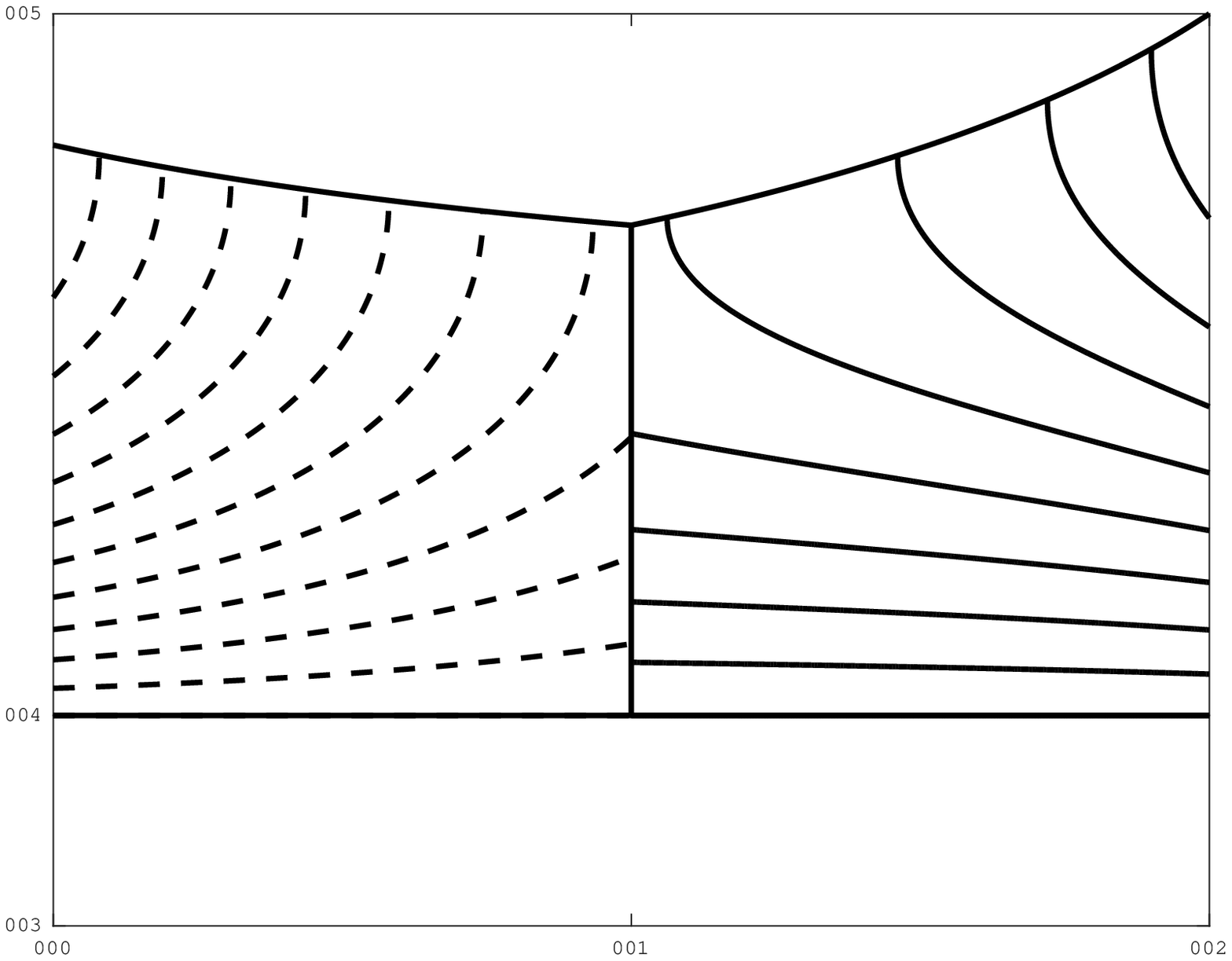}};
		\begin{scope}[x={(image.south east)},y={(image.north west)}]
			\axislabels
    	\end{scope}
	\end{tikzpicture}
	\caption{$k=0$}
	\end{subfigure}
    \begin{subfigure}[t]{\threewide\textwidth}
    \centering
        \begin{tikzpicture}
    	\node[anchor=south west,inner sep=0] (image) at (0,0) {\input{ppUBumpHeightp15.tex}\includegraphics[width=\textwidth]{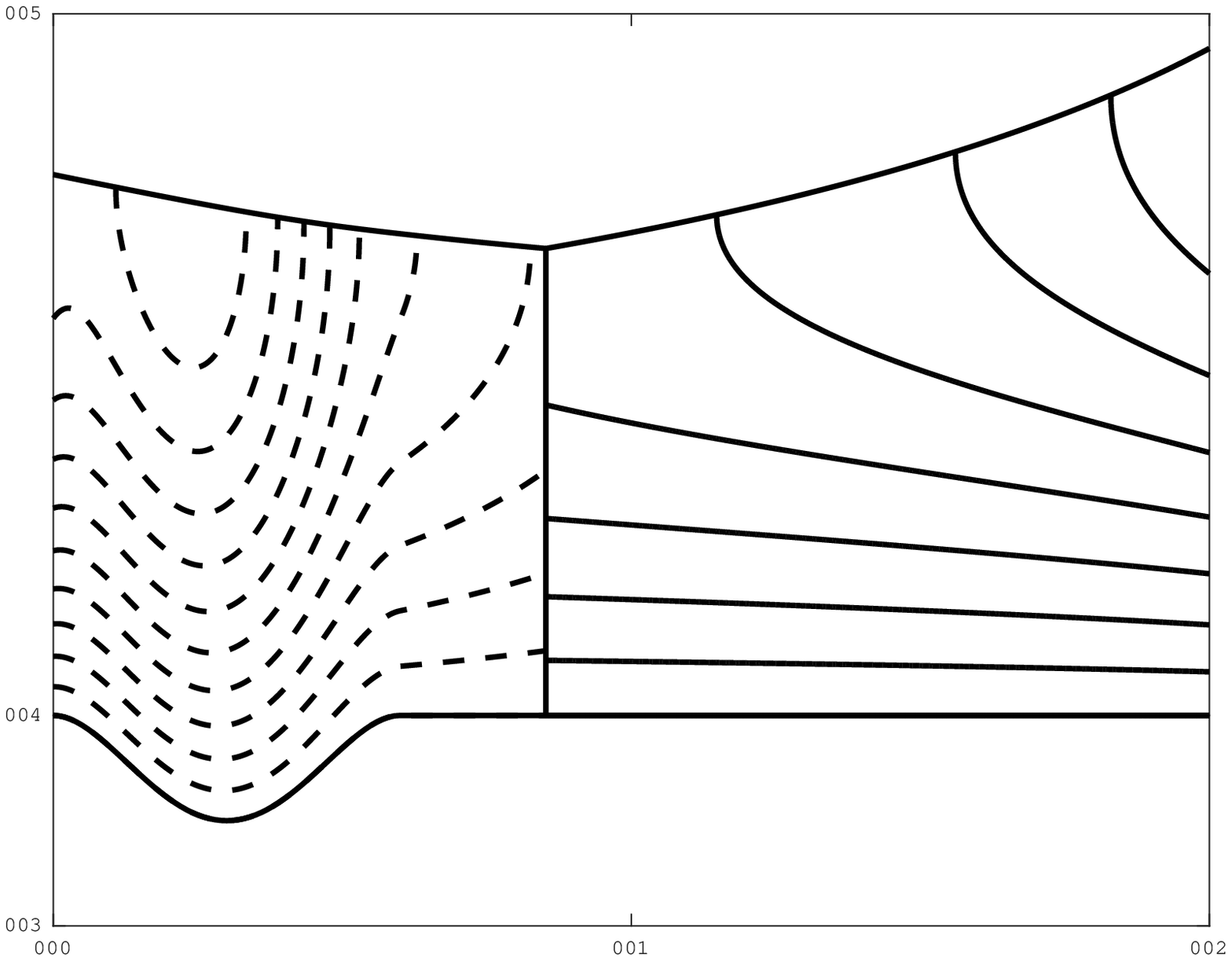}};
		\begin{scope}[x={(image.south east)},y={(image.north west)}]
			\axislabels
    	\end{scope}
	\end{tikzpicture}
	\caption{$k=0.15$}\label{fig:trench1u}
	\end{subfigure}
    \begin{subfigure}[t]{\threewide\textwidth}
    \centering
        \begin{tikzpicture}
    	\node[anchor=south west,inner sep=0] (image) at (0,0) {\input{ppUBumpHeightp3.tex}\includegraphics[width=\textwidth]{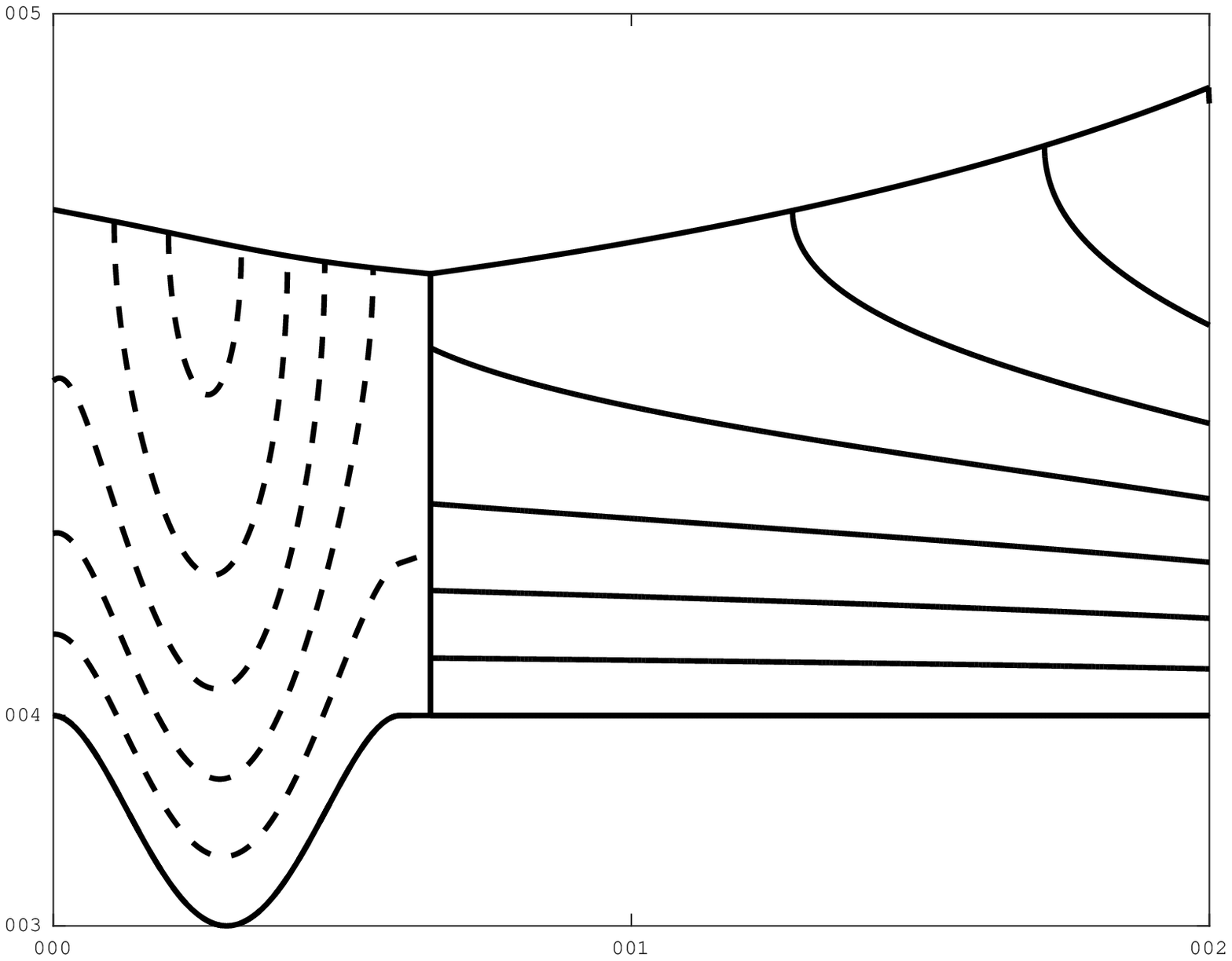}};
		\begin{scope}[x={(image.south east)},y={(image.north west)}]
			\axislabels
    	\end{scope}
	\end{tikzpicture}
	\caption{$k=0.3$}\label{fig:trench2u}
	\end{subfigure}
	\caption{Rectangular channels with $\epsilon=0.3$, $\lambda=0.5$, $Q=0.184$ and $Q_p=0.0035$, with a trench near the inside wall of half-width $b=0.3$ and depth $k$ as indicated. \bl{The top row shows streamlines of the flow in the particle-rich (dashed lines) and particle-free (solid lines) regions, and the bottom row shows contours of the axial velocity. \ys{The scaling of the streamlines and axial-flow contours differs between the regions, with the magnitude of the flow velocity being significantly smaller in the particle-rich region.}}}\label{fig:bumpsDepth}
\end{figure}

Motivated by spiral particle separators, which often feature a trench near the inside wall, we consider a rectangular channel with a trench. The trench is a quartic polynomial that joins smoothly to the flat channel bottom, with maximum depth $k$ and half-width $b$. The equation of the channel bottom is,
\begin{equation}
H(y)=\begin{cases}-kb^{-4}\left(y+1-2b\right)^2\left(y+1\right)^2,\quad &-1\leq y\leq -1+2b,\\0,&-1+2b<y\leq 1,\end{cases}
\end{equation}
and the area of the trench is $(16/15)kb$.
\Cref{fig:bumpsDepth} shows the results for such channels with $\epsilon=0.3$, $\lambda=0.5$, $b=0.3$, $Q=0.184$, $Q_p=0.0035$ and $k=0,0.15,0.3$. The particle-rich region contracts as the trench deepens, and the effect is significant. For the no-trench case, the particle-rich region extends over half the width of the channel (up to $y=0$), but with a trench of depth 0.3 and width 0.6, the interface between the particle-rich and particle-free regions is at $y_r^*\approx-0.35$, \textit{i.e.} particles occupy roughly one third of the width of the channel. The free-surface profile does not change significantly as the trench depth increases, (as seen for clear-fluid flows in \citet{ASG2016}), but the trench means there is more cross-sectional area near the inside wall, and so more particles can fit within a certain distance from the inside wall of the channel. \bl{Since the depth is increased over the trench, the axial velocity (also shown in \cref{fig:bumpsDepth}) is increased \ys{in this region} and, hence, the flux of particles there is magnified. This effect is important in decreasing $y_r^*$ and, hence, the width of the particle-rich region.} The trench has little effect on the secondary flow in the particle-free region, but does entrain a rotating cell of particle-rich fluid. Whilst we expect boundary layers near the vertical channel walls which would cause all the streamlines to connect up and form closed curves (discussed in detail in \citet{ASG2015}), the trench moves the centre of rotation radially outwards from the inside wall.

\begin{figure}
    \centering
	\begin{subfigure}[t]{\threewide\textwidth}
    \centering
        \begin{tikzpicture}
    	\node[anchor=south west,inner sep=0] (image) at (0,0) {\input{ppBumpWidthp15.tex}\includegraphics[width=\textwidth]{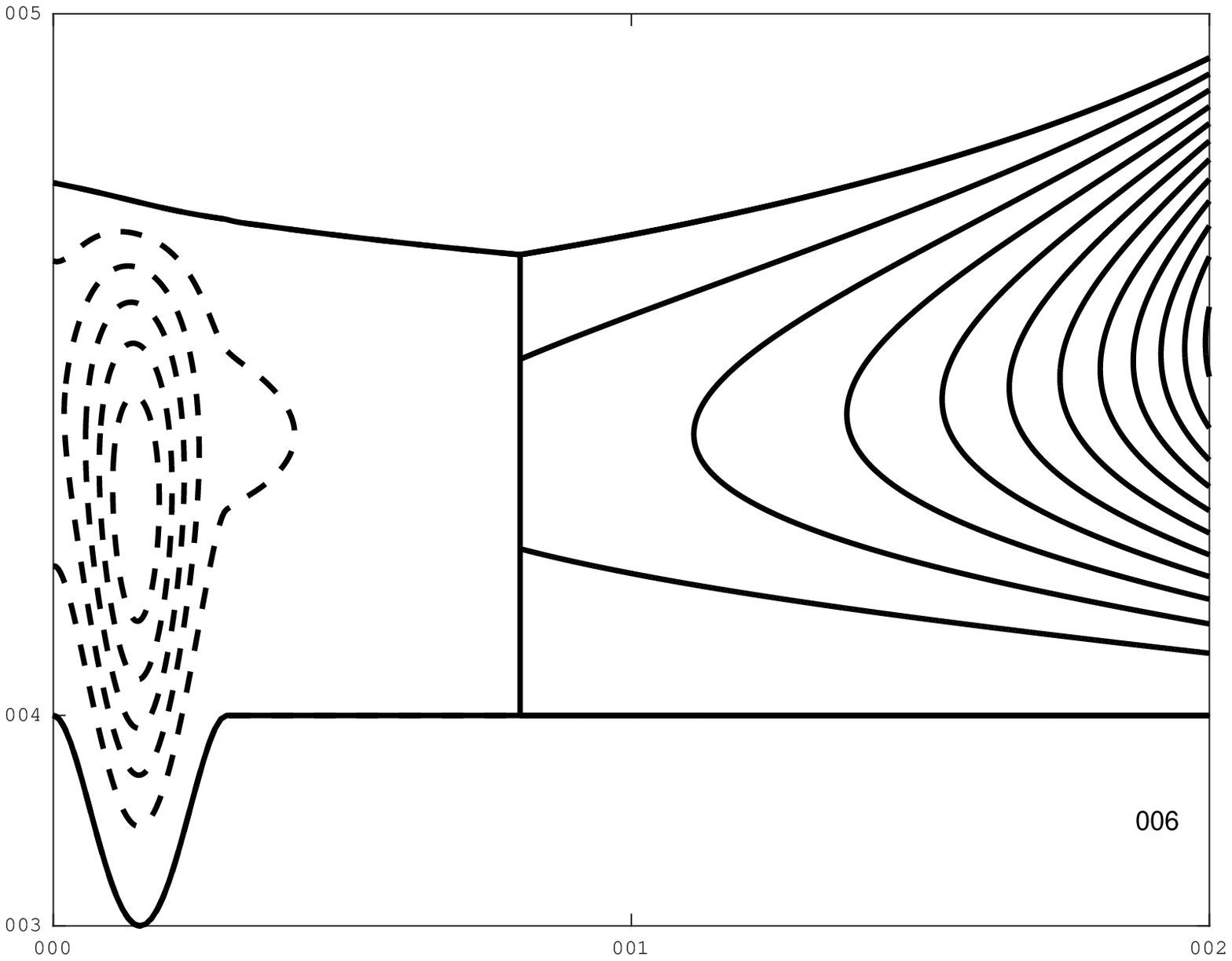}};
		\begin{scope}[x={(image.south east)},y={(image.north west)}]
			\axislabels
    	\end{scope}
	\end{tikzpicture}
	\caption{$b=0.15$}
	\end{subfigure}
    \begin{subfigure}[t]{\threewide\textwidth}
    \centering
        \begin{tikzpicture}
    	\node[anchor=south west,inner sep=0] (image) at (0,0) {\input{ppBumpWidthp3.tex}\includegraphics[width=\textwidth]{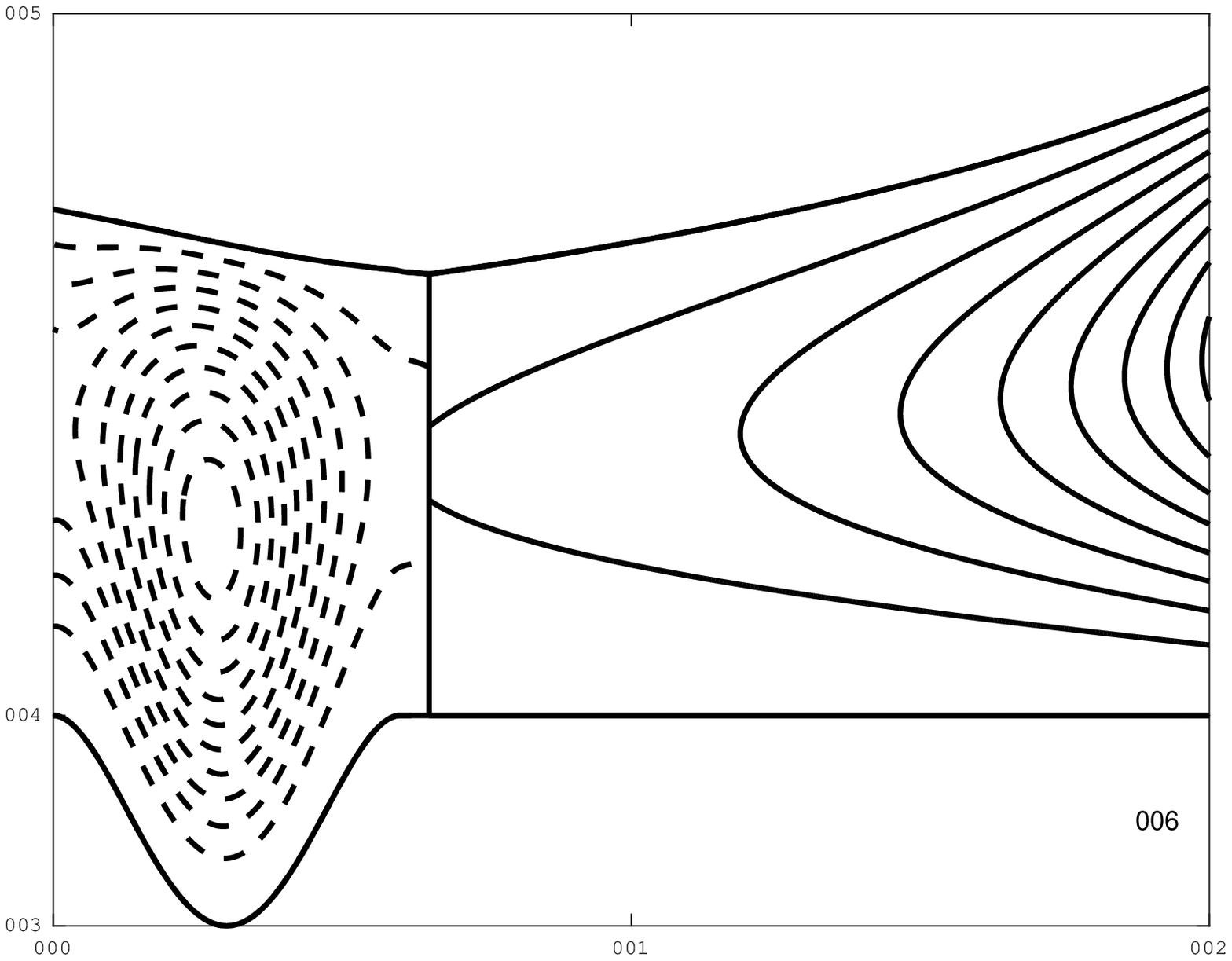}};
		\begin{scope}[x={(image.south east)},y={(image.north west)}]
			\axislabels
    	\end{scope}
	\end{tikzpicture}
	\caption{$b=0.3$}
	\end{subfigure}
    \begin{subfigure}[t]{\threewide\textwidth}
    \centering
        \begin{tikzpicture}
    	\node[anchor=south west,inner sep=0] (image) at (0,0) {\input{ppBumpWidthp45.tex}\includegraphics[width=\textwidth]{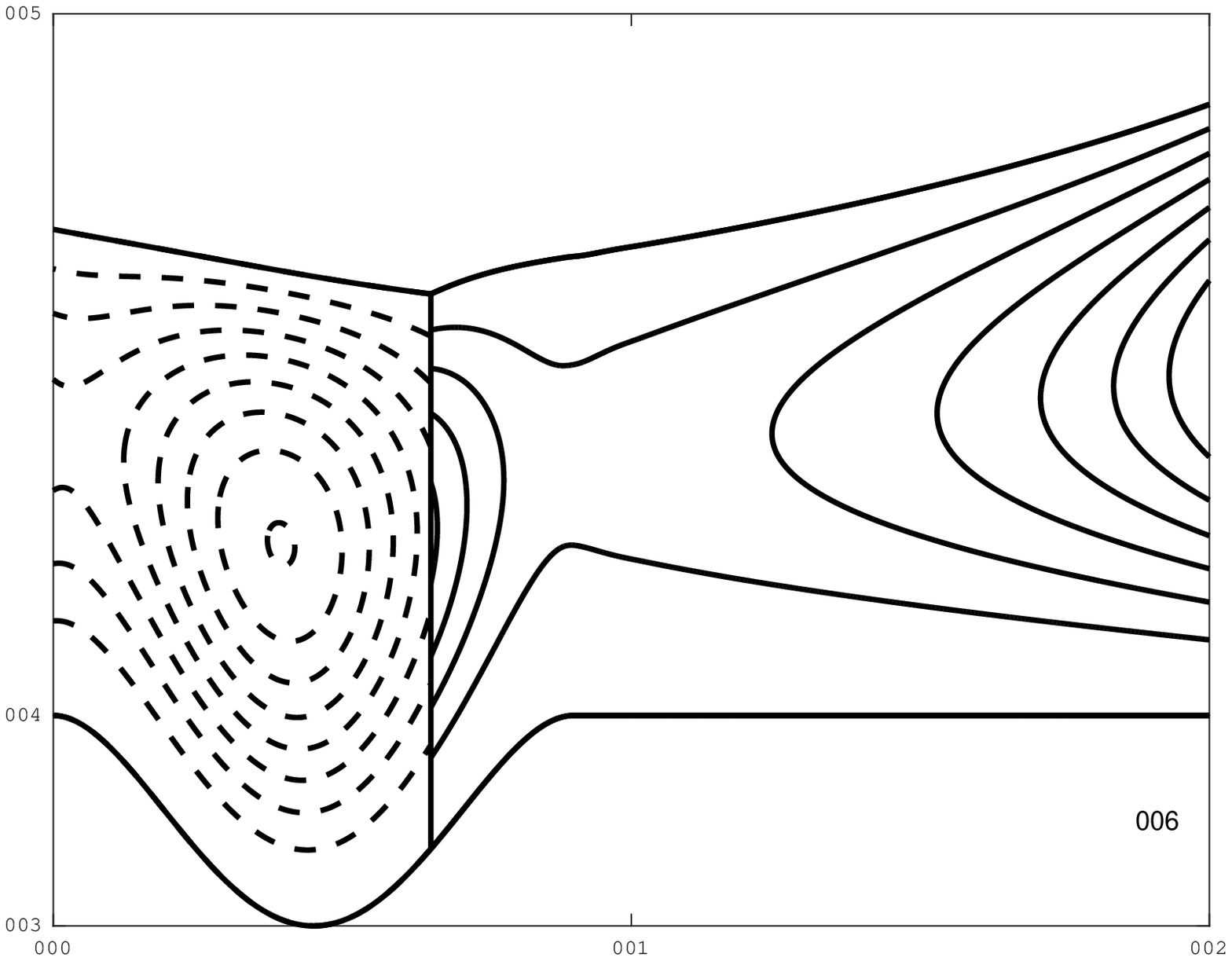}};
		\begin{scope}[x={(image.south east)},y={(image.north west)}]
			\axislabels
    	\end{scope}
	\end{tikzpicture}
	\caption{$b=0.45$}
	\label{fig:bigtrench}
	\end{subfigure}
    \caption{Rectangular channels with $\epsilon=0.3$, $\lambda=0.5$, $Q=0.184$ and $Q_p=0.0035$, with a trench near the inside wall of depth $k=0.3$ and half-width $b$ as indicated. The trenches in (a) and (b) have the same area as the trenches in \cref{fig:trench1,fig:trench2}, respectively, \ys{while that in (c) is larger. The scaling of the streamlines differs between the regions, with the magnitude of the flow velocity being significantly smaller in the particle-rich region.}}\label{fig:bumpsWidth}
\end{figure}

\Cref{fig:bumpsWidth} shows that changing the width of the trench \ys{while holding the depth constant} has a similar effect to increasing the depth \ys{while holding the width constant, namely reducing the extent of particles $y_r^*$ compared to the rectangular channel with no trench}. The trench in \cref{fig:bigtrench} \ys{has a larger area than any in \cref{fig:bumpsDepth} and} is large enough that the particle-rich region does not fill the trench. 
It is possible to create a very wide trench which is mainly filled with particle-free fluid. This is a poor configuration as the benefit 
of the trench \ys{in focussing the particles is reduced once the deepest part of the trench falls} 
outside the particle-rich region. For a given fluid and particle flux, the optimal trench width appears to be the width such that the edge of the trench roughly coincides with $y_r^*$, the edge of the particle-rich region.

\begin{figure}
\centering
\begin{tikzpicture}
    	\node[anchor=south west,inner sep=0] (image) at (0,0) {\input{ppTrenchWidth.tex}\includegraphics[width=\twowide\textwidth]{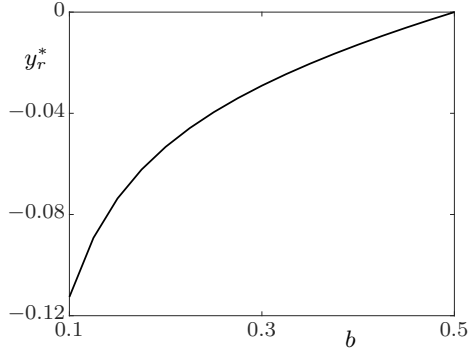}};
		\begin{scope}[x={(image.south east)},y={(image.north west)}]
			\node [align=center] at (0.08,0.8) {$y_r^*$};
			\node at (0.7,0.05) {$b$};
		\end{scope}
	\end{tikzpicture}
\caption{\ys{For} a fixed trench area, increasing the half-width $b$ \ys{(and decreasing the depth $k$)} increases the radial extent \ys{$y_r^*$} of the particle-rich fluid.}
\label{fig:trenchwidth}
\end{figure}

The channels in \cref{fig:bumpsDepth} \ys{and \ref{fig:bumpsWidth} feature} trenches with different areas; comparing trenches with the same area is also interesting. \Cref{fig:trenchwidth} shows the effect of changing the width of the trench on $y_r^*$, whilst keeping the total area of the trench constant. It is clear that decreasing the width of the trench (and hence increasing the depth) decreases $y_r^*$ and, hence, focusses the particles more effectively.

The specific shape of the trench is not of critical importance for particle focussing. A similar trench using the cosine function was tested, with very similar results. The results obtained when adding a trench to the parabolic channel were also qualitatively similar, suggesting that the effect of adding a trench to a channel doesn't depend significantly on the shape of the channel, and the mechanism of increased depth allowing increased axial velocity adequately explains the results.

\subsection{The effect of particle density\label{sec:density}}
\begin{figure}
    \centering
    \begin{subfigure}[t]{\threewide\textwidth}
    \centering
        \begin{tikzpicture}
    	\node[anchor=south west,inner sep=0] (image) at (0,0) {\input{ppRectangleRhos1p2.tex}\includegraphics[width=\textwidth]{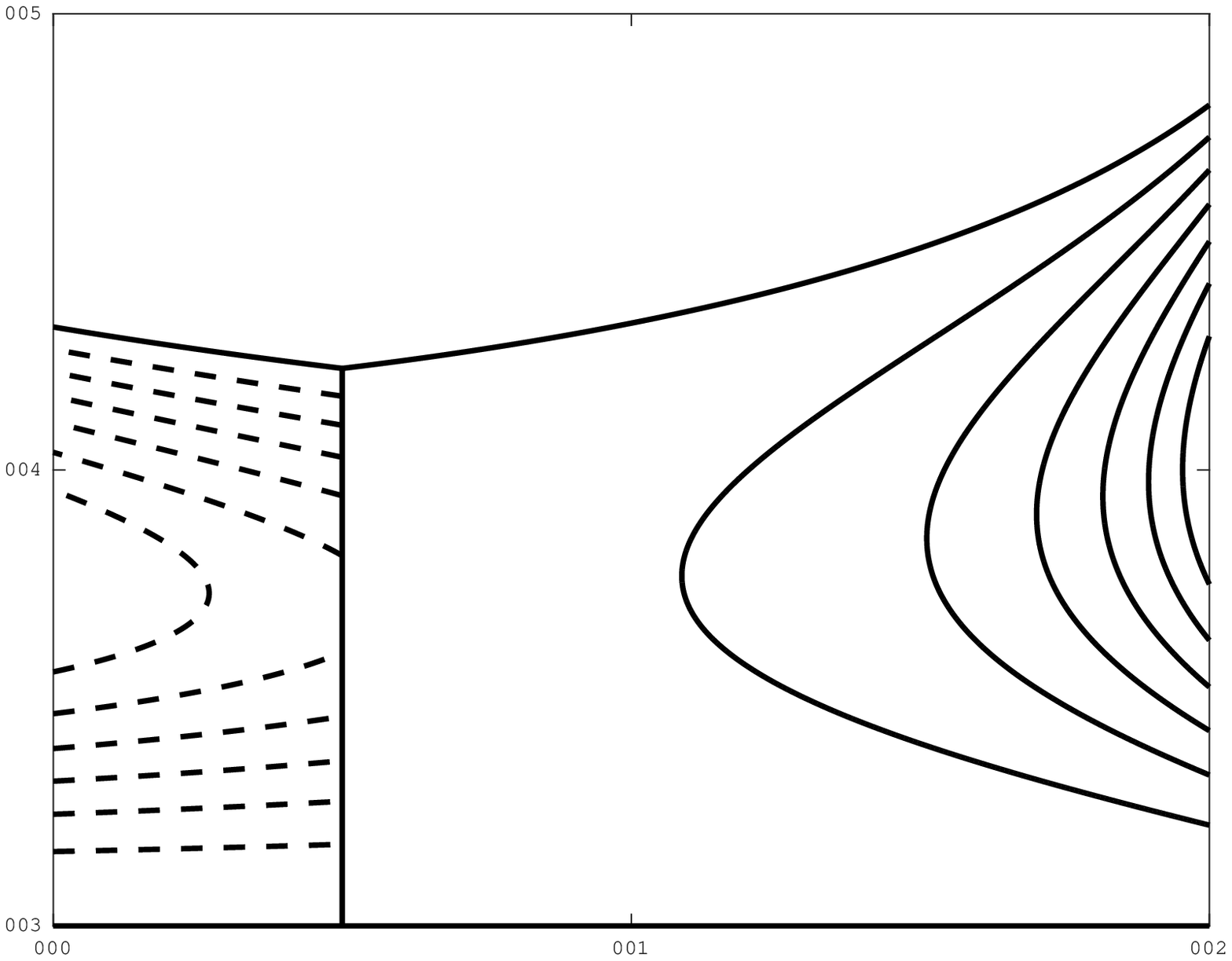}};
		\begin{scope}[x={(image.south east)},y={(image.north west)}]
			\axislabels
    	\end{scope}
	\end{tikzpicture}
	\caption{$\rho_s=1.2$}
	\end{subfigure}
	\begin{subfigure}[t]{\threewide\textwidth}
    \centering
        \begin{tikzpicture}
    	\node[anchor=south west,inner sep=0] (image) at (0,0) {\input{ppRectangleRhos2.tex}\includegraphics[width=\textwidth]{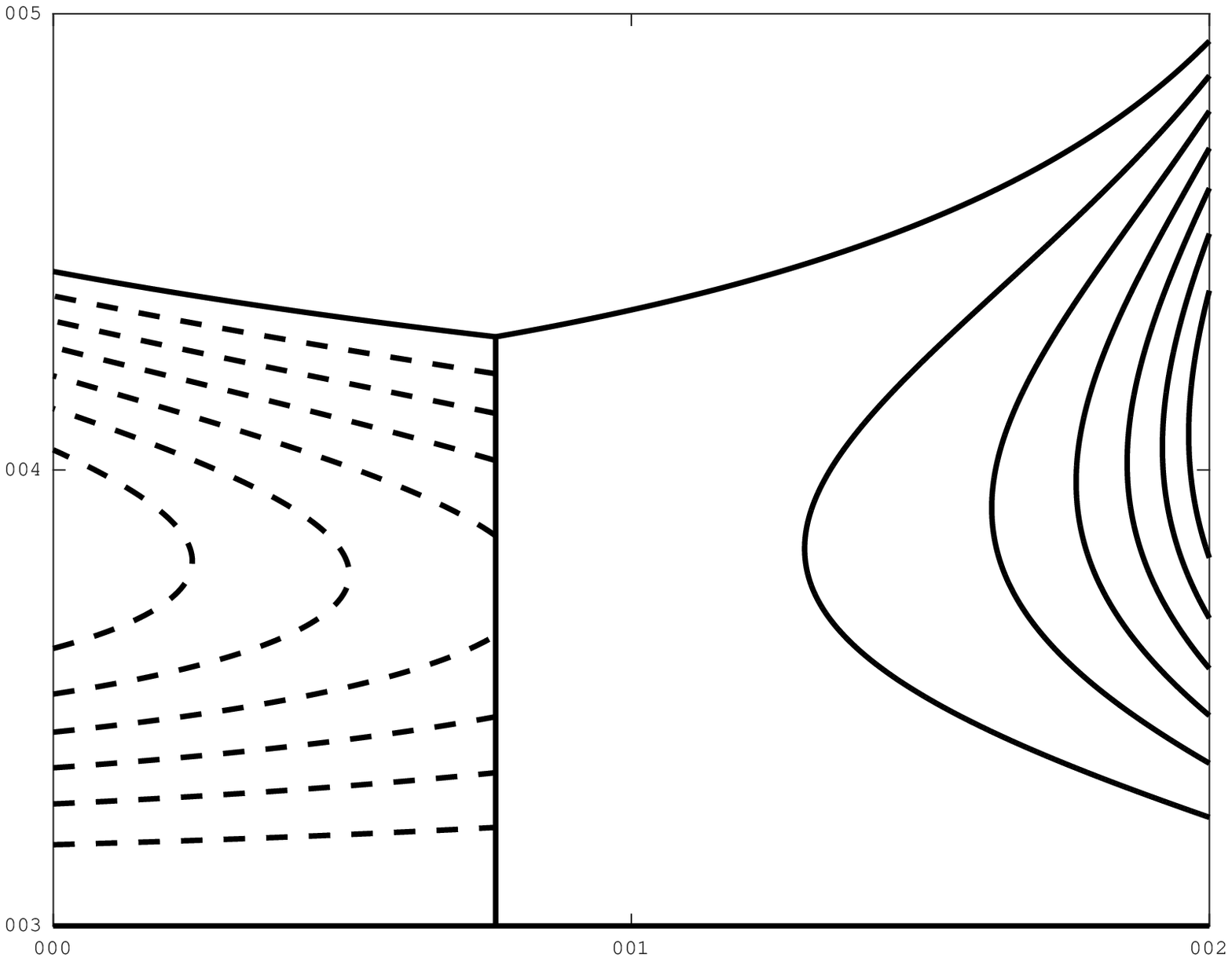}};
		\begin{scope}[x={(image.south east)},y={(image.north west)}]
			\axislabels
    	\end{scope}
	\end{tikzpicture}
	\caption{$\rho_s=2$}
\end{subfigure}
\begin{subfigure}[t]{\threewide\textwidth}
    \centering
        \begin{tikzpicture}
    	\node[anchor=south west,inner sep=0] (image) at (0,0) {\input{ppRectangleRhos2p8.tex}\includegraphics[width=\textwidth]{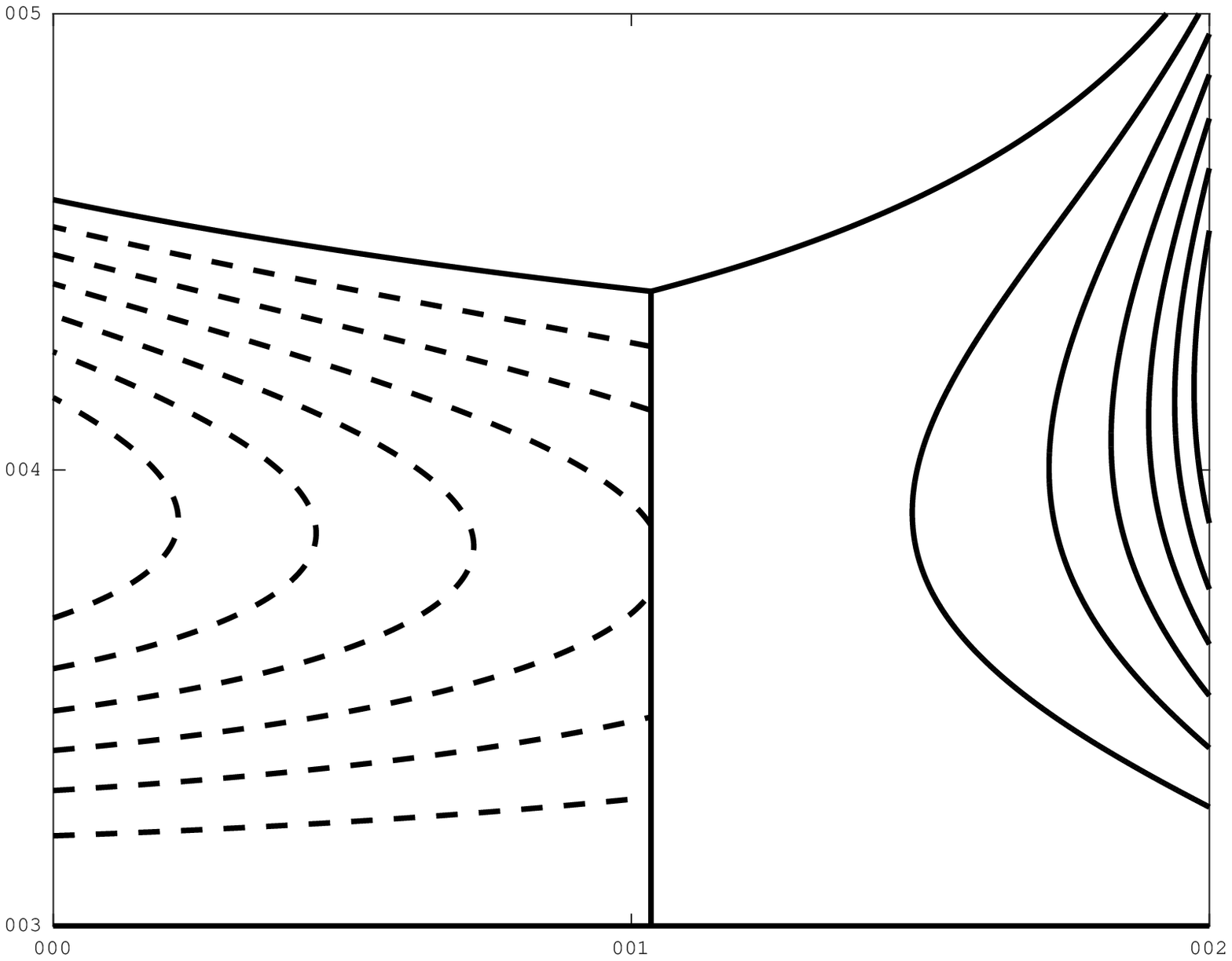}};
		\begin{scope}[x={(image.south east)},y={(image.north west)}]
			\axislabels
    	\end{scope}
	\end{tikzpicture}
	\caption{$\rho_s=2.8$}
	\end{subfigure}
    \caption{Rectangular channels with $\epsilon=0.2$, $\lambda=0.7$, $Q=0.192$ and $Q_p=0.0018$ carrying particles with densities $\rho_s$ as shown.}\label{fig:RectangleDensity}
\end{figure}

\Cref{fig:RectangleDensity} shows a rectangular channel, with $\epsilon=0.2$, $\lambda=0.7$, $Q=0.192$ and $Q_p=0.0018$, carrying particles of increasing density. We first note that $\rho_s$ is not the actual particle density, but is defined in terms of the particle and fluid densities as $\rho_s=(\rho_p-\rho_f)/\rho_f$. The effective nondimensional mixture density is $\rho=1+\phi\rho_s$, so increasing $\rho_s$ increases the density of the mixture for a given concentration $\phi$.

As $\rho_s$ increases, the volume fraction $\phi(y)$ that leads to a well-mixed regime, given by \cref{eq:phiprofile}, increases. \ys{Both $\rho$ and $\mu$ (see \eqref{eq:mu}) increase with $\phi$, with $\mu$ increasing very rapidly as $\phi\rightarrow\phi_m$.} The axial flow velocity $u$ (\cref{eq:u}) includes a factor $\rho/\mu$, and if this factor decreases, the axial velocity will also decrease. For the full range of parameters considered in this section ($\epsilon$ and $\lambda$ restricted so that $\phi(y)<\phi_m$ everywhere in the channel), \ys{$\mu$ grows more rapidly that $\rho$ and} $\rho/\mu$ decreases as $\rho_s$ increases. If the axial velocity decreases the particle-rich region, \ys{hence $y_r^*$,} must grow to accommodate the same particle flux down the channel. This is shown in \cref{fig:RectangleDensity}.

The solutions shown in \cref{fig:RectangleDensity} are quite sensitive to $\rho_s$. As $\rho_s$ increases, the region in the $\epsilon$--$\lambda$ plane where $\phi(y)<\phi_m$ everywhere in the channel gets smaller, so that the choice of centreline geometry becomes increasingly restricted. In these cases, $\phi(y)$ is relatively large throughout the \ys{particle-rich region of the} channel, so we are close to the limit of validity of the model equations \ys{where $\mu\rightarrow\infty$ as $\phi\rightarrow\phi_m$ and the Maron-Pierce equation \eqref{eq:mu} ceases to be a valid representation of the viscosity. As discussed earlier we interpret this to imply beaching and a channel geometry that is not fit for purpose, from which we conclude that our model is only valid for particles with a density $\rho_p$ that is not too much larger than the fluid density $\rho_f$.} 

\section{Conclusions}\label{sec:conc}
We have derived a system of equations \eqref{eq:NSFr}--\eqref{eq:NSFz} governing particle laden flow in helically-wound channels with arbitrary centreline pitch and radius, and arbitrary shallow cross-sectional shape. The thin-film scaling was used to give a much simpler system of equations \eqref{eq:nsR}--\eqref{eq:nsZ}, although these were still too complicated to be solved analytically. Motivated by \ys{available} experimental results, \ys{the specific spiral particle separator application which relies on a radial separation of particles of different densities,} and a previous study \citep{LSB2013}, we restricted our attention to finding solutions with uniform particle volume fraction in the vertical direction. This assumption allowed significant analytic progress to be made, with only the free-surface differential equation requiring numerical solution. \ys{The results we have obtained are entirely consistent with current knowledge on spiral particle separators.}

The solution to the simplified system of equations provides insights into the effect of particles on the fluid flow. The basic features of flows in helical channels that were observed and discussed in \citet{ASG2015,ASG2016} still apply to particle-laden flows, but the change in the density and viscosity of the fluid (due to the presence of particles) changes the balance between gravitational and inertial effects. Increasing $\epsilon$ or $\lambda$ tends to push the fluid closer to the inside wall of the channel, and vice versa.

\ys{We have considered two basic channel cross-sections, rectangular and parabolic. The first is the simplest case that has vertical side walls at which the fluid depth must be determined for given total and particle fluxes, and the second is the simplest case that requires determination of the horizontal extent of the fluid by finding the contact points of the free surface at which the fluid depth is zero. At a vertical side wall the asymptotic model requires zero net flux through the wall but the no-slip boundary conditions cannot be enforced and, in practice, there is a boundary layer near to the wall that is not resolved by the model. For channels with no vertical walls, so that the fluid depth is zero at points of contact of the free surface with the channel bottom, the no-slip boundary conditions are enforced along the entire channel surface. Similar qualitative behaviour was found for rectangular and parabolic cross-sections. We have also shown the effect of a trench in the bottom of an otherwise rectangular channel and examined the effect of trench depth and width. Although not shown, the effect of a trench in an otherwise parabolic cross-section is similar to that seen in the rectangular case.}

\ys{Our model provides insight into the radial distribution of particles, in terms of 
a depth-independent particle volume fraction, and the focussing of the particles within the channel cross-section.} For a given flux of particles and total mixture flux, \ys{we have shown that} the particle-rich region can be focussed into a smaller region of the channel cross-section by changing the channel centreline pitch and radius. The particle rich region can also be focussed by introducing a trench near the inside wall of the channel, essentially increasing the cross-sectional area there. Commercially available spiral particle separators often feature deeper trenches near their inside wall, and our modelling is able to show the benefit in reducing $y_r^*$ that arises from them. A \ys{reported} 
benefit of the trenches in commercial separators is that particles settle into them and are trapped there, however our  assumption that particles are evenly distributed in the vertical direction means we cannot investigate this phenomenon. Our results were relatively sensitive to particle density, something that is also seen in practice, as spiral particle separators are known to be sensitive to the composition of the input slurry.

There is significant scope for further research into flows of particle laden fluids in helical channels. It is not certain that the particle volume fraction in spiral particle separators is independent of depth, and relaxing this assumption in our model would give more widely-applicable results. Numerical solutions of the thin-film system of equations \eqref{eq:nsR}--\eqref{eq:cont}, with $\phi=\phi(y,z)$ would allow a wider range of channel centreline geometries and flow regimes to be studied. Non-helically-symmetric simulations of the thin-film Navier-Stokes equations in the spirit of \citet{Murisicetal2013} could provide significant insight into particle and fluid motion in spiral separators, for example by studying the transition to equilibrium configurations. 

As mentioned in \cref{sec:density}, the particle volume fraction can become very large, and the Maron-Pierce equation for the viscosity may not be realistic. The very large viscosity it predicts could potentially affect the asymptotic expansion of the governing equations. The model used in this chapter treated the slurry as a Newtonian fluid, and at very high particle volume fractions, this is not necessarily the case, and the particle-rich fluid may be better modelled as non-Newtonian or even as a granular material. \ys{Nevertheless, very large particle volume fraction is likely to be associated with beaching and consequently indicates that the spiral separator design is not appropriate to the given feed. Thus our model may be a useful design tool for determining channel geometries that minimise or avoid occurrences of beaching.}

The shear-induced migration model used in this paper is just one particle-laden flow model that can be considered. Different models (such as the suspension balance model of \citet{NB1994}) could be investigated to see whether they would predict different behaviour. Our model is for monodisperse slurries, with a single particle species, however spiral particle separators are intended to separate different species of particles. Models for flows of slurries consisting of two or more species of particles are less well understood than monodisperse slurries, and there is significant scope to improve modelling of such complex mixtures. Combining a recent model for bidensity suspensions \citep{WB2016} with our equations for the helical geometry \ys{is seen as a useful next step towards understanding particle separation in curved channels.} 

\acknowledgments{
We thank B. Harding and J. T. Wong for helpful discussions about some technical details of the model. We gratefully acknowledge funding from an Australian Postgraduate Award to DJA, an Australian Research Council Discovery Early Career Researcher Award (DE130100031) to JEFG and an Australian Research Council Discovery Project (DP160102021) to YMS.
}

\appendix
\section{Governing equations for particle-laden flow in a helical channel}\label{sec:fullparticleequations}
We here give the model equations in our helicoidal coordinate system, derived as in \citet{ASG2016}. We consider steady, helically-symmetric flow, and so quantities are independent of the angular coordinate $\beta$ and time. The components of the velocity vector $\vec v$ are $\vec v=\left(v^r,v^\beta,v^z\right)$, and similarly, the components of the particle-flux vector are $\vec J=\left(J^r,0,J^z\right)$. There is no $J^\beta$ component because it would come from a derivative with respect to $\beta$, and is therefore zero by helical symmetry. \Cref{eq:divvj} becomes,
\begin{equation} \pd{v^r}{r}+\pd{v^z}{z}+\frac{v^r}{r}+\pd{J^r}{r}+\pd{J^z}{z}+\frac{J^r}{r}=0,\label{eq:co1} \end{equation}
and \cref{eq:partrans} yields,
\begin{equation} v^r\pd{\phi}{r}+v^z\pd{\phi}{z}+(\phi-1)\left(\pd{v^r}{r}+\pd{v^z}{z}+\frac{v^r}{r}\right)=0.\label{eq:co2}
\end{equation}

Expressions for the differential operators appearing in \cref{eq:pNS} were found for the constant viscosity and density case in \citet{ASG2016} but there are two significant changes. Firstly, we cannot use $\nabla\cdot\vec v=0$ to simplify the divergence of the rate of strain tensor, and secondly, there are new terms arising from $\nabla\mu\cdot\mathcal S$, which is now non-zero. Hence the Navier-Stokes equations are, in the radial direction,
\begin{align}
&\rho\left[v^{r}\pd{v^r}{r}+v^{z}\pd{v^{r}}{z}-rv^{\beta}v^{\beta}\right] \nonumber\\
&=-\pd{p}{r}+H'(r)\pd{p}{z}+\mu\left[2\pd{^{2}v^{r}}{r^{2}}+\frac{2}{r}\pd{v^{r}}{r}-\left\{\frac{2}{r}H'(r)+H''(r)\right\}\pd{v^r}{z}\right. \nonumber\\
&\left.-3H'(r)\pd{^2v^r}{z\partial r}+\pd{^{2}v^{z}}{r\partial z}+\Phi\pd{^{2}v^{r}}{z^{2}}-H'(r)\pd{^2v^z}{z^2}+\frac{2P}{r}\pd{v^{\beta}}{z}-\frac{2}{r^{2}}v^{r}\right]\nonumber\\
&+\pd{\mu}{r}\left(2\pd{v^{r}}{r}-2H'(r)\pd{v^r}{z}\right)\nonumber\\
&+\pd{\mu}{z}\left(-H'(r)\pd{v^r}{r}+\Phi\pd{v^r}{z}+\pd{v^{z}}{r}-H'(r)\pd{v^z}{z}+H''(r)v^r\right),\label{eq:NSFr}
\end{align}
in the axial direction,
\begin{align}
&\rho\left[v^{r}\pd{v^{\beta}}{r}+v^{z}\pd{v^{\beta}}{z}+\frac{2}{r}v^{r}v^{\beta}\right] \nonumber\\
&=\frac{P}{r^{2}}\pd{p}{z}+\mu\left[\pd{^{2}v^{\beta}}{r^{2}}-\left\{H''(r)+\frac{3}{r}H'(r)\right\}\pd{v^\beta}{z}-2H'(r)\pd{^2v^\beta}{z\partial r}\right.\nonumber\\
&\left.-\frac{P}{r^{2}}\pd{^{2}v^{r}}{z\partial r}+\frac{2P}{r^{3}}\pd{v^{r}}{z}-\frac{P}{r^{2}}\pd{^{2}v^{z}}{z^{2}}+\Phi\pd{^{2}v^{\beta}}{z^{2}}-\frac{2P}{r^{3}}\pd{v^{r}}{z}+\frac{3}{r}\pd{v^{\beta}}{r}-\frac{3P}{r^{3}}\pd{v^{r}}{z}\right] \nonumber\\
&+\pd{\mu}{r}\left(\pd{v^{\beta}}{r}-H'(r)\pd{v^\beta}{z}-\frac{P}{r^2}\pd{v^r}{z}\right)\nonumber\\
&+\pd{\mu}{z}\left(-H'(r)\pd{v^\beta}{r}+\Phi\pd{v^\beta}{z}-\frac{2P}{r^{3}}v^{r}-\frac{P}{r^{2}}\pd{v^{z}}{z}\right),\label{eq:NSFb}
\end{align}
and in the vertical direction,
\begin{align}
&\rho\left[v^{r}\pd{v^{z}}{r}+v^{z}\pd{v^{z}}{z}+rH'(r)v^\beta v^\beta+H''(r)v^rv^r-\frac{2P}{r}v^{\beta}v^{r}\right]\nonumber\\
&=H'(r)\pd{p}{r}-\Phi\pd{p}{z}+\mu\left[\left\{2H''(r)-\frac{H'(r)}{r}\right\}\pd{v^r}{r}-H'(r)\pd{^2v^r}{r^2}+\Phi\pd{^{2}v^{r}}{z\partial r}\right.\nonumber\\
&\left.+\left\{H'''(r)+\frac{H''(r)}{r}+\frac{2H'(r)}{r^2}\right\}v^r+\left\{\frac{2P^{2}}{r^{3}}+\frac{\Phi}{r}-2H''(r)H'(r)\right\}\pd{v^{r}}{z}\right.\nonumber\\
&\left.+\pd{^{2}v^{z}}{r^{2}}-\left\{H''(r)+\frac{H'(r)}{r}\right\}\pd{v^z}{z}-3H'(r)\pd{^2v^z}{z\partial r}+2\Phi\pd{^{2}v^{z}}{z^{2}}+\frac{1}{r}\pd{v^{z}}{r}-\frac{2P}{r}\pd{v^{\beta}}{r}\right]\nonumber\\
&+\pd{\mu}{r}\left(-H'(r)\pd{v^r}{r}+\Phi\pd{v^r}{z}+\pd{v^{z}}{r}-H'(r)\pd{v^z}{z}+H''(r)v^r\right)\nonumber\\
&+\pd{\mu}{z}\left(\frac{2P^{2}}{r^{3}} v^{r}+2\Phi\pd{v^{z}}{z}-2H'(r)\pd{v^z}{r}-2H''(r)H'(r)v^r\right)-\rho g.\label{eq:NSFz}
\end{align}

The no-slip boundary condition is,
\begin{equation}
v^r=v^\beta=v^z=0,\quad\text{at $z=0$.}\label{eq:ans}
\end{equation}
The free surface can be written as the solution to $F(r,z)=0$, and a normal to the free-surface is given by $\vec n=\nabla F=\left(F_{,r},0,F_{,z}\right)$. The no-stress condition at the free-surface gives,

\begin{align}
0=&F_{,r}\left(2\mu \pd{v^r}{r}-2\mu H'(r)\pd{v^r}{z}-p\right)\nonumber\\
&+F_{,z}\left(\mu\left[-H'(r)\pd{v^r}{r}+\pd{v^z}{r}+\Phi \pd{v^r}{z}-H'(r)\pd{v^z}{z}+H''(r)v^r\right]+H'(r)p\right),
\end{align}
\begin{align}
0=&\mu F_{,r}\left(\pd{v^\beta}{r}-H'(r)\pd{v^\beta}{z}-\frac{\Lambda}{r}\pd{v^r}{z}\right)\nonumber\\
&+F_{,z}\left(\mu\left[-H'(r)\pd{v^\beta}{z}+\Phi\pd{v^\beta}{z}-\frac{2\Lambda}{r^2}v^r-\frac{\Lambda}{r}\pd{v^z}{z}\right]+\frac{\Lambda}{r}p\right),
\end{align}
\begin{align}
0=&F_{,r}\left(\mu\left[-H'(r)\pd{v^r}{r}+\pd{v^z}{r}+\Phi \pd{v^r}{z}-H'(r)\pd{v^z}{z}+H''(r)v^r\right]+H'(r)p\right)\nonumber\\
&+F_{,z}\left(\mu\left[\frac{2\Lambda^2}{r}v^r+2\Phi \pd{v^z}{z}-2H'(r)\pd{v^z}{r}-2H''(r)H'(r)v^r\right]-\Phi p\right).
\end{align}\label{eqs:PARTnossslip}

The kinematic condition is $\vec v\cdot\vec n=0$ at the free surface $F(r,z)=0$, and gives
\begin{equation}
v^rF_{,r}+v^zF_{,z}=0.\label{eq:PARTkinematic}
\end{equation}

\section{Notation}\label{notation}
\setlength{\tabcolsep}{8pt}
\begin{tabular}{rp{0.4\textwidth}rp{0.4\textwidth}}
$A$ & radius of channel centreline &
$\vec x(r,\beta,z)$ & position vector \\
$a$ & channel half-width &
$y$ & scaled radial coordinate \\
$b$ & trench half-width &
$y_b$/$y_t$ & maximum limits of particle-rich region \\
$d$ & particle diameter &
$y_l$/$y_r$ & inner/outer extents of slurry \\
$\Fr$ & Froude number &
$y_l^*$/$y_r^*$ & inner/outer extents of particle-rich region \\
$H(r)$ & channel cross-sectional shape &
$z$ & vertical coordinate \\
$h(r)$ & fluid depth &
$\beta$ & angular coordinate \\
$h_l$/$h_r$ & depth at $y_l$/$y_r$ &
$\dot\gamma$ & shear rate \\
$\vec J$ & particle flux vector &
$\delta$ & fluid aspect ratio, thin-film parameter \\
$k$ & trench depth &
$\epsilon$ & centreline curvature parameter \\
$K_c$ & shear-induced migration coefficient & 
$\Lambda(r)$ & slope of channel bottom at radius $r$ \\
$K_v$ & viscosity-gradient diffusion coefficient &
$\lambda$ & slope of channel centreline \\
$P$ & pitch of helical centreline divided by $2\pi$ &
$\mu$ & slurry viscosity \\
$p$ & pressure &
$\mu_f$ & fluid viscosity \\
$Q$ & total flux &
$\rho$ & slurry density \\
$Q_p$ & flux of particles &
$\rho_f$ & density of clear fluid \\
$r$ & radial coordinate &
$\rho_p$ & density of particles \\
$\Re$ & Reynolds number &
$\rho_s$ & dimensionless density parameter \\
$U$ & axial velocity scale &
$\Upsilon$ & $1+\Lambda^2$ \\
$u$ & axial velocity &
$\phi$ & local particle volume fraction \\
$v$ & radial velocity &
$\phi_m$ & maximum particle volume fraction \\
$v^i$ & velocity component in $i$-direction &
$\omega$ & hindrance function \\
$w$ & vertical velocity & &
\end{tabular}

\section*{References}
\bibliographystyle{plainnat}

\bibliography{referencelist}

\end{document}